\documentclass[12pt,onecolumn,a4paper]{article}

\def\reference{\parskip 0pt\par\noindent\hangindent 0.5 truecm}

\usepackage{epsfig}

\newcommand{\affil}[1]{$^{\rm #1}$}

\newcommand{\micron}{\mbox{$\,\mu$m}}

\usepackage{epsfig}

\def\plotone#1{\centering \leavevmode
\epsfxsize=\textwidth \epsfbox{#1}}

\def\plottwo#1#2{\centering \leavevmode
\epsfxsize=.45\textwidth \epsfbox{#1} \hfil
\epsfxsize=.45\textwidth \epsfbox{#2}}

\date{} 
\title{\bf Science Programs for a 2\,m-class Telescope at Dome C, Antarctica: 
\\ PILOT, the Pathfinder for an International Large
Optical Telescope}
\author{
To appear in \\ {\it Publications of the Astronomical Society of Australia} \\
Submitted November 2004, Revised Version April 2005 \\
Accepted April 12, 2005 \\ ~ \\
{\it M.G. Burton\affil{A,Z}, J.S. Lawrence\affil{A}, 
M.C.B. Ashley\affil{A}, J.A. Bailey\affil{B,C},}\\
{\it C. Blake\affil{A}, T.R. Bedding\affil{D}, J. Bland-Hawthorn\affil{B}, I.A. Bond \affil{E},}\\
{\it K. Glazebrook\affil{F}, M.G. Hidas\affil{A}, G. Lewis\affil{D}, S.N. Longmore\affil{A},} \\
{\it S.T. Maddison\affil{G}, S. Mattila\affil{H}, V. Minier\affil{I}, S.D. Ryder\affil{B},}\\
{\it R. Sharp\affil{B}, C.H. Smith\affil{J}, J.W.V. Storey\affil{A}, C.G. Tinney\affil{B},}\\
{\it P. Tuthill\affil{D}, A.J. Walsh\affil{A}, W. Walsh\affil{A}, M. Whiting\affil{A},}\\ 
{\it T. Wong\affil{A,K}, D. Woods\affil{A}, P.C.M. Yock\affil{L}}\\\\
{\small
\affil{A}\,School of Physics, University of New South Wales, Sydney, 
           NSW 2052, Australia}\\
{\small
\affil{B}\,Anglo Australian Observatory, Epping, Australia}\\
{\small
\affil{C}\,Centre for Astrobiology, Macquarie University, Australia}\\
{\small
\affil{D}\,University of Sydney, Sydney, Australia}\\
{\small
\affil{E}\,Massey University, Auckland, New Zealand}\\
{\small
\affil{F}\,John Hopkins University, Baltimore, USA}\\
{\small
\affil{G}\,Swinburne University, Melbourne, Australia}\\
{\small
\affil{H}\, Stockholm Observatory, Stockholm, Sweden}\\
{\small
\affil{I}\, CEA Centre d'Etudes de Saclay, Paris, France}\\
{\small
\affil{J}\, Electro Optics Systems, Queanbeyan, Australia}\\
{\small
\affil{K}\, CSIRO Australia Telescope National Facility, Epping, Australia}\\
{\small
\affil{L}\, University of Auckland, Auckland, New Zealand}\\
{\small
\affil{Z}\,E-mail: M.Burton@unsw.edu.au}
}
\begin{document}
\maketitle
%
%
%
\begin{minipage}{.9\textwidth}
{\bf Abstract}\\
%
The cold, dry and stable air above the summits of the Antarctic
plateau provides the best ground-based observing conditions from
optical to sub-mm wavelengths to be found on the Earth.  PILOT is a
proposed 2\,m telescope, to be built at Dome C in Antarctica, able to
exploit these conditions for conducting astronomy at optical and
infrared wavelengths.  While PILOT is intended as a pathfinder towards
the construction of future grand-design facilities, it will also be
able to undertake a range of fundamental science investigations in its
own right.  This paper provides the performance specifications for
PILOT, including its instrumentation.  It then describes the kinds of
science projects that it could best conduct.  These range from
planetary science to the search for other solar systems, from star
formation within the Galaxy to the star formation history of the
Universe, and from gravitational lensing caused by exo-planets to that
produced by the cosmic web of dark matter.  PILOT would be
particularly powerful for wide-field imaging at infrared wavelengths,
achieving near-diffraction limited performance with simple tip-tilt
wavefront correction.  PILOT would also be capable of near-diffraction
limited performance in the optical wavebands, as well be able to open
new wavebands for regular ground based observation; in the mid-IR from
17 to 40\micron\ and in the sub-mm at 200\micron.

\medskip
\end{minipage}

%
%

\section{Introduction --- the Antarctic plateau}
The highest regions of the Antarctic plateau provide a unique
environment on the earth for conducting observational astronomy.  This
is because of the extreme cold, the dryness and the stability of the air
column above these locations --- leading to a lower sky background,
greater transparency and sharper imaging than at temperate sites.

The Antarctic high plateau includes an area about the size of
Australia, all of which is above 3,000\,m elevation.  With a
year-round average temperature of $-50^{\circ}$\,C, falling as low as
$-90^{\circ}$\,C at times, the sky thermal emission, which dominates
at wavelengths longer than 2.2\micron, is far less than at temperate
sites (at shorter infrared wavelengths the emissivity is primarily due
to OH airglow from the upper atmosphere).  A reduced concentration of
particulates in the atmosphere lowers the sky emissivity
(predominantly arising from dust and aerosols at temperate sites), further
lowering the background at these wavelengths.  Columns of precipitable
water vapour are less than 250\micron\ for much of the year, opening
atmospheric windows across the infrared and sub-millimetre bands.
Wind speeds are low at the summits of the plateau, with violent storms
non-existent.  The thinness of the surface inversion layer, combined
with the minimal turbulence above it, provides conditions of
extraordinary stability, with the lowest levels of seeing on earth.
These conditions are also particularly suitable for wavefront
correction.  The plateau also has the lowest levels of seismic
activity on the planet.  Together with the low wind, this reduces
constraints on the required strength and stiffness of large
structures.

Taken together, these conditions create an important opportunity for
observational astronomy, from the optical to the millimetre wavebands.
Indeed, given the relative ease of access compared to space, they may
provide the best environment from which to conduct some grand-design
experiments, such as the search for exo-earths, for the next several
decades.  Nevertheless, the astronomy so far conducted in Antarctica
has been largely confined to just a few of the competitive niches
(see, for example, Indermuehle, Burton \& Maddison, 2005).  These
include a series of successful cosmic microwave background experiments
and sub-mm astronomy with modest aperture telescopes, together with,
from particle astrophysics, the installation of networks of cosmic ray
facilities, the building of the first neutrino telescope, and the
collection of meteorites from blue-ice fields where they have been
transported to after falling onto the plateau.  Aside from site
testing, no astronomy has yet been conducted from any of the summits
of the Antarctic plateau.  While there is no doubt that the
performance of large Antarctic telescopes that operate in the optical
and infrared would be significantly better than that of comparable
facilities at temperate sites, so far the largest telescope to observe
in these wavebands has been the 60\,cm SPIREX telescope at the South
Pole (Hereld 1994, Fowler et al.\ 1998).  The South Pole, however, at 2,835\,m, is on the
flank of the plateau and suffers from the katabatic air flow off the
summit at Dome A, which disturbs the seeing in the surface inversion
layer.  Better sites than South Pole are to be found on the summits of
the plateau, in particular at the accessible site of Dome C\@. An
intermediate-sized telescope at Dome C is an important next step in
Antarctic astronomy, prior to investing in major optical/IR
facilities.  Its successful operation would demonstrate that the gains
inferred from the site testing campaigns can in fact be realised.  It
would also allow the logistical and engineering requirements of
running such a facility through the Antarctic winter to be appraised.

Operating such an intermediate-sized telescope as a technology
demonstrator is only part of the requirement, however.  The demands of
scientific enquiry also mean that it is essential that such a
telescope be able to undertake competitive science as well, even if
its primary purpose is as a step towards more powerful facilities to
follow.  It is the purpose of this paper to consider the scientific
case for such an intermediate facility --- a 2\,m-class telescope
capable of diffraction-limited imaging from optical wavebands to the
mid-infrared (i.e.\ from 0.5 to 40\micron).  We discuss below some of
the scientific programs that such a facility, which we have dubbed
PILOT---the Pathfinder for a International Large Optical
Telescope---could tackle if it were built at Dome C\@.  This 3,250\,m
elevation site is the location of the new Concordia scientific station
(75S, 123E), built by the French and Italian national Antarctic
programs (Candidi \& Lori 2003, Storey et al.\ 2003) and opened
for winter operations in 2005.  This document also builds upon two
earlier science cases for Antarctic astronomy, the first when the
program was beginning in Australia (Burton et al.\ 1994), and the
second when the emphasis was on building the 2\,m Douglas Mawson
Telescope, which focussed on wide-field thermal-IR imaging (Burton,
Storey \& Ashley, 2001).

\section{The advantages of Antarctica for astronomy}
\label{sec:conditions}
As the result of extensive site-testing programs that have been
conducted at the South Pole for over two decades, and at Dome C since
1996, it has been established that there are a number of major
advantages that an Antarctic plateau observatory would have over the
same facility operating at temperate latitudes.  These include:

\begin{itemize}
\item Low temperature.  At wavelengths shortward of the blackbody-like peak 
in the sky emission spectrum, the flux drops considerably for a small
fall in temperature.  Above the Antarctic plateau the background
reduction relative to temperate sites, for a typical mid-winter
temperature of $-60^{\circ}$\,C, is $\sim 20$ times at near-infrared
wavelengths (2.2--5\micron) (Ashley et al.\ 1996, Nguyen et al.\ 1996,
Phillips et al.\ 1999, Walden et al.\ 2005).  This is equivalent to
obtaining the same sensitivity using a telescope of several times the
diameter at a temperate site (see \S\ref{sec:sensitivity}).  Between
2.27 and 2.45\micron\ the background drop is even greater, around 50
times.

\item Low water vapour.  With the precipitable water vapour content
averaging $\sim 250$\micron\ above the plateau in winter (Chamberlin,
Lane \& Stark 1997, Lane \& Stark 1997, Lane 1998), the atmospheric
transmission is considerably improved, particularly at mid-IR and
sub-mm wavelengths (Chamberlain et al.\ 2000, Hidas et al.\ 2000,
Calisse et al.\ 2004), over temperate locations.  New windows become
accessible for ground based observation between 20 and 40\micron\, and
at 200\micron.  In addition, the low water vapour also lowers the
emissivity of the atmosphere, further reducing the sky flux.  At the
very highest location on the plateau, the 4,200\,m Dome A, the water
vapour content may drop below 100\micron\ at times, further opening
new windows right across the far-infrared spectrum (see Lawrence
2004a).

\item Low aerosol contribution.  The lack of dust and other particulates 
in the atmosphere greatly reduces the contribution to sky emissivity
from aerosols (Chamberlain et al.\ 2000, Walden et al.\ 2005),
so further reducing the background emission in the mid-infrared bands
(from 7--40\micron).

\end{itemize}

\begin{figure}
\plotone{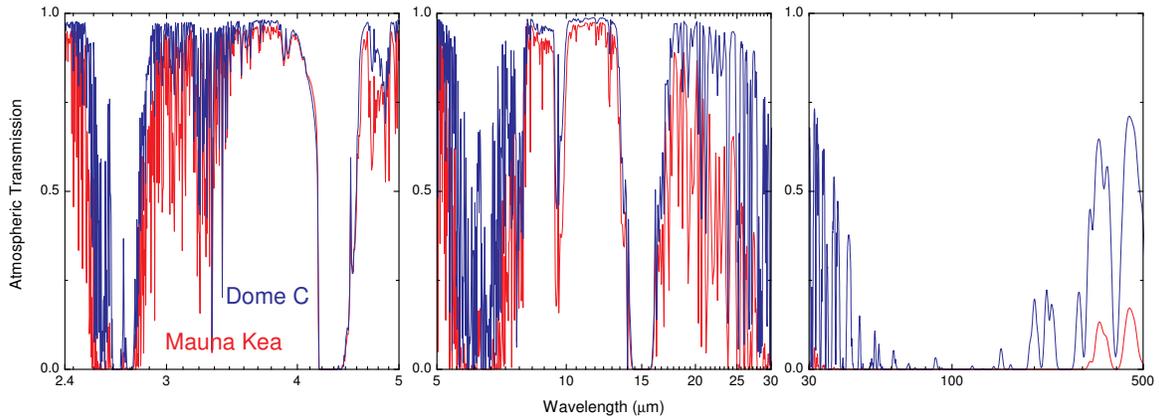}
\caption{Atmospheric transmission calculated for Mauna Kea (red) and Dome C
(blue) using the LBLRTM code, with the measured atmospheric parameters
for the sites as input.  The three panels show the start of the
thermal-IR bands (2.4--5.0\micron), the mid-IR (5--30\micron) and
the far-IR and sub-mm bands (30--500\micron), respectively.}
\label{fig:trans}
\end{figure}

\begin{figure}
\plotone{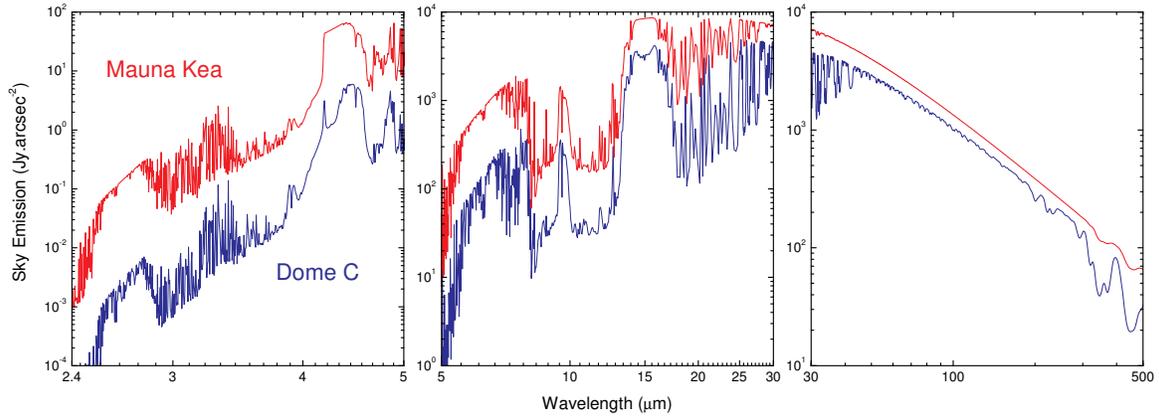}
\caption{Atmospheric emission calculated for Mauna Kea (red) and Dome C (blue),
using the LBLRTM code.  The sky emission is in Jy arcsec$^{-2}$, and
the three panels show the start of the
thermal-IR bands (2.4--5.0\micron), the mid-IR (5--30\micron) and
the far-IR and sub-mm bands (30--500\micron), respectively.}
\label{fig:emiss}
\end{figure}

The Line-By-Line Radiative Transfer Model (LBLRTM) (Clough \& Iacono,
1995) was used to model the atmospheric transmission and sky emission
at Dome C across the infrared and sub-mm bands (see Lawrence 2004a
for details).  This code makes use of a model atmosphere whose input
is the temperature, water vapour, pressure and gaseous constituent
profiles with height, based on data from comprehensive measurements
made at the South Pole and from balloons launched at Dome
C\@. Figs.~\ref{fig:trans} \&
\ref{fig:emiss} show the results from this modelling, and compare the
results calculated for Mauna Kea Observatory in Hawaii, generally
regarded as the best temperate observatory site in the world. The
background drop across the near- and mid-IR is clear, as are the new
windows that open up in the mid-IR and sub-mm. The advantage is even
greater towards the edges of the windows.

There are further advantages to be gained for the Antarctic
plateau sites because of the nature of the atmospheric turbulence that
causes astronomical seeing:

\begin{itemize}
\item Superb seeing.  This results from the exceedingly stable air flow
over the summits of the plateau. The major contribution to the wind on
the plateau is katabatic in origin, and this implies calm conditions
will exist on the summits of the plateau (Marks et al.\ 1996, 1999,
Marks 2002, Travouillon et al.\ 2003a, 2003b).  At Dome C, the average
wind speed is just 2.8\,m/s, one of the lowest values of any
observatory on earth (Aristidi et al.\ 2005).  The highest ever
recorded wind speed there is just 20\,m/s.  With the high-altitude jet
stream rarely intruding, the remaining turbulence which causes
astronomical seeing is driven by the wind speed gradient within a
narrow boundary layer, confined to no more than 30\,m in height above
Dome C\@.  This leads to extraordinarily good seeing, with the mean
value of $0.27''$ for the V band determined from the first 6 weeks of
winter-time data to be obtained from this site (Lawrence et al.\
2004).  This is less than half the value measured on Mauna Kea.  For
more than 25\% of the time the seeing is below $0.15''$, values never
attained from temperate locations.  The daytime seeing is also very
good, being less than $1''$ for much of the time (Aristidi et al.\
2003), and falling as low as $0.2''$. Since the seeing varies as
$\lambda^{-1/5}$, a 2\,m telescope in Antarctica would have near
diffraction-limited performance longward of the H band (1.65\micron)
in average seeing conditions.

\item Wide isoplanatic angle.  With small contributions to the seeing
originating from the upper atmosphere, the isoplanatic angle is 2--3
times larger than at temperate sites ($\sim 6''$ at V; Lawrence et
al.\ 2004).  This is because the angular size of turbulent cells, as
seen by the telescope, is larger due to their proximity.  The
isoplanatic angle also depends on the Fried parameter ($r_0$, the
effective maximum aperture size for diffraction limited performance).
Since $r_0$ increases as $\lambda^{6/5}$, the isoplanatic angle can
attain arcminute-sized values in the near-IR\@.

\item Long coherence times.  The slow drift of turbulent cells through
the telescope field, resulting from the slow wind speed, serves to
increase the coherence time to $\sim 8$ milliseconds at V band
(Lawrence et al.\ 2004).  This is of particular importance when
wavefront correction is being performed, as it means longer
integration times and reduced AO system bandwidths.  Combined with the
larger isoplanatic angle, guide stars for wavefront correction will
invariably be available for any source under study (see Marks et al.\
1999).  In addition, the coherence time also increases with the Fried
parameter as $\lambda^{6/5}$, thus simplifying the operation of
wavefront correction systems at longer wavelengths.

\item Low scintillation noise.  In addition to a wider isoplanatic angle, 
the shallower curvature on the wavefront of incident radiation reduces
the scintillation noise.  Since scintillation provides the principle
limit to high-precision photometry, the reduced noise offers the
prospect of more accurate measurement of the flux of stars,
particularly when small variations are sought.  The precision
obtainable should be several times better than at temperate sites
(Fossat 2003).

\item Astrometric interferometry.  The
precision with which relative positions of objects can be
determined over a narrow angle using an interferometer depends on
$\int h^2 C_N^2(h) dh$, where $C_N^2(h)$ is the refractive index
structure function at height $h$.  Since these fluctuations are
largely confined to the same narrow inversion layer as the previous
four properties discussed above, rather than arising from a
high-altitude jet stream, this leads to large gains in the
precision of measurement of source positions.  Calculations by Lloyd,
Oppenheimer \& Graham (2002) suggest that a given accuracy could be
obtained 300 times faster at the South Pole than from a temperate
site, with precisions of a few micro-arcseconds accuracy attainable.

\item Continuous observation.  At the South Pole any source above the
horizon can be observed continuously, weather permitting, providing
the opportunity for long time-series measurements.  In addition, at
thermal infrared wavelengths (i.e.\ $\lambda > 3$\micron) daytime
observations can readily be undertaken, still with reduced backgrounds
compared to temperate sites at night time.  At Dome C, while there can
be up to a $30^{\circ}$ variation in elevation during the course of
the day, sources generally remain accessible for longer time periods
than at temperate sites.

\item Stability.  While limited data currently exist on the sky stability
at Dome C (aside from the turbulence discussed above), indications are
that the sky conditions are very stable.  For instance, measurements with
an all-sky camera in 2001 showed clear skies for 74\% of the time
during winter (Ashley et al.\ 2004a).  The infrared sky during a period
of clear sky in January 2004 (Walden et al.\ 2005) showed only a 10\%
variation in flux at 11\micron\ over 5 days, an exceptionally stable
value compared to the variations experienced at temperate sites.

\item Costs.  Increased costs are not a major factor in Antarctic
operations, a facet which often comes as a surprise for scientists new
to the field.  While some costs are higher than at temperate sites,
others are far lower.  For instance, the need for a protective dome
are minimal in the absence of storms and strong winds.  The smaller
instrument and telescope sizes required, for the same performance as a
temperate site, also reduces the cost.  While logistics considerations
determine where a telescope may be placed (i.e.\ at an existing base),
they then make transportation relatively simple -- Antarctica provides
the only observatory sites in the world adjacent to airports with
heavy-lift capability, for example.  Other cost gains are more
indirect, but significant nevertheless.  For instance, there are no
indigenous species at the site whose habitat may be disturbed.

\end{itemize}

\section{The disadvantages of Antarctica for astronomy}
There are, of course, certain disadvantages to working in Antarctica.
The most notable is that the telescope requires winterization.
Repairing a telescope during mid-winter is particularly difficult.
Among the practical difficulties of installation and operation, the
dry air increases the risk of static damage to electronic components.
The use of liquid helium over the winter months is a major problem
because supply lines must be maintained across inter-continental
distances.  For nine months of the year the site will be physically
isolated from the rest of the world.  See Ashley et al.\ 2004b for a
further discussion of these issues.

The South Pole, lying under the auroral circle, suffers from frequent
aurorae.  Dome C, however, lies close to the geomagnetic south pole,
so that that aurorae are generally below the horizon.  Auroral
emission is largely concentrated in a few spectral lines, particularly
atomic oxygen and bands from molecular oxygen and nitrogen.  Auroral
emission has not yet been measured at Dome C, but based on
extrapolation from the South Pole, is estimated to have median values
in B band of between 22.9 and 24.0 magnitudes per square arcsecond,
and in V band between 23.7 and 24.7 magnitudes per square arcsecond
(Dempsey et al.\ 2005).  These estimates are about 3 magnitudes
fainter than at the South Pole, and less than the anticipated night
sky emission in these bands at Dome C (22 mags per square arcsecond at
V; see Table~\ref{table:point}).

The amount of astronomical dark time (when
the Sun is more than $18^{\circ}$ below the horizon, in the absence of
the Moon) at Dome C is only 50\% of that at temperate sites. Fortunately this has no
affect on observations beyond 1\micron.  There is also less of the sky
accessible for viewing -- the converse of being able to view other
parts of the sky for longer periods.  However, several important
targets are well placed; the Magellanic Clouds pass overhead and the
Galactic Centre is readily accessible, for instance.

Perhaps the biggest issue to contend with regarding working on the
Antarctic plateau is that of human psychology.  Spending a winter on
the plateau, in several months of continuous darkness, is a
challenging experience, and there are relatively few skilled people
who would consider doing this.  However, with nearly five decades of
such experiences from the Amundsen--Scott South Pole station, it is
clear that the challenges can be met.  A harder challenge may be that
of human perceptions of the continent.  Most peoples' knowledge of
Antarctica is based on stories from the `heroic age' of Antarctic
exploration in the early 20$\rm^{th}$ century, where events were
shaped by the adverse weather conditions faced on the coastal fringes
of the continent.  The climate in the interior is vastly different, as
discussed in \S\ref{sec:conditions}, and modern technology now allows
humans to work there in comfort.

\section{PILOT --- the Pathfinder for an International Large Optical Telescope}
PILOT is envisaged as a 2\,m-class telescope, to be sited at Dome C,
Antarctica.  It would have an alt-azimuth mount, with an f/1.5 primary
figured to a wavefront error below 40nm\footnote{This high precision,
better than normally demanded of optical telescopes, is required to
reach the diffraction-limited image quality ($0.05''$ resolution at
550\,nm) that the site makes possible.}. There would be a Cassegrain
focus and two Gregorian-fed Nasmyth focii with an f/20 beam, providing
an unvignetted field of view of at least one degree. One of these
focii would be capable of feeding the beam to a future multi-telescope
interferometer.  A 45-element adaptive secondary would be used for
wavefront correction.

\subsection{Performance specifications}
\label{sec:performance}
\subsubsection{Sensitivity}
\label{sec:sensitivity}
The low infrared background and good seeing at Dome C implies that a
telescope placed there should be an order of magnitude more sensitive
than a mid-latitude telescope of the same size for many kinds of
observations.  In this section we quantify this statement by making a
detailed examination of the performance that PILOT would have, in
comparison to current 8\,m class telescopes on the best temperate
sites, as well as to possible future 8\,m telescopes in Antarctica.

\begin{figure}
\plotone{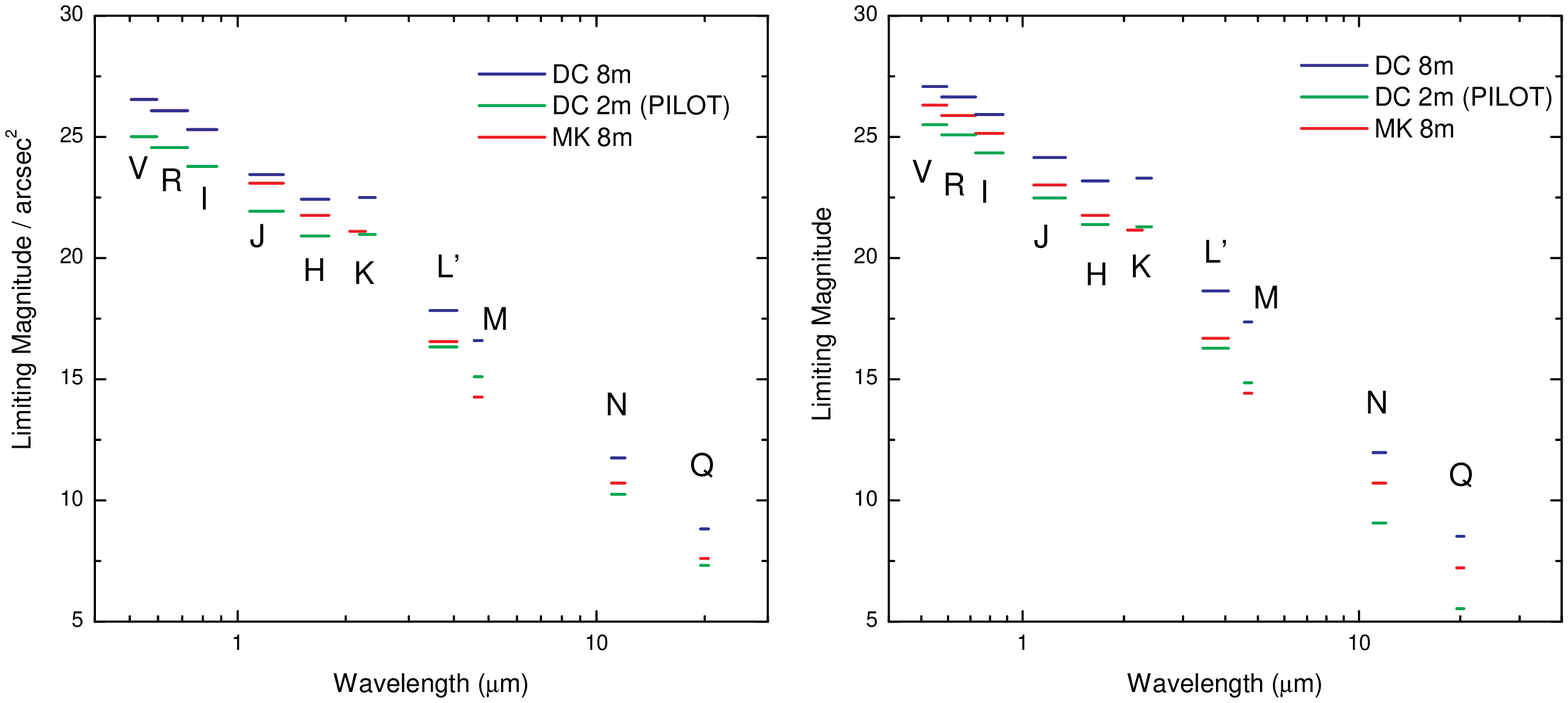}
\caption{Comparison of extended source (left) and point source sensitivities 
(right) (in magnitudes per square arcsecond and magnitudes,
respectively) for an 8\,m telescope on Mauna Kea (red), a 2\,m
telescope on Dome C (i.e.\ PILOT; green) and an 8\,m telescope at Dome
C (blue), as a function of waveband, for a SNR=10 after 1 hour of
integration.  The width of each bar indicates the filter bandpass
adopted. Further details regarding the assumptions made in these
calculations are given in Table~\ref{table:point}.}
\label{fig:sens}
\end{figure}

In comparing the relative sensitivity of different-sized telescopes we
must take account of whether point sources or extended objects are
being measured, as well as the waveband of interest for a particular
study.  The comparison also depends on the capabilities of any
adaptive optics systems being used.  We have thus calculated
sensitivities for both point and extended sources and show these in
Fig.~\ref{fig:sens} and Tables
\ref{table:point}\footnote{Note that the standard notation $2 (-7)
\equiv 2 \times 10^{-7}$, etc.\ is used in all Tables in this paper.} and
\ref{table:extended}.  We compare these to the values for an 8\,m
telescope on Mauna Kea, as well as to an 8\,m telescope built at Dome
C, for the same instrument parameters.  The Figure shows the limiting
magnitudes for SNR=10 in an hour of observation in the principal
filter bands, for background limited operation.  Seeing-limited
spatial resolution (or diffraction-limited, if that is larger), as
indicated in Table~\ref{table:point}, is assumed (see also
Fig.~\ref{fig:seeing}).

\begin{table}
\caption[]{Calculated Point Source Sensitivities}
\label{table:point}
\begin{center}
\begin{tabular}{ccccccccccc}
\hline
{Band} & {$\lambda$} & $\Delta\lambda$  & {MK} & {Ant} & {Ant} & \multicolumn{3}{c}{Spatial Resolution} & \multicolumn{2}{c}{Background} \\
       & ($\mu$m)  & ($\mu$m) & 8m          & 2m          & 8m          & MK 8m  & Ant 2m & Ant 8m & MK & Ant \\
       &             &        & \multicolumn{3}{c}{(PILOT)}    &\multicolumn{3}{c}{(arcseconds)} & \multicolumn{2}{c}{(Jy arcsec$^{-2}$)}\\
\hline
V & 0.55 & 0.09 & 26.3 & 25.5 & 27.1 & 0.50 & 0.25 & 0.24 & 6 (-6) & 6 (-6) \\
  &      &      & 1 (-7) & 2 (-7) & 5 (-8) \\
R & 0.65 & 0.15 & 25.9 & 25.1 & 26.7 & 0.48 & 0.24 & 0.23 & 1 (-5) & 1 (-5) \\
  &      &      & 1 (-7) & 3 (-7) & 6 (-8) \\
I & 0.80 & 0.15 & 25.1 & 24.3 & 25.9 & 0.46 & 0.24 & 0.23 & 2 (-5) & 2 (-5) \\
  &      &      & 2 (-7) & 4 (-7) & 9 (-8) \\
J & 1.21 & 0.26 & 23.0 & 22.5 & 24.1 & 0.42 & 0.24 & 0.21 & 9 (-4) & 5 (-4) \\
  &      &      & 1 (-6) & 2 (-6) & 4 (-7) \\
H & 1.65 & 0.29 & 21.8 & 21.4 & 23.2 & 0.40 & 0.26 & 0.20 & 3 (-3) & 1 (-3) \\
  &      &      & 2 (-6) & 3 (-6) & 6 (-7) \\
K & 2.16 & 0.22 & 21.2 & 21.3 & 23.3 & 0.38 & 0.30 & 0.19 & 2 (-3) & 1 (-4) \\
  & 2.30 & 0.23 & 2 (-6) & 2 (-6) & 3 (-7) \\
L & 3.76 & 0.65 & 16.7 & 16.3 & 18.6 & 0.35 & 0.42 & 0.19 & 2 (+0) & 2 (-1) \\
  &      &      & 5 (-5) & 8 (-5) & 9 (-6) \\
M & 4.66 & 0.24 & 14.4 & 14.9 & 17.4 & 0.34 & 0.52 & 0.20 & 4 (+1) & 5 (-1) \\
  &      &      & 3 (-4) & 2 (-4) & 2 (-5) \\
N & 11.5 & 1.0  & 10.7 &  9.1 & 12.0 & 0.40 & 1.2 & 0.32  & 2 (+2) & 2 (+1) \\
  &      &      & 2 (-3) & 8 (-3) & 6 (-4) \\
Q & 20   & 1.0  & 7.2  & 5.5  & 8.5  & 0.57 & 2.1 & 0.53  & 3 (+3) & 5 (+2) \\
  &      &      & 4 (-2) & 2 (-1) & 1 (-2) \\
\hline
\end{tabular}
\end{center}
Comparison of point source sensitivities in Vega--magnitudes (top line
of pair) and Janskys (bottom line of pair) for an 8\,m telescope on
Mauna Kea (MK; e.g.\ Gemini), a 2\,m telescope on Dome C (Ant; i.e.\
PILOT) and an 8\,m telescope at Dome C\@, as a function of waveband
(listed are central wavelengths and bandpasses, with for K band
different choices applicable to Mauna Kea and Dome C).  The SNR=10
after 1 hour of integration, for the spatial resolution listed in the
relevant columns.  This is calculated by adding, in quadrature, the
seeing to the diffraction limit, for the relevant combination of site
and telescope parameters (see also Fig.~\ref{fig:seeing}).  For the
background determination, the pixel scale is taken to be half this
value, with the object flux assumed to be summed over 25 such pixels.
The telescope emissivity has been taken as 3\% with a temperature of
$0^{\circ}$\,C at Mauna Kea and $-60^{\circ}$\,C at Dome C\@.  An
overall system efficiency of 50\% is assumed in all cases, together
with the spectral resolution as listed for each bandpass. The sky
fluxes, in Jy arcsecond$^{-2}$, adopted for each site are listed in
the last two columns.  Further details are provided in Lawrence
(2004a).  These calculations have not taken into account the further
gains possible when an AO system is operating.  This is discussed
further in \S\ref{sec:resolution}.
\end{table}

\begin{table}
\caption[]{Calculated Extended Source Sensitivities}
\label{table:extended}
\begin{center}
\begin{tabular}{cccccc}
\hline
{Band} & {$\lambda$} & Width  & {Mauna Kea} & {Antarctic} & {Antarctic}  \\
       & ($\mu$m)  & ($\mu$m) & 8\,m        & 2\,m        & 8\,m           \\
       &           &          &             & (PILOT)     &              \\
\hline
V & 0.55 & 0.09 & 26.5 & 25.0 & 26.5 \\
  &      &      & 9 (-8) & 4 (-7) & 9 (-8) \\
R & 0.65 & 0.15 & 26.1 & 24.6 & 26.1 \\
  &      &      & 1 (-7) & 4 (-7) & 1 (-7) \\
I & 0.80 & 0.15 & 25.3 & 23.8 & 25.3 \\
  &      &      & 2 (-7) & 7 (-7) & 2 (-7) \\
J & 1.21 & 0.26 & 23.1 & 21.9 & 23.4 \\
  &      &      & 1 (-6) & 3 (-6) & 7 (-7) \\
H & 1.65 & 0.29 & 21.8 & 20.9 & 22.4 \\
  &      &      & 2 (-6) & 5 (-6) & 1 (-6) \\
K & 2.16 & 0.22 & 21.1 & 21.0 & 22.5 \\
  & 2.30 & 0.23 & 2 (-6) & 3 (-6) & 6 (-7) \\
L & 3.76 & 0.65 & 16.6 & 16.3 & 17.8 \\
  &      &      & 6 (-5) & 7 (-5) & 2 (-5) \\
M & 4.66 & 0.24 & 14.3 & 15.1 & 16.6 \\
  &      &      & 3 (-4) & 1 (-4) & 3 (-5) \\
N & 11.5 & 1.0  & 10.7 &  10.2 & 11.8  \\
  &      &      & 2 (-3) & 3 (-3) & 7 (-4) \\
Q & 20   & 1.0  & 7.6  & 7.3  & 8.8  \\
  &      &      & 3 (-2) & 4 (-2) & 9 (-3) \\
\hline
\end{tabular}
\end{center}
Comparison of extended source sensitivities in Vega--magnitudes per
square arcsecond (top line of pair) and Janskys per square arcsecond
(bottom line of pair), for an 8\,m telescope on Mauna Kea, a 2\,m
telescope on Dome C (i.e.\ PILOT) and an 8\,m telescope at Dome C\@,
as a function of waveband. The SNR=10 after 1 hour of integration.
Other details are as for Table~\ref{table:point}.
\end{table}

At short wavelengths, telescope aperture is the most important
parameter determining sensitivity, and this is reflected in the
performance figures.  When the thermal background dominates, however,
the sensitivity of an Antarctic 2\,m is similar to that of a temperate
latitude 8\,m.  An Antarctic 8\,m is typically an order of magnitude
more sensitive than a temperate latitude 8\,m in all observing bands.
In the thermal infrared ($\lambda > 3$\micron), the diffraction limit
exceeds the seeing, and so the best spatial resolution is then
achieved with the larger facility.  Again, an Antarctic 8\,m would
have superior performance in all wavebands.

Because of the superb seeing, the spatial resolution attained with a 2\,m
Antarctic telescope is superior to the temperate 8\,m, for all
wavelengths less than 2.3\micron.  A gain of a factor of two in
resolution is typical in the optical bands.

The sensitivity calculations in Table~\ref{table:point} do not take
into account of the gains possible through the further use of a
tip-tilt or an AO system.  This serves to concentrate the flux into a
smaller angular distribution, so improving the sensitivity for
background limited operation.  As discussed in
\S\ref{sec:resolution}, an AO system could achieve a Strehl ratio as
high as 0.8 in V band at Dome C, whereas only seeing-limited
resolution is possible in this band from Mauna Kea.  This can lead to
another gain of up to $\sim 2$~magnitudes in the point source
sensitivity at V for PILOT, though this would only apply within the
isoplanatic angle (of $\sim 6''$ at V).

Finally, it should be noted that the raw sensitivity figures do not
demonstrate the full gain attainable in mid-IR wavelengths, since the
background flux is considerably more stable in Antarctica as well.
Fluctuations in the background provide the primary limitation to
photometry at temperate sites.  These calculations also do not show
the greater wavelength range where observations are possible in
Antarctica---the result of the opening of new windows because of the
drier atmosphere.

\subsubsection{Spatial resolution and isoplanatic angle}
\label{sec:resolution}
\begin{figure}
\plotone{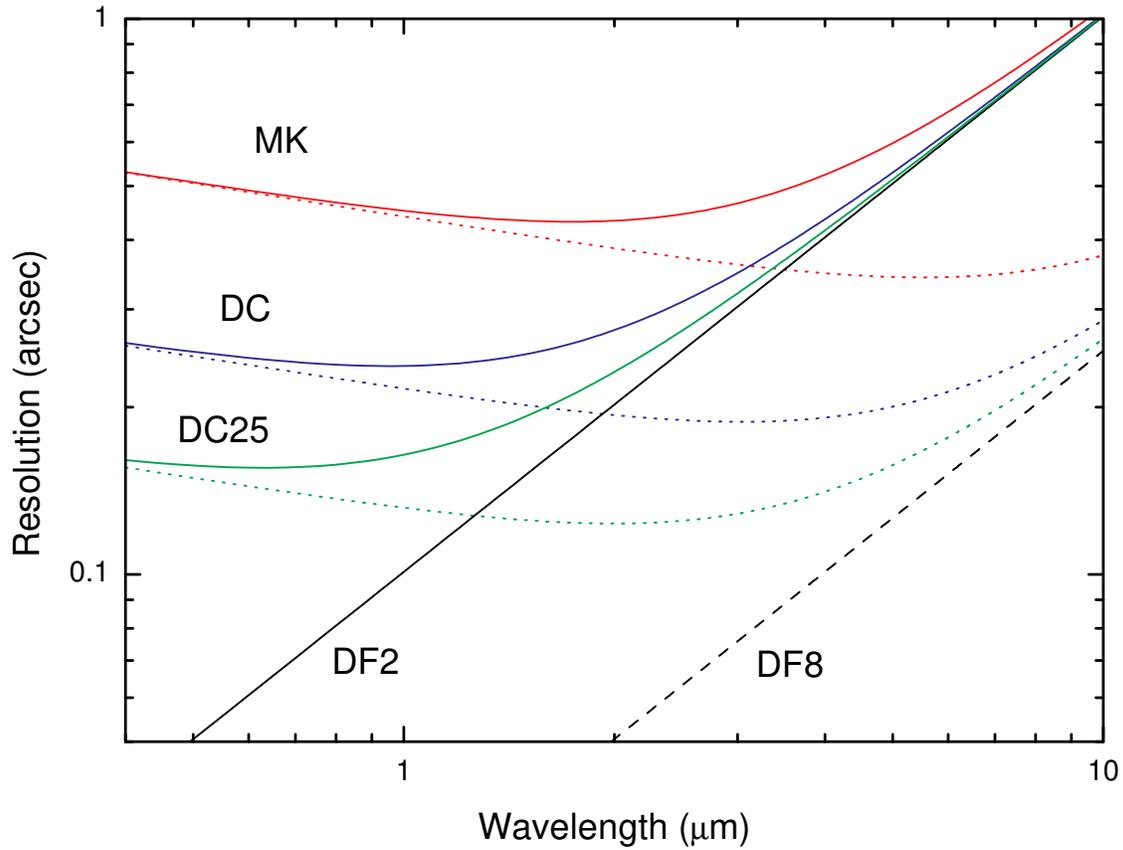}
\caption{Resolution attainable with a 2\,m telescopes (solid lines)
and an 8\,m telescope (dotted lines) at both Dome C and Mauna Kea, as
a function of wavelength.  The resolution is calculated by adding, in
quadrature, the seeing to the diffraction limit.  Curves are shown for
Mauna Kea (red), median seeing conditions at Dome C (blue) and for the
best 25\% conditions at Dome C (green).  Also shown in black, for
comparison, is the diffraction-limited performance for a 2\,m and an
8\,m telescope.}
\label{fig:seeing}
\end{figure}

As discussed in \S\ref{sec:conditions}, the data obtained so far suggest that the
median seeing at Dome C is
more than a factor of two better than at the best temperate sites. In
natural seeing conditions, without the application of any
wavefront correction, a 2\,m telescope can therefore achieve better
resolution in the visible and near-IR than much larger facilities at
temperate sites. Of course, since the seeing improves with wavelength
as $\lambda^{-0.2}$, eventually the larger facility will provide
superior resolution on account of its smaller diffraction
limit. Figure~\ref{fig:seeing} illustrates the seeing-limited
resolution attainable at Dome C and Mauna Kea as a function of
wavelength for 2\,m and 8\,m-sized telescopes.  For the best quartile
conditions, near-diffraction limited performance for a 2\,m telescope
at Dome C is obtained in H band (and longer wavelengths), whereas in
median seeing conditions this occurs in K band. An 8\,m telescope on
Mauna Kea, however, has superior resolution to a 2\,m telescope at
Dome C for wavelengths greater than 3\micron\ (L band), for then the
diffraction limit exceeds the seeing for the 2\,m telescope.

\begin{figure}
\plotone{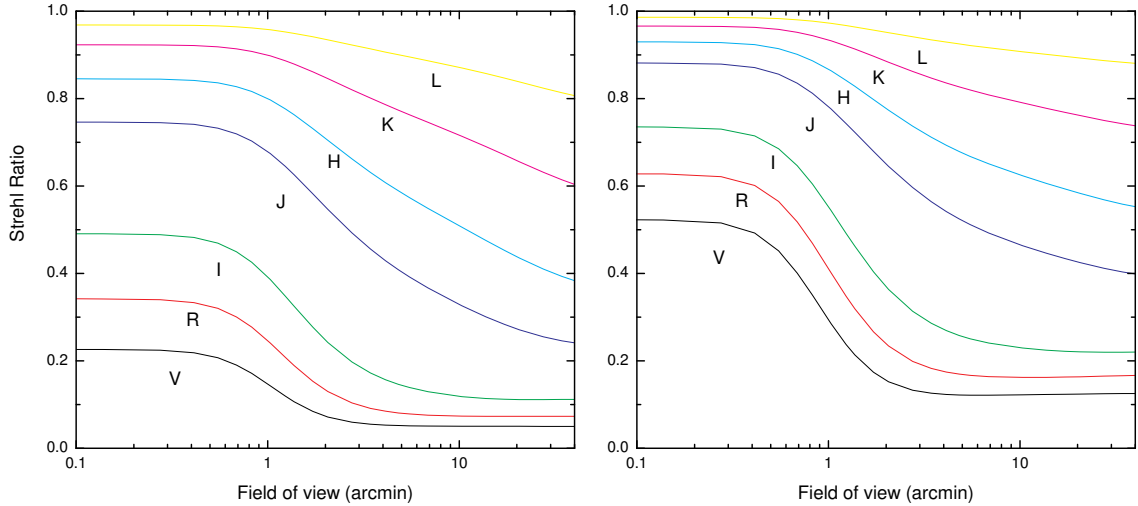}
\caption{Strehl Ratio (the ratio of the peak flux to the flux obtained
with diffraction limited imaging), as a function of angular distance
from the field centre, for a tip-tilt correction system operating on a
2\,m telescope at Dome C, for wavebands from optical (V) to the
thermal-IR (L). The left-hand plot shows the values in the median
seeing conditions, the right-hand plot in the best 25\% seeing
conditions. For longer wavebands (M, N, Q), the Strehl ratio is
effectively unity over all feasible fields of view.  The right hand
axes of these plots reflects the Strehl ratio obtained in the natural
seeing conditions.}
\label{fig:tiptilt}
\end{figure}

The turbulence profile above Dome C is unique, being largely confined
to a layer close to the surface.  This, combined with a lack of high
altitude winds, results in significant improvements for the
performance of any wavefront correction system compared to a typical
mid-latitude site.  Based on the measurements of the seeing and
turbulence profile at Dome C, as measured with a SODAR and MASS
(Lawrence et al.\ 2004), we have calculated in Fig.~\ref{fig:tiptilt}
what the Strehl ratio (the ratio of the peak flux to that obtained
with diffraction-limited performance) would be for a tip-tilt system
on the PILOT telescope (see Lawrence 2004b for further details).

From these Figures it can be seen that at Dome C tip-tilt correction
recovers most of the diffraction limit in the J, H \& K bands (i.e.\
1.2--2.4\micron), even in median seeing conditions.  This occurs over
a wide field of view of several arcminutes (the tilt isoplanatic
angle).  It can also be achieved over the entire sky, as there will
always be a star available for wavefront correction within the tilt
isoplanatic angle.

Using a higher order AO system than tip-tilt will, of course, provide
superior angular resolution (within the isoplanatic angle) on a
temperate 8\,m telescope than with PILOT\@.  In the J band
(1.25\micron), for instance, this is equivalent to having a guide star
bright enough to be used for AO correction within $\sim 10''$ of the
source on Mauna Kea. However, no correction can be made for larger
angles, resulting in seeing-limited performance over larger fields of
view on Mauna Kea.  In addition, only a small proportion of the sky
can be covered with such a system as it relies on the existence of the
nearby guide star.

\begin{figure}
\plotone{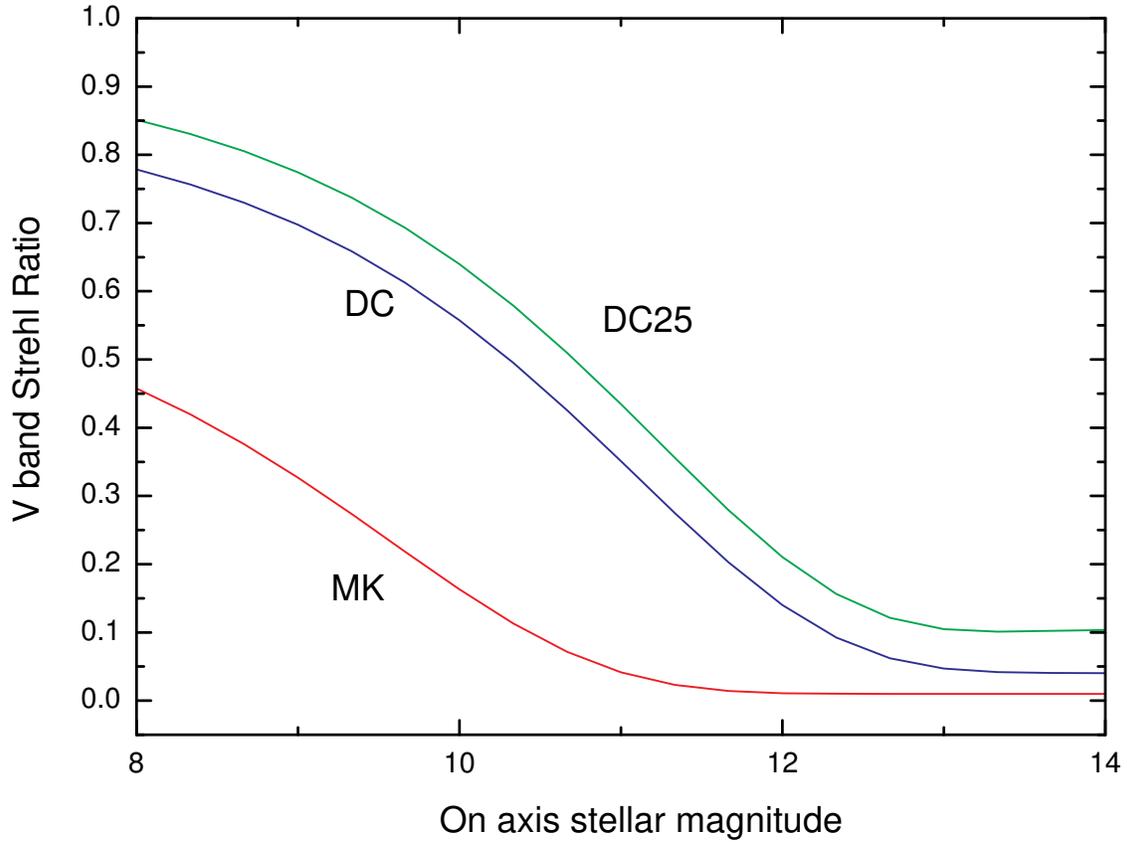}
\caption{Strehl ratio obtained in the visible (V band), as a function
of the magnitude of an on-axis guide star used for AO correction, for
2\,m aperture telescopes operating with 45 actuators at Dome C and at
Mauna Kea. For Dome C, the performance is shown for both the median
and best 25\% seeing conditions, whereas for Mauna Kea it is for the
median seeing.  In practice, the guide star must also be found within
the isoplanatic angle, which is $6''$ in V band at Dome C, about 9
times larger in area than at Mauna Kea. }
\label{fig:aostrehl}
\end{figure}

AO systems provide the best spatial resolution possible for a
telescope, albeit over much smaller fields of view than a tip-tilt
system. The performance of an AO system is a function of many
parameters, including the number of actuators in use, the size of the
telescope, the brightness of the guide star being used for correction
and its angular distance from the object under study, in addition to
the site seeing and isoplanatic angle. In Fig.~\ref{fig:aostrehl} we
compare the Strehl ratio obtainable with an on-axis AO system with 45
actuators on a 2\,m telescope at Dome C to that of the same-sized
telescope on Mauna Kea (see Lawrence 2004b for details). The Strehl
ratio is shown as a function of the magnitude of the star being used
for the correction; i.e.\ within the isoplanatic angle, which is $6''$
at V for Dome C and $2''$ at Mauna Kea. As can be seen, significant
correction would be possible in the visible when guide stars brighter
than $\rm 10^{th}$ magnitude exist within $6''$ of a source at Dome
C\@.  In contrast, from Mauna Kea a star brighter than $\rm 8^{th}$
magnitude, less than $2''$ from the source, is required for correction
to be possible. This tight constraint effectively makes AO systems
unusable in optical wavebands at Mauna Kea, whereas they become
feasible at Dome C\@.  Furthermore, with a ground layer AO (GLAO)
system at Dome C, correction should be obtainable in the visible over
a wide field of view (albeit with a slightly lower Strehl ratio; e.g.\
see Ragazzoni, 2004)\footnote{With PILOT in a Gregorian configuration,
an adaptive secondary conjugates to an atmospheric layer some 20\,m
above the telescope.  Such a GLAO system is ideally suited for removal
of any ground layer turbulence, and would make it possible to achieve
near-diffraction limited images in the optical over a field of view of
a degree or more.  This would open further science opportunities
beyond those implied in this paper using the strawman instrument suite
in Table~\ref{table:strawman}.}.

\subsection{Strawman instrument suite}
\label{sec:strawman}
Based on the above resolution and sensitivity calculations, and the
detector technologies associated with the different wavebands, we
suggest six instrument concepts where PILOT provides unique
capabilities that could be exploited with simple camera systems.

\begin{itemize}
\item High spatial resolution imaging in the visible (V, R \& I bands) with a 
45-element ground layer adaptive optics system, providing correction
over a moderate field, near to the diffraction limit.

\item Wide-field optical imaging (V, R \& I bands), fully-sampled 
for the natural seeing conditions, so also providing good spatial
resolution ($\sim 0.25''$).

\item Near-diffraction-limited imaging in the near-infrared (J, H \& K bands) 
over moderately wide-fields, using a tip-tilt system.

\item Wide-field, near-infrared (J, H \& K bands) imaging
over the whole sky, using a tip-tilt system, but under-sampled with
$0.3''$ pixel scale.

\item High sensitivity, wide-field, diffraction-limited imaging in the 
thermal part of the near-infrared (K, L \& M bands), without the need
for wavefront correction in L and M bands.

\item Wide-field, high-sensitivity, diffraction-limited imaging in the
mid-infrared (N \& Q bands), without the need for any wavefront
correction.

\end{itemize}

In addition, PILOT could also operate in the terahertz wavebands, and
would be particularly effective if equipped with a (multi-beam)
spectrometer or bolometer operating in the 200, 350 and 450\micron\
windows.

Table~\ref{table:strawman} gives strawman instrument specifications
for imaging cameras to take advantage of these capabilities, as well
as for a single-element terahertz-frequency spectrometer.  We assume
arrays of the largest sizes typically available commercially in their
respective wavebands, in order to provide simple camera systems.
However, one can readily envisage butting several arrays together (or
using a multi-beam sub-mm instrument) if the scientific motivation for
covering wider fields is sufficiently high.  The wide-field near-IR
JHK camera is under-sampled\footnote{It is also possible to recover
the full resolution through the technique of `drizzling' -- taking
multiple exposures with half pixel offsets between them.  This
technique has been demonstrated using the wide-field cameras of the
HST, which have $0.1''$ pixel scale for a $0.1''$ telescope point
spread function.}, to give a wide field of view across these
wavebands, whereas all the other cameras are Nyquist-sampled at the
diffraction limit (or the seeing limit for the wide-field optical
imager) for the relevant wavelength (0.5, 1.25, 3.8 and 11\micron,
respectively).  Each instrument is also listed with a figure of merit
determined by $A \Omega / \Delta \Omega$, where $A$ is the telescope
aperture, $\Omega$ the field of view of the camera and $\Delta \Omega$
the spatial resolution.

\begin{table}
\caption[]{Strawman Instrument Suite for PILOT}
\label{table:strawman}
\small
\begin{center}
\begin{tabular}{ccccccc}
\hline
{Wavebands} & {Wavelength} & {Array} & {Detector} & {Pixel Scale} & 
{Field of View} & {Figure of Merit} \\
& (microns) & Size & Material & (arcsec) & (arcmin) & $A \Omega / \Delta \Omega$ (m$^2$)\\
\hline
{Optical: VRI} & 0.5--1.0 & 4K & Si     & $0.03''$ & $2.0'$ & 1 (+7)\\
(High Resolution)\\ 
{Optical: VRI} & 0.5--1.0 & 4K & Si     & $0.1''$ & $6.8'$ & 1 (+7)\\
(Wide Field)\\ 
{Near-IR: JHK} & 1.2--2.4 & 4K & HgCdTe & $0.08''$ & $5.3'$ & 1 (+7) \\ 
(High Resolution) \\
{Near-IR: JHK} & 1.2--2.4 & 4K & HgCdTe & $0.30''$  & $20.5'$ & 1 (+7) \\
(Wide Field) \\
{Thermal-IR: KLM} & 2.2--5.9 & 1K & InSb   & $0.23''$  & $4'$   & 8 (+5) \\
\\
{Mid-IR: NQ}  & 7--40    & 0.5K & SiAs & $0.7''$  & $6'$  & 2 (+5) \\
\\
{Sub-mm} & 200--450 & 1 & NbN & 25--60$''$ & 0.5--1.0$'$ & 8 (-1) \\
\hline
\end{tabular}
\end{center}
\normalsize {Outline specifications for seven instrument concepts that
could take advantage of the capabilities offered by PILOT\@. Aside
from the wide-field optical and near-IR imagers, all cameras sample at
half the diffraction limit -- to take advantage of the good natural seeing
(the wide-field optical camera samples at half the seeing limit).
Their fields of view are then determined by the largest, commercially
available array for the relevant wave-band.  Larger fields would be
possible by butting several such arrays together.  While the
performance has been calculated for broad band filters, a suite of
narrow band filters (1\% spectral resolution) would also be included
in any instrument.  If a multi-beam instrument were built for the
sub-mm, rather than a single-element bolometer, this would of course
greatly increase the capabilities in this waveband too.  The Figure of
Merit is determined by $A \Omega / \Delta \Omega$, where $A$ is the
telescope aperture (2\,m diameter), $\Omega$ the field of view of the
camera and $\Delta \Omega$ twice the pixel scale (i.e.\ the effective
spatial resolution).}
\end{table}

\subsection{Wide field surveys}
As a relatively small telescope, wide-field instruments can be built
for PILOT cheaply in comparison to instruments with the same
field of view on 8\,m class telescopes.  Combined with the sensitivity
in the infrared due to the low background and the high spatial
resolution due to the good seeing, this makes PILOT a particularly
powerful facility for undertaking wide-field infrared surveys.  The
field sizes that can be achieved easily surpass those obtainable on
larger telescopes, at least without large financial investments.  In
the thermal infrared, while the sensitivity cannot compare with
surveys conducted from space, the spatial resolution is several times
higher than that of, for example, the space infrared telescope
Spitzer.  Near-IR surveys conducted from space, for instance with
NICMOS on the HST, suffer from small fields of view.  Hence PILOT can
readily probe large parts of parameter space not being tackled with
other facilities, lying between the deep, pencil-beam surveys
being conducted with some 8\,m telescopes, and the shallow,
low-resolution all-sky surveys conducted using facilities such as
2MASS\@.  Table~\ref{table:surveys} lists some prospective
surveys that could be undertaken using PILOT with the strawman
instrument suite listed in Table~\ref{table:strawman}, and
sensitivities adopted from Tables~\ref{table:point} and
\ref{table:extended}.  These surveys are still relatively modest in
observing time, requiring 1 month of data (assuming a 25\% data
gathering efficiency over the period).  Depending on the importance
assessed to a particular science objective, it would be relatively
easy to extend them for additional instrumentation costs, for instance
by providing dichroic beamsplitters and additional arrays to survey at
two wavelengths simultaneously, or by butting several arrays together
to increase the field of view, or simply by conducting the survey
over an entire observing season, rather than just one month, to
increase the survey area.

We now compare these prospective surveys using PILOT to some other
planned wide-field surveys. The UKIDSS survey on the 4\,m UKIRT on
Mauna Kea\footnote{See www.ukidss.org.} will cover somewhat larger
areas in the northern sky in K band as listed for the wide-field
camera in Table~\ref{table:surveys}, but to one magnitude shallower in
depth (18.4 mags for 4,000 square degrees and 21 mags for 35 square
degrees for SNR=5).  The planned survey on the dedicated 4\,m VISTA
telescope on Cerro Paranal in Chile\footnote{See www.vista.ac.uk.} 
will reach a magnitude deeper in K band for its widest field component
(20.5 mags for 5,000 square degrees, SNR=5), but will reach a similar
limit for its intermediate field (21.5 mags for 100 square degrees).
Both these planned survey facilities, however, will dedicate several
years of telescope time to these programs, compared to a few months
for a similar survey with PILOT\@. In the case of the VISTA telescope
it requires constructing a 1 square degree field of view camera to
perform the task, and the survey will take over a decade to complete.  The spatial
resolution will also be more than a factor of two lower than could be
achieved with PILOT\@.

The 2MASS near-IR sky survey\footnote{See www.ipac.caltech.edu/2mass.} 
reached magnitude limits of J=16.7, H=15.9 and K=15.1 (SNR=5), but
with $2''$ spatial resolution, about 4 magnitudes worse than would be
achieved with PILOT, but covering virtually ($\sim 95\%$) of the
sky. The GLIMPSE survey being conducted with the cryogenically cooled
Spitzer space infrared telescope (Benjamin et al.\ 2003), also with
$\sim 2''$ resolution, will survey $2^{\circ} \times 120^{\circ}$
degrees of the Galactic plane in the thermal-IR, achieving SNR=5 for
magnitudes of 15.5, 14.6, 12.7 and 11.4 at $\lambda =$~3.6, 4.5, 5.8
\& 8.0\micron, respectively.  The shortest two of these wavelengths
are similar to those of the L and M bands, for which PILOT's
equivalent limits for a wide-field survey are $\sim 1$~magnitude
worse, although with four times higher spatial resolution.

No wide-field, high spatial resolution surveys are planned at all for
the thermal-IR L and M bands. While the depths that could be reached
with PILOT would be shallower than at K band, this area of parameter
space has so far barely been explored because of the poor sensitivity
obtainable from temperate sites.  Spitzer/GLIMPSE will provide deeper
surveys over selected areas of the sky, albeit at significantly worse
spatial resolution, but there is a clear role here for the kind of
survey that PILOT could undertake at L and M bands.

\begin{table}
\caption[]{Prospective Wide Field Infrared Surveys with PILOT}
\label{table:surveys}
\begin{center}
\begin{tabular}{ccccccccc}
\hline
Band & $\lambda$ & Resolution & \multicolumn{2}{c}{Point Source} &
\multicolumn{2}{c}{Extended Source} & Frame Time & Survey Area \\ &
$\mu$m & arcsec& Mags & Jy & Mags/arc$^2$ & Jy/arc$^2$ & minutes &
square deg. \\
\hline
\multicolumn{9}{l}{\it JHK High Resolution Camera, $0.08''$ pixel scale, $5.3'$ FOV} \\
J    & 1.25      & 0.24  & 21.0 & 8 (-6) & 20.4 & 1 (-5) & 1  & 90 \\
     &           &       & 23.3 & 1 (-6) & 22.7	& 2 (-6) & 60 & 1  \\
H    & 1.65      & 0.26  & 19.9 & 1 (-5) & 19.4 & 2 (-5) & 1  & 90 \\
     &           &       & 22.2	& 2 (-6) & 21.7 & 3 (-6) & 60 & 1  \\
K    & 2.3       & 0.3   & 19.8 & 8 (-6) & 19.5 & 1 (-5) & 1  & 90 \\
     &           &       & 22.1	& 1 (-6) & 21.8 & 2 (-6) & 60 & 1  \\
\\
\multicolumn{9}{l}{\it JHK Wide Field Camera, $0.3''$ pixel scale, $20'$ FOV} \\
J    & 1.25      & 0.3   & 21.0 & 8 (-6) & 20.4	& 1 (-5) & 1  & 1200 \\
     &           &       & 23.3	& 1 (-6) & 22.7 & 2 (-6) & 60 & 20   \\
H    & 1.65      & 0.3   & 19.9 & 1 (-5) & 19.4 & 2 (-5) & 1  & 1200 \\		
     &           &       & 22.2 & 2 (-6) & 21.7 & 3 (-6) & 60 & 20   \\
K    & 2.3       & 0.3   & 19.8 & 8 (-6) & 19.5 & 1 (-5) & 1  & 1200 \\
     &           &       & 22.1 & 1 (-6) & 21.8 & 2 (-6) & 60 & 20   \\
\\
\multicolumn{9}{l}{\it KLM Camera, $0.23''$ pixel scale, $4'$ FOV} \\
K    & 2.3       & 0.3   & 19.8 & 8 (-6) & 19.5 & 1 (-5) & 1  & 50 \\
     &           &       & 22.1 & 1 (-6) & 21.8 & 2 (-6) & 60 & 1  \\
L    & 3.8       & 0.42  & 14.8 & 3 (-4) & 14.8 & 3 (-4) & 1  & 50 \\
     &           &       & 17.1 & 4 (-5) & 17.1 & 4 (-5) & 60 & 1  \\
M    & 4.8       & 0.52  & 13.4 & 8 (-4) & 13.6	& 4 (-4) & 1  & 50 \\
     &           &       & 15.7	& 1 (-4) & 15.9 & 5 (-5) & 60 & 1  \\
\\
\multicolumn{9}{l}{\it NQ Camera, $0.7''$ pixel scale, $6'$ FOV} \\
N    & 11.5      & 1.2   & 7.6  & 3 (-2) & 8.7  & 1 (-2) & 1  & 100 \\
     &           &       & 9.9  & 4 (-3) & 11.0 & 2 (-3) & 60 & 2   \\
Q    & 20        & 2.1   & 4.0  & 8 (-1) & 5.8  & 2 (-1) & 1  & 100 \\
     &           &       & 6.3  & 1 (-1) & 8.1  & 2 (-2) & 60 & 2   \\
\hline
\end{tabular}
\end{center}
Prospective wide-field surveys that could be undertaken with PILOT
using the strawman instruments listed in Table~\ref{table:strawman},
each with one month of telescope time devoted to it.  Sensitivities
are given for point source and extended sources, for a $5 \sigma$
detection in the time per frame listed in column 8.  The area
surveyed, in square degrees, assumes a 25\% efficiency over the
observing period.  The spatial resolution listed is the sum in
quadrature of the seeing and diffraction limit, as in
Table~\ref{table:point}.
\end{table}

\subsection{Sub-millimetre science with PILOT}
The principal areas of sub-mm research are the study of broadband
thermal radiation, from either the cosmic microwave background or cool
(5--50\,K) dust, and atomic and molecular spectral line emission from
gas at temperatures from about 10\,K to several hundred K\@.  While
such sub-mm wavelength science is not the prime focus for PILOT, the
telescope will work perfectly well in these wavebands.  That a 2\,m
diameter primary mirror is sufficient for state of the art sub-mm
research is evidenced by the many publications arising from the
AST/RO, VIPER and DASI telescopes (e.g.\ see Stark 2002). PILOT's
optics and design could be arranged so as to accommodate one of the
multi-pixel heterodyne or bolometer arrays currently being developed
by numerous groups (e.g.\ Payne 2002).

At the South Pole it has been shown that terahertz windows open for
some 50 days per year (Chamberlin et al.\ 2003), a figure likely to be
significantly bettered at Dome C, where the water vapour content is
lower still.  Considering the huge reductions in cost, there remain
many sub-mm projects best engaged with ground-based telescopes rather
than in space. For example, the USA is developing a 10\,m telescope
(the South Pole Telescope, Ruhl et al.\ 2004) for
the sole purpose of exploiting the South Pole's sub-mm sky to survey
for the Sunyaev-Zeldovich effect in galaxy clusters.

\section{Operating regimes for Antarctic telescopes}
In this section we consider how the conditions in Antarctica translate
best into performance gains for telescopes sited there, in comparison
with temperate-latitude facilities.

Assuming equal telescope performance (and ignoring atmospheric
transparency differences), the integration time, $T$, needed to reach
a given sensitivity limit scales as $B (\theta/D)^2$ between two
telescopes placed at different sites, where $B$ is the background
emission, $\theta$ is the spatial resolution and $D$ the telescope
diameter.  This then provides for three different operating regimes
that should be considered when comparing the performance of an
Antarctic telescope to that of a temperate-latitude facility:

\begin{itemize}
\item When the diffraction limit is obtained,
$\theta \propto 1/D$, so that $T$ is proportional to $B/D^4$.  In
Antarctica, this regime applies in the thermal-IR, at 3\micron\ and
longer wavelengths, when the diffraction limit for a 2\,m telescope
exceeds the median seeing.

\item For observing extended-sources, when the source size $\theta$ is the
same between sites.  Then $T$ is proportional to $B/D^2$.  This
applies in any regime where $\theta$ is greater than the seeing or
diffraction limit. Since the background drop is typically $\sim 20$
times in the 3--5 and 8--13\micron\ windows, an Antarctic 2\,m is more
sensitive than a temperate 8\,m telescope for imaging extended
sources.

\item The seeing-limited regime.  Here performance comparison depends 
on the achievable spatial resolution, after wavefront correction.
This varies significantly with both wavelength and site, as well as on
the performance of the adaptive optics system being used.  It also
depends critically on the isoplanatic angle within which any wavefront
correction may be made.  For instance, in the near-IR, diffraction
limited operation can be obtained from both Mauna Kea and Dome C, but
the isoplanatic angle is significantly larger at Dome C\@.  In the
optical regime, performance close to the diffraction limit can only be
obtained using an AO system on an Antarctic telescope; for the
temperate latitude telescope only the seeing limit will be achieved,
since the site conditions will not allow an AO system to operate.  In
this case, a 2\,m telescope at Dome C can be more sensitive for point
source imaging in the optical than an 8\,m at a temperate site,
within the isoplanatic angle.  The science focus in this regime will
be on projects where the angular resolution is critical to achieve a
new result.

\end{itemize}

There is also a fourth regime which applies:

\begin{itemize}

\item In wavebands where the atmospheric transmission makes observations 
virtually impossible from the temperate site, but feasible from
Antarctica, then completely new kinds of investigation can be
considered.  This applies in the mid-infrared regimes, from
17--40\micron\, and in the far-infrared terahertz windows, especially
near 200\micron.  From the very highest site on the Antarctic plateau,
Dome A, further windows open up, for instance, near 60 and
150\micron.

\end{itemize}

The power of an Antarctic 2\,m for science, then, is not predicated on
its ability to perform just a single experiment well.  An Antarctic
2\,m telescope provides a versatile tool, able to tackle science
programs in a variety of regimes, each using relatively simple (and
inexpensive) instrumentation.  The science program for PILOT thus
revolves around its operation as a community facility, and not just
that of a dedicated survey facility.  In the sections ahead we outline
some of the programs that could be undertaken with such a telescope,
taking advantage of the range of performance gains that we have
quantified.

\section{Science programs for PILOT}
\label{sec:scienceprograms}
In this section we discuss science programs that could be carried out
using the PILOT telescope, operating under the conditions prevalent at
Dome C (discussed in \S\ref{sec:conditions}), with the performance
specifications described in \S\ref{sec:performance} and the instrument
suite described in \S\ref{sec:strawman}.  These fall into three
principal categories:

\begin{itemize}
\item our solar system and others,
\item our galaxy and its environment,
\item our universe and its evolution.
\end{itemize}

\subsection{Our solar system and others}
\subsubsection{The characterization of orbital debris around the Earth}
Since the launch of the first artificial satellite into orbit in 1957,
vast numbers of objects have been placed in earth orbit. Over time
there have been collisions and explosions of various rocket bodies and
casings, ejected hatches, bolts, fuel and coolants virtually filling
the orbital sphere with debris, ranging in sizes from metres to just a
few microns. These objects now present a significant hazard to manned
and unmanned space flight, yet the nature and distribution of the
global debris population remains poorly quantified.

It is therefore of great interest to characterize debris objects,
their sizes, heights and altitudes.  By observing these objects at
various sun-illumination phase angles their albedo can be calculated,
and when combined with their brightness this yields their
size. Variability in brightness also provides information about the
object spin rate and stability, and thereby drag and moment of
inertia.

Observing satellite and debris from the Antarctic plateau provides a
number of potential advantages to temperate locations. As the majority
of satellites and debris of interest lie in polar orbits, there is a
natural concentration of objects at the highest latitudes. This
provides higher observing efficiency for may programs and allows
objects of high interest to be observed on every orbit.  Since the
orbits precess in longitude, at mid-latitude sites some targets are
only sporadically available. Furthermore, the extended terminator
period at Antarctic latitudes means that satellites can be tracked
every orbit for long periods of time. This capability opens new
avenues for detecting, tracking and characterizing orbital debris. The
excellent seeing and infrared performance of an Antarctic based
telescope also allows smaller and fainter targets to be tracked with
multi-spectral capability, further enhancing the ability to determine
the composition of the debris. Knowing the material type (from
spectral information) makes it possible to more accurately assess the
cross-sectional area of the object, providing for accurate orbit
predictions and longevity of predictions.

\subsubsection{High resolution planetary imaging}
By using the selective imaging (or ``lucky imaging'') technique
(Baldwin et al.\ 2001, Dantowitz, Teare \& Kozubal 2000), it should be
possible to obtain diffraction limited planetary images at any
wavelength using a 2\,m telescope, giving results comparable
with the Hubble Space Telescope. The images can be used for general
monitoring of changes to planetary atmospheres and surfaces. Some
specific projects that could be carried out include:

\begin{itemize}
\item Studies of the atmospheric circulation of Venus in the lower cloud layer.
By observing Venus at 1.7 or 2.3\micron\ the cloud structure can be
seen, silhouetted against the thermal radiation from the lower
atmosphere. Continuous monitoring of cloud motions at high spatial
resolution would provide valuable data that could be compared with the
predictions of atmospheric general circulation models (GCMs) for
Venus.  Currently such observations (e.g.\ Chanover et al.\ 1998) are
limited by the short duration that Venus can be observed from any one
site, and by the limited spatial resolution of the images.

\item Studies of the Venus oxygen airglow.  Imaging of Venus at 1.27\micron\
will enable the strong (mega-Rayleigh), and highly variable, molecular
oxygen airglow emission to be observed from the Venusian upper
atmosphere. Continuous monitoring of variations in its intensity and
spatial structure would place constraints on the dynamics and
chemistry of the upper atmosphere.

\item Mars surface pressure imaging. By observing Mars in narrow-band 
filters that isolate the 2\micron\ CO$_2$ bands it should be possible
to image the surface pressure distribution (Bailey et al.\ 2004,
Chamberlain, Bailey \& Crisp 2004). This would allow monitoring of weather
systems on Mars and provide data that can then be used to test Martian
atmospheric GCMs (Forget et al.\ 1999).

\end{itemize}

This program of planetary imaging takes advantages of a number of
facets of the Antarctic environment: good seeing, the slow seeing
variation timescale, 24 hour coverage, low daylight sky brightness,
especially near the Sun, combined with good daytime seeing (as Venus
must be observed in daylight), and low water vapour.

It requires imaging in broadband filters at any wavelength from 0.3 to
2.5\micron, and through selected narrow-band IR filters.  Diffraction
limited resolution is essential for its success (i.e.\ $0.06''$ at
0.5\micron\ and $0.25''$ at 2\micron), but is obtained automatically
by the selective imaging technique, without even the need for
wavefront correction. Continuous short (10--50\,ms) exposures are
needed, with real time processing to select and stack frames.
Continuous monitoring is needed for several days, at time resolutions
of around a minute.  For Venus and Mars, in particular, observations
are restricted to favourable opportunities when the planets are in the
south and close to conjunction and opposition, respectively. Venus has
to be observed in daylight, which in Antarctica means the summer
time. The project is likely to require a specially designed instrument
that can provide short exposure times and fast frame rates.

\subsubsection{Imaging and photometric follow-up of transiting planet 
candidates}
Transit searches attempt to detect close-orbiting extrasolar planets
by detecting the periodic dip in the host star's lightcurve as the
planet transits it.  They are generally performed with wide-field
cameras looking at crowded fields.  However, $\sim 90$\% of candidates
detected by such searches will actually be eclipsing binary stars,
either undergoing grazing eclipses, or with the depth of the eclipse
diluted by a third, blended star (Brown 2003).

High spatial resolution imaging and multi-colour photometry provide
efficient ways to determine the nature of the transiting system for
two reasons. Firstly, stars blended in the wide-field search images
can be resolved, allowing the star hosting the transiting object to be
identified and so the true (i.e.\ undiluted) transit depth measured.
Secondly, the host star's colours yield estimates of its size and
mass, and thus constrain models of the transiting system (Drake \&
Cook 2004). Many binary systems can be identified by the colour
dependence of the eclipse shape, even when diluted (Tingley 2004).

PILOT provides a suitable facility for such observations because of
the high spatial resolution obtainable at Dome C and the ability to
provide guaranteed follow-up of candidates with periods of up to a few
weeks. The low scintillation and small variation in airmass also
improves the photometric precision. It should be possible to improve
from the current 2--5 milli-magnitude rms errors of wide-field transit
searches to significantly better than 1 milli-magnitude precision. The
scintillation and Poisson-noise limits for a 1-minute exposure on a
V=13 star with PILOT are each $\sim 0.3$~mmag (based on equations in
Kjeldsen \& Frandsen, 1992). The higher S/N of light curves thus
obtained would provide tighter constraints on the physical properties
of the transiting system.  The good access to the Galactic plane is
also a benefit as that is where most transits searches are looking.
Candidate sources will also generally be bright enough to be used as
their own reference stars for AO correction.

Imaging would be primarily in the V and I bands, followed later
possibly by imaging in the near-IR J, H and K bands to constrain
stellar types.  The science requires the highest possible spatial
resolution that can be achieved in the visible.  To be truly
advantageous over other telescopes, this requires achieving better
than $0.5''$ resolution.  A field of view of at least 1--2$'$
across around each candidate star is needed. Candidates will generally
be near the Galactic plane or in other crowded fields, such as
clusters.

For each candidate, a minimum of two complete transit events would
need to be observed.  Observation at half-period intervals can also be
used to record shallow secondary eclipses.  A transit event typically
lasts 2--5 hours, and typical periods will be for 1--7 days.  Thus, a
total of $\sim$10--20 hours of observing per candidate will be needed,
spread over a period of a few days to weeks.  The target stars will be
relatively bright ($\rm V \le 14$\,mags), so individual exposures will
be less than a few minutes.

There are several transit search programs currently underway which are
also well located to provide candidates for follow-up by PILOT.  These
are Vulcan-South (being conducted at the South Pole; Caldwell et al.\
2004), the UNSW/APT Planet Search (at Siding Spring Observatory; Hidas
et al.\ 2004) and the All-Sky Automated Survey (at Las Campanas,
Chile; Pojmanski 2002).

\subsubsection{Disks and the earliest stages of planet formation}
\begin{figure}
\plotone{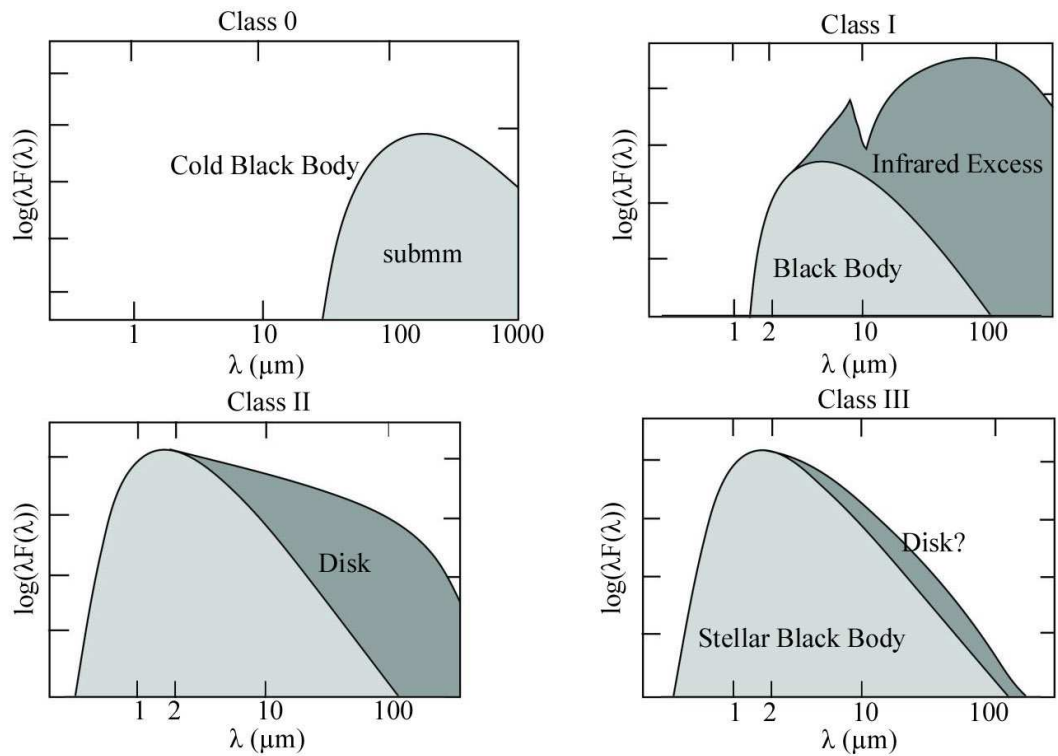}
\caption{Illustrative sketches show the evolution of the spectral energy
distribution during low mass star formation.  Initially the core is
cold, $T \sim$~20--30\,K, peaking in the sub-millimetre (Class 0). An
infrared excess appears and peaks in the far-IR, the emission from a
warm envelope heated by the accretion luminosity (Class I). The peak
shifts to the mid- and near-IR as a disk forms (Class II). Its
spectral shape depends on whether the disk is passive (merely
re-processing the radiation from the central star) or active (also
kept hot by ongoing accretion).  Finally, the disk dissipates (Class
III).  As is apparent, all stages of disk evolution are best studied
at wavelengths longer than 3\micron; i.e.\ in the thermal infrared
(adapted from Lada 1999b). }
\label{fig:sed}
\end{figure}

As low mass stars form, they pass through a sequence of stages as an
accreting disk first forms around a protostar in a molecular cloud,
and then is dissipated as the star emerges onto the main sequence
(i.e.\ the familiar Class 0, I, II and III stages of star formation;
see Fig.~\ref{fig:sed}).  The disks not only appear to play an
essential role in the star formation process, funneling accreting
material and retaining angular momentum, they also provide the raw
material for planetary formation. If the dust particles which came
together to constitute the Earth were spread throughout the
proto-solar system as 1\micron-sized particles in a disk, their
emitting area would be $\sim 10^{13}$ times larger!  Disks are thus
far more readily observed that the planetary systems they evolve
into. Identifying disks, and studying their evolution, therefore provides
both a probe of the star formation process, and insights into the
production of planetary systems.

\begin{figure}
\plotone{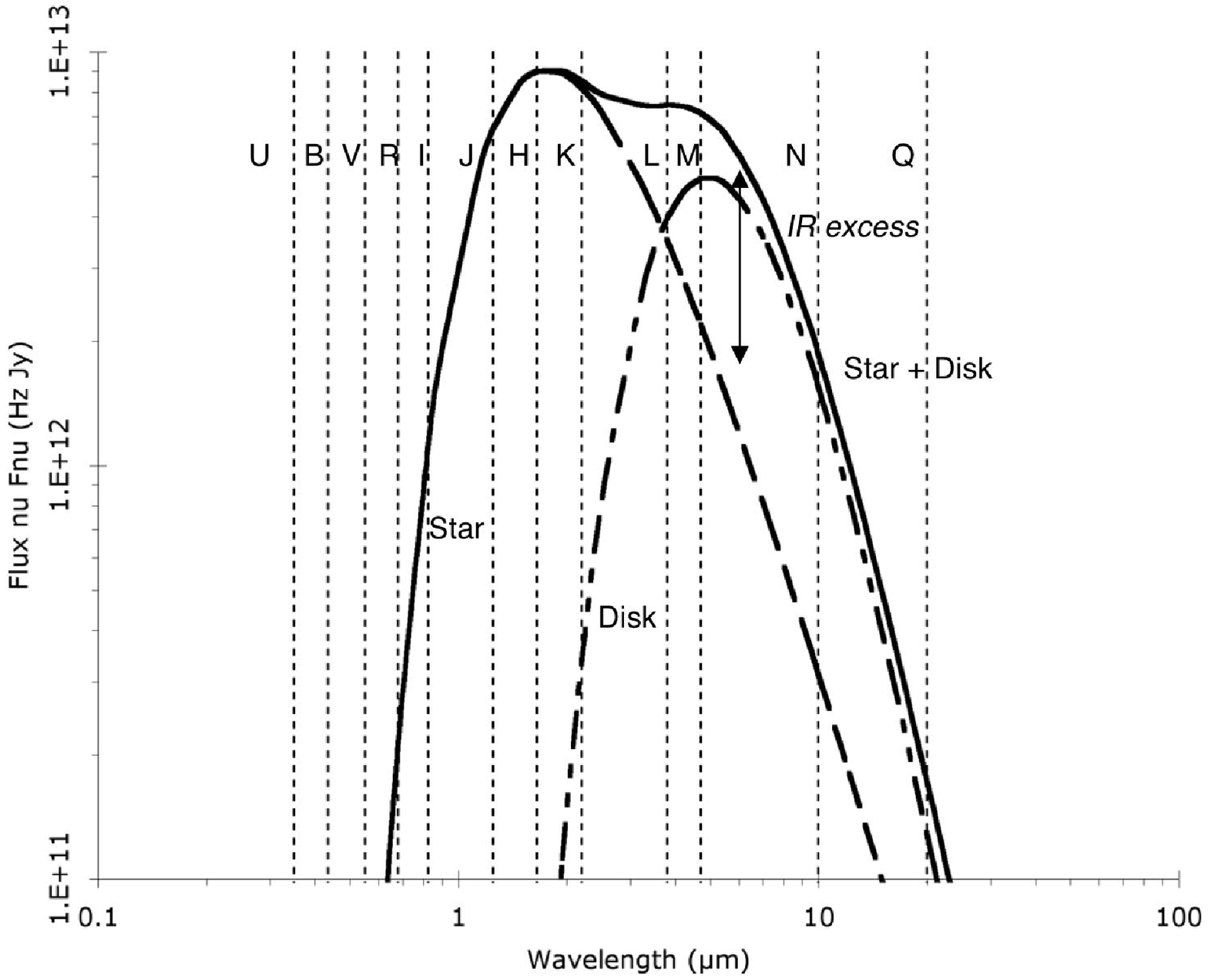}  
\caption{An illustrative example of IR-excess.  The figure shows the 
spectral energy distribution from a 6,000\,K star the same size as the
Sun and 150\,pc away, surrounded by an edge-on 500\,K disk, of radius
5\,AU and height 0.1\,AU\@. There is 10 magnitudes of visual
extinction to the star, and the thermal emission from the disk is
presumed to become optically thin at $\lambda = 2\mu$m (with a
$\lambda^{-1.7}$ extinction law and emissivity index $\beta = 2$ for
the grains).  The curves show the emission from the star, the disk and
their sum.  The vertical dashed lines indicate the centres of the
photometric passbands from the UV to the mid-IR (i.e.\ U to Q bands).
The excess emission from the disk is only weakly discernible at K band
(2.2$\mu$m), but is clearly distinguished for observations in L and M
bands (3.8 and 4.7$\mu$m). Adapted from an illustration in Lada
(1999a).}
\label{fig:irexcess}
\end{figure}

Disks are most readily studied in the thermal infrared ($\lambda >
3$\micron; see Fig.~\ref{fig:irexcess}).  These wavelengths not only
penetrate to the depths of cloud cores, but also enable the embedded
population to be distinguished from background stars.  The warm (few
hundred Kelvin) disks emit strongly at $\lambda > 3\,\mu$m, and thus
are readily distinguished from reddened stars in colour-colour
diagrams that extend into the thermal infrared (e.g.\
$\frac{[H-K]}{[K-L]}$).  Near-IR colour-colour diagrams (e.g.\
$\frac{[J-H]}{[H-K]}$), while relatively easy to construct because of
the better sensitivities available, show only small IR excesses from
these disks, which are readily confused with reddening.  Near-IR
surveys also fail to identify the most deeply embedded sources.

\begin{figure}
\plottwo{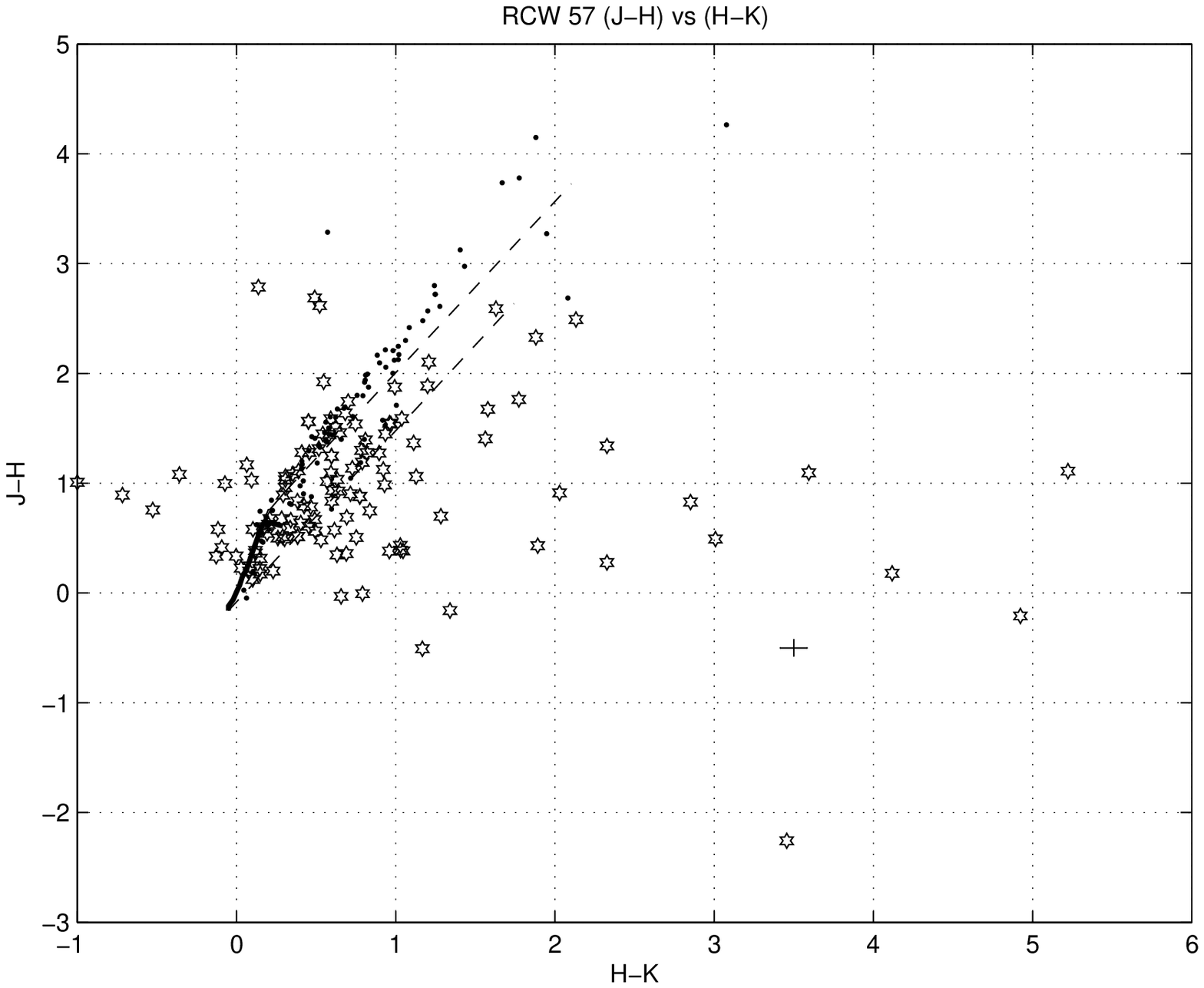}{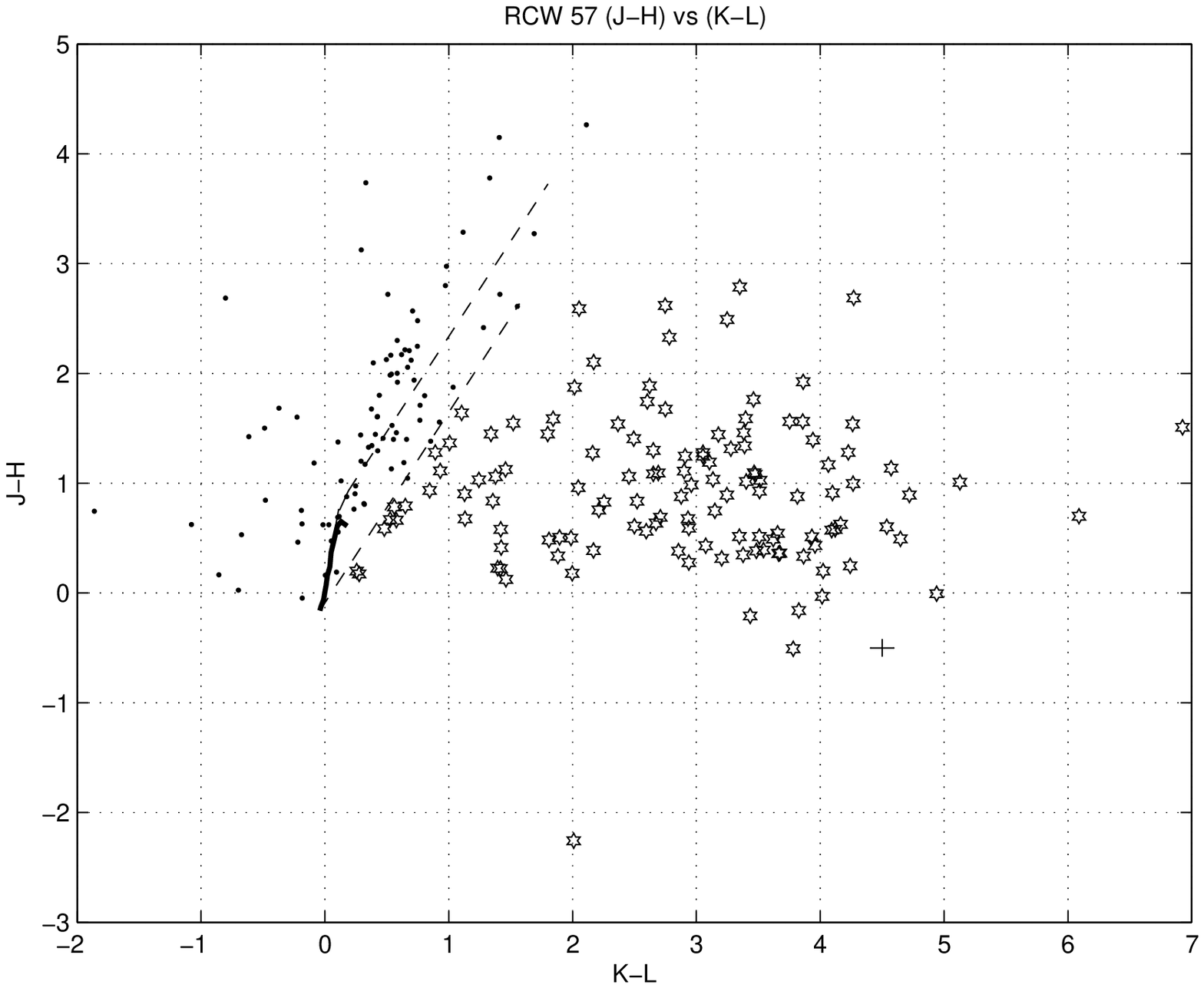}
\caption{Two infrared colour-colour diagrams for the star forming 
region RCW~57 constructed from near-IR data obtained with 2MASS
(JHK) and SPIREX (L band) (Maercker, Burton \& Wright 2005).  The
near-IR colour-colour plot is shown on the left (J--H {\it vs.}\ H--K)
and repeated for the same sources using thermal-IR data on the right
(J--H {\it vs.}\ K--L). The main sequence is shown by the solid line,
with reddening vectors up to $A_V = 30$~mags by the dashed lines.
Star-shaped symbols show sources identified as having a disk from the
L band data, whereas dots indicate sources which do not show evidence
for disks.  While a few sources show strong excesses at K band, it is
clear that this comprehensive separation of disks from reddened stars
could not be carried out using the JHK data alone.}
\label{fig:rcw57}
\end{figure}

Limited work has so far been undertaken at L--band (3.8$\mu$m) as
achievable sensitivities are typically 4--5 magnitudes worse than at
K--band (2.2$\mu$m) from temperate observing sites.  However this
defect can be remedied using an Antarctic telescope because of the
reduced thermal background.  As an illustration of this,
Fig.~\ref{fig:rcw57} shows how the two colour-colour diagrams differ
for the galactic star forming region RCW~57 in the (from Maercker,
Burton \& Wright, 2005).  These data were obtained using the 60\,cm
SPIREX telescope at the South Pole. There is a clear improvement
between using just near-IR (1--2.2\micron) data and also using
thermal-IR data (3.8\micron) in the ability to discern the presence of
disks.  Only a few of the stars identified as having an IR excesses
through use of the L band data would be so identified if just the
near-IR data were available for this task.  PILOT could be used to
survey regions of star formation to search for IR-excesses in this
manner, using the fluxes measured in the J, H, K and L--bands to
determine which sources likely had disks around them.

\subsubsection{Planetary microlensing in the inner Galaxy}
\begin{figure}
\plotone{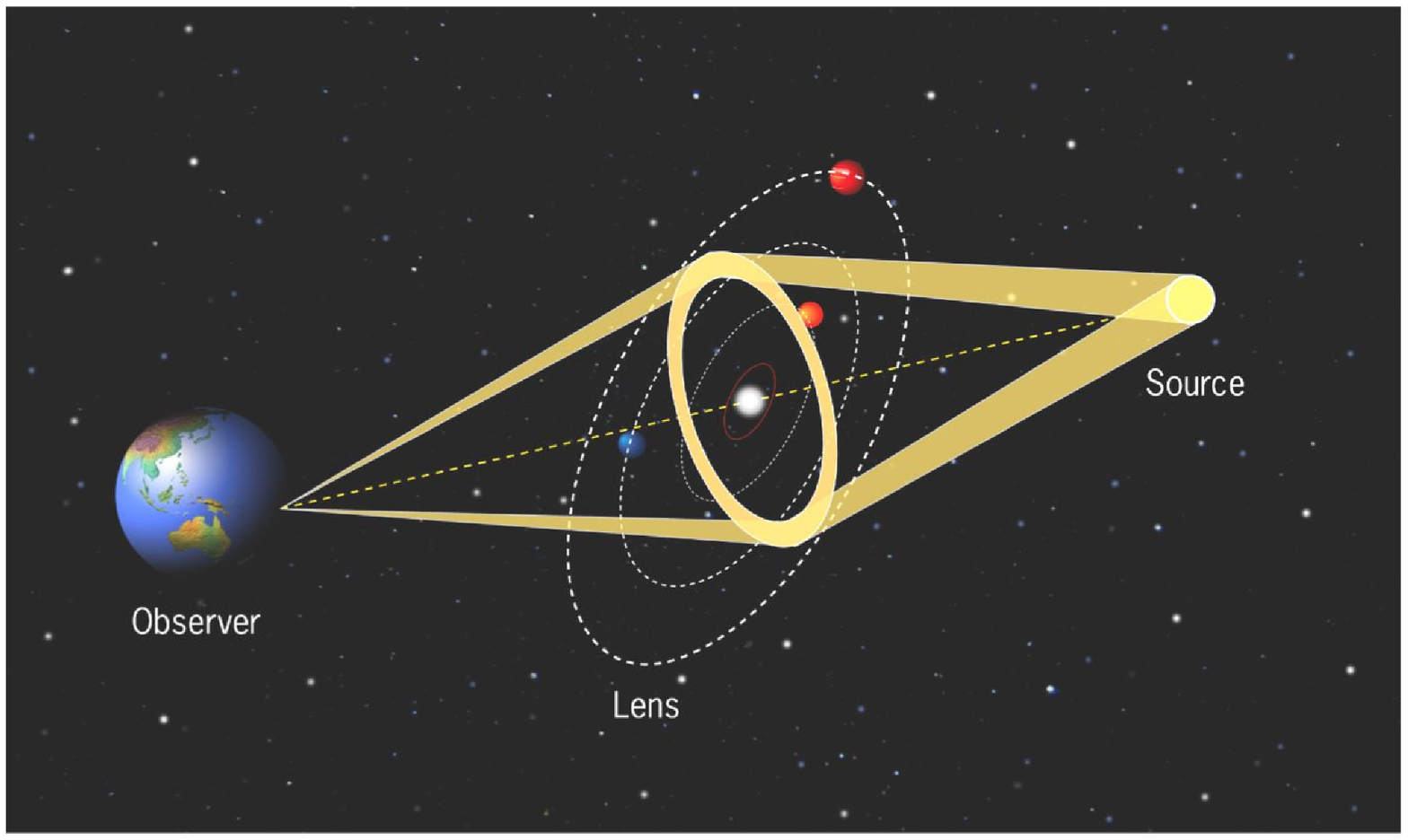}
\caption{Illustration of gravitational microlensing. In this example, the alignment 
of the lens and the source is perfect, leading to the formation of a
complete Einstein ring. For Galactic microlensing, the radius of the
Einstein ring is typically $\sim$2~AU\@. Planets orbiting the lens star
may therefore perturb the Einstein ring.  Several microlensing events
are detected annually with alignments of $\sim$0.01~AU\@. These events
provide significant sensitivity to the presence of earth-mass
planets.}
\label{fig:lens}
\end{figure}

Gravitational microlensing occurs if the light ray from a ``source''
star passes sufficiently close to a massive foreground ``lens'' object
so that its path is bent, or lensed, into multiple images (see
Fig.~\ref{fig:lens}). These images cannot be directly resolved but, as
the lensing star moves across the line of sight, the total amplified
light from the source star can be measured and this generates a
symmetrical light curve profile that is now well recognized (Paczynski
1986). If the lens star has a planetary companion, additional lensing
may occur, producing a perturbation in the light curve (Mao \&
Paczynski 1991, Gould \& Loeb 1992). A positive demonstration of this
effect was recently reported by the MOA and OGLE groups, where an
observed 7 day perturbation in the light curve of a microlensing event
was attributed to a 1.5 Jupiter mass planet in an orbit around a 0.3
solar mass main sequence star with an orbital radius of $\sim$3~AU
(Bond et al.\ 2004).

\begin{figure}
\plotone{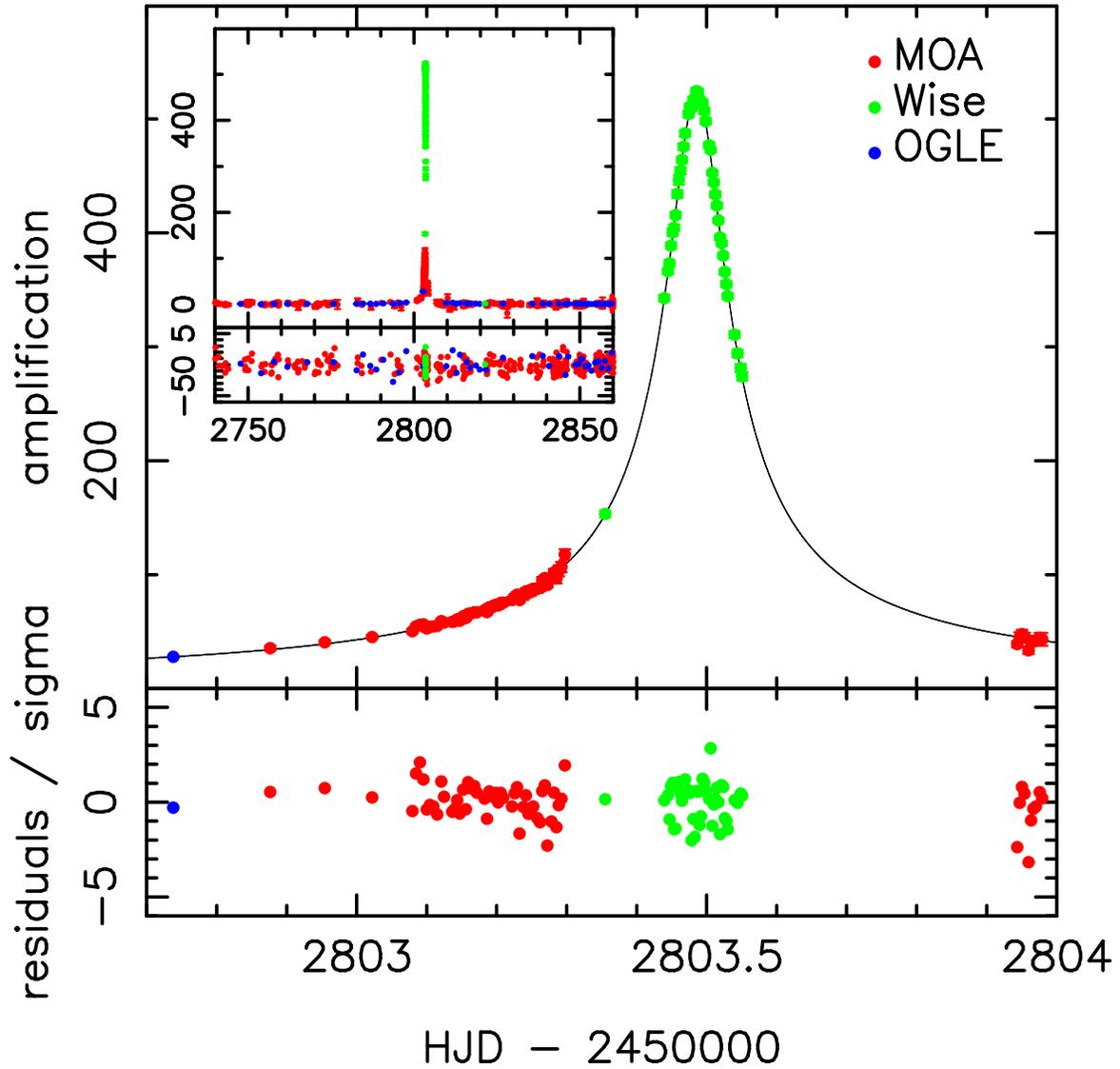}
\caption{Light curve of the high magnification microlensing event 
MOA 2003--BLG--32 obtained by MOA and OGLE and the Wise observatory
(Abe et al.\ 2004). Observations of similar events from Dome C would
allow one to obtain a seamless uninterrupted light curve using a
single instrument.}
\label{fig:MOA2003}
\end{figure}

Planetary microlensing is extremely difficult to detect in
practice, even though the detection rate of microlensing events 
in the galactic bulge is high. In general, the planetary perturbations in the light curve
are very short lived (occurring on timescales ranging from a few days
to a few hours), and can occur anytime during the typically 40 day
timescale of a microlensing event. The chances of catching these
perturbations can be improved by focussing on the the special class of
events where the lens star moves into near perfect alignment with the
background star resulting in very high magnifications, in excess of
50. In these events, the planetary perturbations occur within a 24--48
hour period centred on the time of peak amplification (Griest \&
Safizadeh 1998). Furthermore, the sensitivity to planets is greatly
enhanced, as was demonstrated in actual observations of high
magnification events. In MACHO 1998--BLG--35, which peaked at an
amplification of 80, a small perturbation was observed just on the
threshold of detectability, probably due a planet of between 0.5 and
1.5 earth masses (Rhie et al.\ 2000, Bond et al.\ 2002). More recent
observations of the event MOA 2003--BLG--32 (see
Fig.~\ref{fig:MOA2003}), which peaked at an extremely high
magnification of more than 500, also demonstrated the high planetary
sensitivity in these events (Abe et al.\ 2004). In this case, no
planetary perturbations were seen, but earth-mass planets were excluded from about
2.3--3.6\,AU, and mars-mass planets close to the Einstein ring.


To fully realize the sensitivity to planets in high magnification
events, it is necessary to obtain densely sampled uninterrupted
observations during a 24--48 hour period. This is currently attempted
using a network of collaborating telescopes around the globe. The
microlensing survey groups, MOA and OGLE, find around 6--12 high
magnification events each year. However, most of these do not receive
the required dense sampling -- usually because of a combination of
unfavourable weather and the lack of a telescope at a suitable
longitude at the critical time. Indeed, the event MOA~2003--BLG--32 is the
only event to date for which the peak was reasonably fully monitored.
In this event, the main observations were made (see Fig.~\ref{fig:MOA2003})
with a 1\,m telescope towards the galactic bulge from a site, at latitude $+30^{\circ}$N,
with average seeing, yet good sensitivity to earth-mass, and some sensitivity to mars-mass, 
planets was achieved.

PILOT would be ideally suited as a follow-up telescope for high
magnification events. The clear skies would greatly improve
the ``catchment'' of high magnification events alerted by the survey
groups. Also, for each event, ``baseline'' measurements could be taken
at times when the background star is unamplified.  Most high
magnification events have very faint source stars and the good
seeing at Dome C would allow exposures to achieve a depth greater than
could be obtained at similar sized telescopes elsewhere. The source
stars typically have I$\sim$22 and V$\sim$23 mags, well within the
capability of PILOT\@. Also, because of the small seeing disk, the
baseline flux measurements would be much less affected by blending
from contamination of nearby stars.

Over a 3--4 year period, 30--40 high magnification events could be
monitored at sensitivities comparable to, or better than,
MOA~2003--BLG--32.  Approximately 300 hours of telescope time would be
required per annum. This would provide good statistics on the
abundances of gas giants and ice giants, as well as rough statistics
on how common terrestrial planets may be.  Such results would not be
possible using telescopes elsewhere on the globe.

If a dedicated wide-field telescope were built for microlensing
studies at Dome C, equipped with a large format array camera, this
project could be extended to determine how common all these different
types of planets are.  Such a telescope would ideally provide both
optical (V \& I band) and near-IR (J, H \& K band) coverage (see
Bennett et al.\ 2004). A field-of-view of $\sim$1.5 deg$^2$ would
suffice, with two fields of $\sim$0.75 deg$^2$, continuously sampled
in the two passbands in a chopping mode mode at intervals of 20
minutes or less. This could be achieved with a giga-pixel camera with
a pixel size $\sim 0.1''$.  Several thousand microlensing events could
be monitored over a 3 to 4 year period -- all with high density,
uninterrupted sampling measurements. If systems similar to our own
solar system are common, such a survey could detect $\sim$50
terrestrial planets, $\sim$100 ice giant planets, and $\sim$1000 gas
giant planets.

\subsection{Our Galaxy and its environment}
\subsubsection{Stellar oscillations}
\label{sec:oscillation}
A star is a gaseous sphere and will oscillate in many different modes
when suitably excited.  The frequencies of these oscillations depend
on the sound speed inside the star, which in turn depends on
properties such as density, temperature and composition.  The Sun
oscillates in many modes simultaneously.  By comparing the mode
frequencies to theoretical calculations ({\em helioseismology}),
significant revisions to solar models have been made (e.g.\
Christensen-Dalsgaard 2002).  For example, the Sun's convection zone
has turned out to be 50\% deeper than previously thought; the helium
abundance cannot be as low as is required to reproduce the apparent
neutrino flux (a result vindicated by recent evidence from the Sudbury
Neutrino Observatory for neutrino flavour oscillations; SNO 2002); the
angular velocity does not increase rapidly with depth, thereby
removing the inconsistency between planetary orbits and general
relativity; and the opacities had been underestimated immediately
beneath the convection zone.  Measuring oscillation frequencies in
other stars ({\em asteroseismology}) will allow us to probe their
interiors in exquisite detail and study phenomena that do not occur in
the Sun.  We expect asteroseismology to produce major advances in our
understanding of stellar structure and evolution, and of the
underlying physical processes.

Thanks to the steadily improving Doppler precision provided by modern
spectrographs, the field of asteroseismology has finally become a
reality.  For a recent review, see Bedding \& Kjeldsen (2003).
However, to make further gains requires that stars be observed as
continuously as possible.  It would also be valuable to observe
clusters of stars photometrically with extremely high precision.
Antarctica offers both of these possibilities and is second only to
space as the site for an asteroseismology program.  The low
scintillation at Dome C, combined with the long time period available
for observations, presents an opportunity to carry out
asteroseismology on solar-like stars and to achieve a breakthrough in our
understanding of the role played in stellar evolution by processes
such as convection, rotation and mixing.

\subsubsection{The earliest stages of massive star formation}
Massive star formation drives the star-gas cycle of the Galaxy.  It is
the crucible through which Galactic evolution is played out, creating
the most luminous objects in the Galaxy and driving the material
cycles of the elements. It is of particular interest to identify the
very earliest stages of massive star formation.  However, despite
being manifest across the Galaxy on account of their luminosity, the
process of massive star formation remains something of a mystery.
There are no nearby examples to dissect, unlike low mass star
formation, and the processes at work take place on short timescales
and in obscured environments that have not been readily accessible to
observation. 

Massive protostars are so cold they can only be detected at far-IR and
sub-mm wavelengths.  As they warm up, their spectral energy
distributions evolve, becoming visible in the mid-IR and then near-IR
wavebands.  The earliest stages have been nearly inaccessible to
study, since the emission from 20--30\,K objects peaks around
100\micron. From Dome C, however, measurements could be made at
200\micron, close to this emission maximum.  Moreover, they can then
be combined with measurements in the sub-mm, at 350, 450 and
850\micron, spectral windows which are open virtually continuously at
Dome C, allowing the energy distributions to be determined in the
Rayleigh-Jeans part of the spectrum.  Such measurements, since they
are at optically thin wavelengths, allow the dust mass to be
estimated.  The slope of the energy distribution then provides a probe
of the dust emissivity, which is a function of the nature of the dust
grains (e.g.\ their size, composition and distribution).  Hence a
terahertz frequency spectrometer at Dome C will provide a new tool for
investigating the earliest stages of massive star formation across the
Galaxy.

\begin{figure}
\plotone{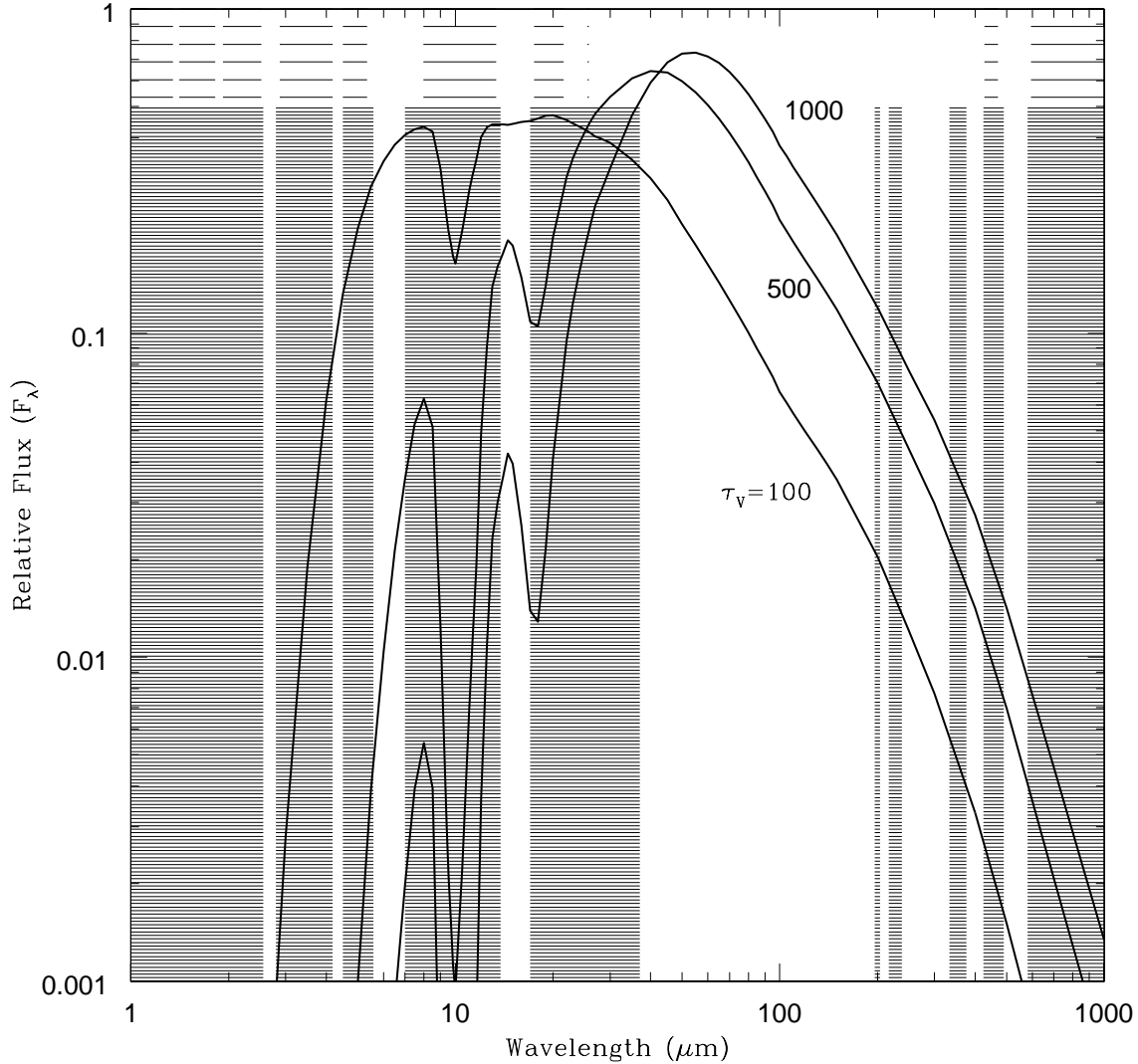}
\caption{Model spectral energy distributions (SEDs) representative of embedded 
massive protostars, as a function of the optical depth of the
obscuring material to the central star ($\tau_V = 100$, 500 \& 1000).
It is clear that infrared observations are needed to study such
objects.  These SEDs have been calculated using the DUSTY code
(Ivezi\'{c} \& Elitzur, 1997), for a central source temperature of
30,000\,K and an $r^{-1.5}$ density distribution of silicate
grains. The features in the spectrum at 10 and 18\micron\ are due to
absorption from the silicates. The shaded overlays show the windows
accessible from Dome C, and at the top, in lighter shading, from Mauna
Kea.  They illustrate the new windows that open at Dome C in the
mid-IR and sub-mm that will facilitate this program.}
\label{fig:msfseds}
\end{figure}

As the cores heat up, an embedded object appears at mid-IR
wavelengths.  It heats the surrounding dust grains, from which
molecules are evaporated, driving a rich organic chemistry in the gas
phase, all of which is amenable to investigation through sub-mm and
mm-wave spectroscopy. All but the very coldest massive protostars are
visible in the mid-IR, and the spectral features (particularly the
silicate dust feature) can be used to measure the column density. The
colour variations across the mid-IR window, from 8--30\micron,
indicate the evolutionary state of the sources (i.e.\ how deeply
embedded they are).  Hence multi-wavelength imaging in the mid-IR
window, combined with measurements of the spectral energy distribution
of the sub-mm emitting cores, provides a method of determining the
initial embedded population of massive star forming cores, the
luminosity and mass of the dominant sources within them, and their
differing evolutionary states.  There does not yet exist a paradigm
for massive star formation equivalent to the Class 0--III states
associated with low mass star formation, and one of the goals of such
studies is to identify whether there are characteristic observational
signatures associated with each stage. For example,
Fig.~\ref{fig:msfseds} shows model plots of the relative energy
distributions of embedded massive protostars, as calculated using the
DUSTY code (Ivezi\'{c} \& Elitzur, 1997), overlaid with the
atmospheric windows available at Dome C\@.  Such objects will become
accessible for study with a telescope such as PILOT, equipped with
mid-IR and sub-mm instrumentation.

\paragraph{Spectral features in the infrared and their evolution during
star formation} The infrared window from 3--30\micron\ contains a
plethora of spectral features, many of which trace the physical
conditions in the circum-protostellar environment.  The main modes of
excitation seen through these features are: atomic fine structure
lines (diagnostics of ionization), gas-phase rotational/vibrational
bands, PAH emission and solid-state vibrational bands (from dust and
ice).

\begin{figure}
\plotone{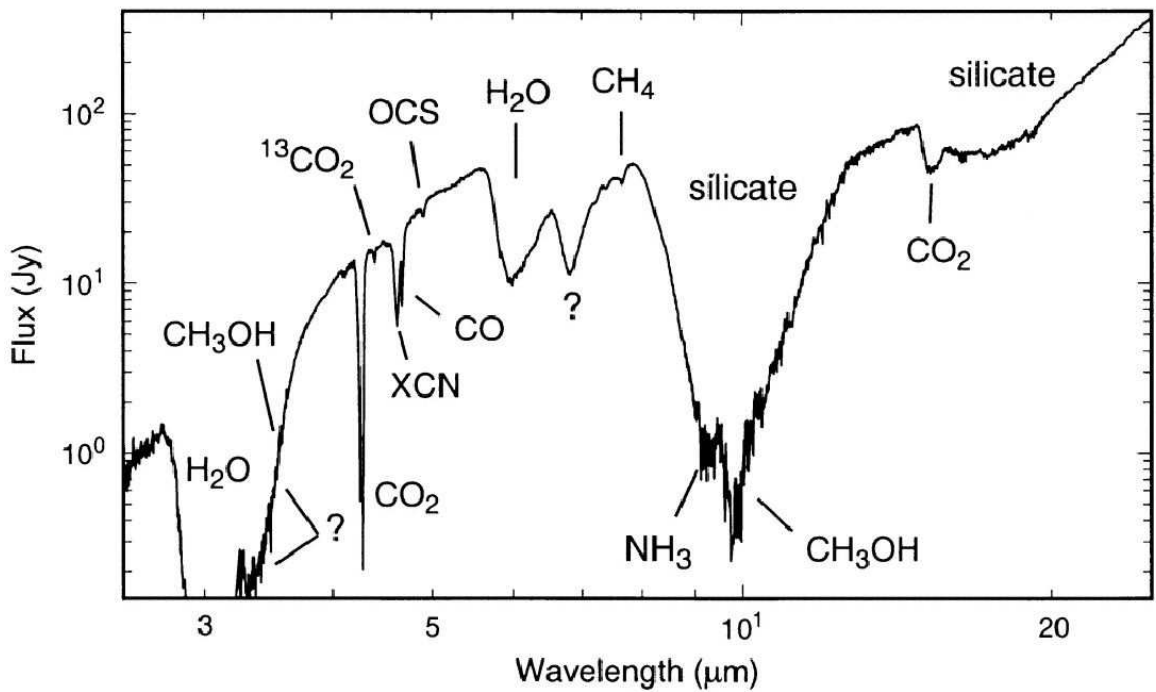}
\caption{A composite infrared spectrum of W33A, a
massive young stellar object, obtained with a resolving power of
500--1000 using several instruments on the ISO satellite (from Gibb et
al.\ 2000).  Many of these features are present in the accessible part of the
spectrum from Dome C\@.  For instance, ices in the grain mantles
produce absorption from H$_2$O at 3.1 and 6 \micron, from CO at
4.7\micron, from CH$_4$ at 7.7\micron, from NH$_3$ at 9\micron\ and
from CH$_3$OH at 10\micron, as well as silicate absorption at 9.7 and
18\micron\ (see van Dishoeck, 2004 for a complete list).  There may
also be PAH emission features at 3.3, 6.2, 7.7, 8.6, 11.3 and
12.7\micron.}
\label{fig:ices}
\end{figure}

Observations from the ISO (Infrared Space Observatory) satellite have
shown that there is a wealth of information to be reaped from mid-IR
spectroscopic observations (see Fig.~\ref{fig:ices}). The sharply
rising continuum flux, with absorption along the line of sight by dust
and ice mantle components, is characteristic of heavily embedded
massive protostars.  These ices provide the raw material for the
chemical evolution of the dense cores. As a protostar begins to heat
up, it evaporates the ices into the gas phase, beginning the hot
molecular core phase of massive star formation.  Observation of the
ice features therefore probes the state of the coldest and densest
regions of molecular clouds, prior to the onset of star formation in
them.

The depletion of species onto the dust grains is also an important
process in the early evolution of the cores. This is partly because
the chemistry of star forming regions is greatly changed as a result,
but also because depletion interferes with our ability to study the
earliest stages of star formation. If a species depletes onto dust
grains it cannot then be used as spectral line probe of the state of the
gas where this occurs. Hence, as a result of depletion, the kinematics of the
innermost, coldest, densest regions of molecular clouds have not been
probed.  For instance, even N$_2$H$^+$, which had been considered as a
non-depleter, has recently been found to indeed deplete in B68 (Bergin
et al.\ 2002), so rendering it unsuitable for probing the core.

There appear to be two remaining species which might be used to follow
the kinematics of the earliest stages of star formation, the
transitions of para-H$_2$D$^+$ at 218\micron\ and ortho-D$_2$H$^+$ at
202\micron.  These ions should be the last to deplete out of the gas
in the coldest and densest cores (Bergin, private
communication). The two transitions are accessible from Dome C\@.

The variation in spectral lines features between embedded protostars
traces the time-dependent chemistry of the regions, providing an
evolutionary clock to classify the earliest stages of star formation.
Combined with the derived physical conditions, mid-IR spectral line
observations provide a powerful and relatively unexplored tool for
probing regions of star formation.

The ISO observations shown in Fig.~\ref{fig:ices} are limited by the
spatial resolution of the 60\,cm telescope aperture, providing
information on only the large scale core properties.  Recent high spatial
resolution mid-IR observations (e.g.\ Longmore et al.\ 2005), show
many of the cores that appear as point sources with MSX (a 35\,cm
satellite operating in broad bands from 8 to 21\micron) are resolved
into multiple sources, altering the interpretation of large scale
data. There is a lacuna for larger, ground based telescopes to extend
the spectroscopic observations to higher spatial resolution.

Dome C is an attractive site for such a telescope, not only because of the
significant decrease in the thermal background and the improved
transmission and sky stability. The spectrally quiet 30--40\micron\
window that becomes available provides a good continuum baseline,
free from the solid state absorption lines that hinder derivation of
the optical depth from spectra that only extend to 25\micron\ (i.e.\
as obtainable from mid-latitude sites).

\subsubsection{A wide-field sub-millimetre survey of star forming 
regions: from low mass protostellar cores to the progenitors of OB
stars}

Star formation is a key astrophysical mechanism driving the evolution
of the Galaxy. Significant progress has been achieved in our
understanding of the formation of low and intermediate mass
stars. These stars form from the collapse of dense cloud cores in the
molecular interstellar medium of the Galaxy. Two very early
evolutionary stages have recently been identified as part of this
process: gravitationally-bound, starless, pre-stellar cores (e.g.\
Ward-Thompson et al.\ 2002), and cold protostars or Class 0 protostars
(e.g.\ Andr\'e, Motte \& Bacmann 1999).  Yet, several fundamental
questions remain open.  For instance, how do pre-stellar cores form in
the molecular clouds and what governs their collapse and evolution to
protostars? What determines the distribution of the initial
distribution of stellar mass (i.e.\ the IMF)?

The formation of high-mass (M $>8$ M$_{\odot}$) stars is, in contrast,
less well understood. Massive stars are born in heavily obscured
regions ($A_v>20$~mags) of cold molecular gas and dust, far away from
us (typically a few kpc) and within dense stellar clusters
($\sim10^4$~stars~pc$^{-3}$). They also experience short lives
($\sim10^6-10^7$ years).  These observational constraints have limited
our knowledge about the processes involved in high-mass star formation
(HMSF). An empirical evolutionary sequence has been proposed: massive
stars form in hot molecular cores and then evolve to ZAMS stars, that
ionize their environment to produce several classes of HII regions
(hyper compact, ultra compact, etc.). Recent (sub-)millimetre
observations also claim the discovery of high-mass protostellar
objects, which are likely to be protoclusters of embedded young
stellar objects (e.g.\ Minier et al.\ 2005).

Theoretically, high-mass star formation is problematic.  Increasing
outward radiative acceleration becomes important for high-mass stars
and will ultimately stop accretion. The infall accretion rate must
then exceed the outflow rate to produce a high-mass star, i.e.\ the
gravity must overcome the radiative pressure. Two scenarios have been
proposed to explain this: HMSF could proceed either through
protostellar mergers (Bonnell \& Bate 2002) or through the collapse of
a single, supersonically turbulent core with a sufficiently large
accretion rate (McKee \& Tan 2003). High-mass, pre-stellar cores are
only expected in the second scenario.  The main physical processes
leading to high-mass stars are then still unclear. We might ask
whether massive stars also form through the collapse of a pre-stellar
core. Do ``Class 0'' high-mass protostars exist?

To address the unanswered questions in these fields of low and high
mass star formation, we propose a sub-mm survey to find and classify
the cores. Unbiased surveys of Giant Molecular Clouds (GMCs) in the
far-IR/sub-mm continuum are necessary in order to make a complete
census of the pre- and proto-stellar population within star-forming
regions. This is because pre-stellar cores, Class 0 protostars and
HMSF protoclusters emit the bulk of their luminosity between 60 and
400\micron\ (Ward-Thompson et al.\ 2002; Andr\'e et al.\ 1999; Minier
et al.\ 2005). The coldest objects have the peak of their spectral
energy distribution around 200\micron.

An unbiased and wide-field survey at 200, 350 \& 450\micron\ of nearby
regions of star formation will allow us to potentially detect all the
pre-stellar cores, protostars and HMSF protoclusters in a given
star-forming region. Their mass and luminosity can then be derived via
spectral energy distribution modelling, obtaining the early stellar
population (i.e.\ the IMF). To do so, a sample of star forming regions
from low mass (e.g.\ Chamaeleon) to OB star progenitor complexes
(e.g.\ NGC~6334) is needed.  A sensitivity to reach a 0.1~M$_{\odot}$
protostar at 0.5\,kpc at 200\micron\ is desirable.

Such an investigation benefits considerably from an Antarctic location
because observations at 200, 350 and 450 \micron\ are possible. The
opening of the 200\micron\ window is an exceptional feature as this
wavelength is generally unobservable from the ground. The ESA Herschel
Space Observatory will observe in the continuum from 60 to 500\micron\
with the PACS and SPIRE instruments. Using a filled bolometer array, as
developed for PACS, will greatly facilitate such a survey in
Antarctica.

\subsubsection{The Galactic ecology}
The environment of star forming complexes, which dominate the southern
Galactic plane, can be studied in the thermal infrared from
2.4--4.1\micron\ through the spectral features emitted from excited
gas in these regions.  The comparison can be made through studying the
emission from three spectral features: H$_2$ emission at 2.4\micron\
in the v=1--0\,Q--branch lines, the 3.3\micron\ PAH emission feature
and Br-$\alpha$ emission at 4.05\micron.  This allows the conditions
and extent of the molecular, neutral (i.e.\ the photodissociation or
PDR interface) and the ionized gas to be probed.  For
example, the SPIREX images obtained at the South Pole of the NGC~6334
massive star formation complex (Burton et al.\ 2000), show shells of
fluorescently excited molecular gas surrounding several sites of
massive star formation, each $\sim 1$\,pc apart and spread along a
dense molecular ridge.  Complex PAH organic molecules pervade the
region, and are fluorescently excited by the far-UV radiation that
escapes from these stars. These images are displaying the Galactic
ecology at work -- the interaction between young stars and the
interstellar medium, in its ionized, neutral and molecular phases.

The spectral features in the images also provide information on the
excitation processes at work in the clouds. Any H$_2$ line emission
results from either shocks or UV-fluorescence.  The v=1--0 Q--branch
lines at 2.4$\mu$m are both stronger and suffer less extinction than
the commonly used 1--0 S(1) line at 2.12$\mu$m.  They emit in the
K--dark band, the region of lowest atmospheric thermal emission from
Antarctica, and so a particularly sensitive window for
observation. Several solid state absorption features are also present,
for instance the ice band at 3.1$\mu$m. Polycyclic Aromatic
Hydrocarbons (PAH), organic molecules that are fluoresced by far-UV
radiation from the young stars and trace the edge of photodissociation
regions, are visible through a spectral feature at 3.3$\mu$m (as well
as having several features in the mid-IR window).  Through these
thermal-IR lines, the PDR structure can be imaged at high spatial
resolution, unlike other prime PDR tracers, such as the far-IR [CII]
158$\mu$m line, which can only provide low resolution maps.  Finally,
HII and ultra-compact HII regions can be traced in the Br\,$\alpha$
4.05$\mu$m line, even when deeply embedded.  These can be compared
directly to radio interferometric maps of the same gas to determine
the extinction, and to estimate the ionizing luminosity.

\subsubsection{Spectral line sub-millimetre astronomy}
The formation, structure, dynamics, abundance, composition and
energetics of molecular clouds in the Milky Way and nearby galaxies
are all active areas of research. Questions regarding how molecular
clouds form, what triggers their collapse, why the process of star
formation does not halt as the collapsing gas heats up, which species
dominate the cooling and the role played by magnetic fields, all remain
significant problems. These issues are best addressed using mm and
sub-mm wavelength observations, since the radiation in these bands
contains large numbers of indicative spectral lines which, unlike
optical lines, can escape from the dense gas and dust clouds from
which they originate.

When molecules in a cloud collide, energy can be transformed from
kinetic into rotational, vibrational and electronic potential
energy. It is then released through the emission of photons with
wavelengths and intensities characteristic of the species involved, so
revealing the abundance, temperature and density in the gas. The most
abundant elements in molecular clouds -- hydrogen and helium -- are
however poor tracers of these conditions as they cannot be
collisionally excited to radiate at the low temperatures found in the
clouds. On the other hand, carbon, the fourth most common element, can
be readily observed in its ionised, neutral and molecular states, as
well as in dust grains, thus sampling the full range of physical
conditions found. Nitrogen, the sixth most abundant element, has an
important atomic line at 205\micron, but this has never been observed
from the ground.

Neutral carbon has fine structure transitions at 610\micron\ ($^3P_1
\rightarrow\, ^3P_0$) and 370\micron\ ($^3P_2 \rightarrow\, ^3P_1$).  The
610\micron\ transition is only 23~K above ground and has a critical
density of 1000~cm$^{-3}$. This implies that it is easily excited and
so the [CI] lines will be readily detectable, even when emitted by
moderate density interstellar gas exposed only to typical ambient
Galactic radiation fields. Excellent atmospheric transmission is
required to measure these important lines, however, and so there has
been relatively little work done on the neutral lines of carbon 
compared to the optical (i.e.\ ionized) lines of
the atom. The first detection of [CI] $^3P_1\rightarrow\, ^3P_0$ in
another galaxy (the Large Magellanic Cloud) was made with a 1.7\,m
radio-telescope from the South Pole (AST/RO; Stark et al.\ 1997), and
the most extensive mappings of [CI] $^3P_2\rightarrow\, ^3P_1$ in any
objects have also been made with this instrument (NGC6334 and the
Galactic Centre; Martin et al.\ 2004). 

Carbon monoxide (CO), the most abundant molecule after H$_2$, needs to
be observed in several rotational ($J$) levels if the physical
conditions are to be inferred from the line strengths. The low--$J$
states are close to local thermodynamic equilibrium (LTE), so their
excitation temperatures, $T_{\mathrm{ex},J}$, are all close to the
kinetic temperature, $T_{\mathrm{kin}}$.  Hence the ratio of line
brightnesses for transitions between such states is near unity and so
independent of $T_{\mathrm{kin}}$. However, for any temperature and
density there will be higher--$J$ states that will fail to be in LTE,
and so must be sub-thermally excited (i.e., $T_{\mathrm{ex},J} <<
T_\mathrm{kin}$, with line ratios $T_{J\rightarrow{J-1}} /
T_{1\rightarrow0}$ departing from unity.  Thus, measurements of
optically thin transitions of CO from rarer isotopomers (e.g.\
$^{13}$CO, C$^{18}$O), and from high--$J$ states, are needed to probe
the hottest and densest regions in molecular cloud cores.  Such
measurements require sub-mm wavelength observations, and so are
well-suited for an Antarctic telescope.

\subsubsection{Continuum wavelength sub-millimetre astronomy}
A wide range of astrophysical phenomena can be studied using far-IR
and sub-mm bolometers, including dust in molecular clouds, hot
star-forming cores and stellar remnants, synchrotron emission in our
Galaxy, the origin of the far-IR background and the evolution of
infrared-luminous galaxies at high redshifts. A wide field instrument
on PILOT would complement the new generation of mm and sub-mm
interferometers being built, such as the Atacama Large Millimetre
Array (ALMA) in northern Chile. Bolometer arrays with thousands of
pixels can now be envisaged (e.g.\ Staguhn et al.\ 2003), and a strong
science case for such arrays on small telescopes has been presented
(Stark, 2003).  By virtue of the extremely dry conditions, a telescope
at Dome C would be well suited for utilising these bolometer arrays
for continuum mapping in the 200, 350 and 450\micron\ atmospheric
windows.

\subsubsection{A survey for cool Brown Dwarfs and extrasolar giant planets}
\label{ref:bds}
Brown dwarfs are sub-stellar mass objects, whose mass is too low for
the nuclear fusion of hydrogen to sustain their luminosity over the
bulk of their lifetime (though the fusion of lithium, deuterium and
even hydrogen is possible in a brown dwarf, depending on its mass and
age).  Brown Dwarfs are born hot, shining primarily through the
release of gravitational potential energy and accretion luminosity,
and spend the rest of their lives cooling, emitting in the infrared
(e.g.\ Burrows et al.\ 2001).  As they cool, their spectra change
radically, quite unlike stars.  Spectroscopically, they can at first
look similar to late-type M dwarfs, they then pass through a L dwarf
stage where warm ($\sim$1300--2100\,K) dust emission dominates, and
they finally end in the T dwarf stage where absorption bands from
methane, water and ammonia dominate (c.f.\ the infrared emission from
Jupiter) as they cool further.  This means that the spectral type of a
brown dwarf depends on both its mass and the age, unlike a main
sequence star.  Modelling is needed to separate the degeneracy between
mass and age for a star of a given surface temperature.  Infrared
colours (JHKLM bands from 1-5\micron) can be used to spectrally
identify brown dwarfs, though the presence of strong methane
absorption bands means that while the K--L colour is red (as expected
from a cool object), the near-infrared colours (e.g.\ H--K) of T
dwarfs may be neutral or even blue (as the radiation is forced to
shorter wavelengths to escape through spectral windows in the brown
dwarf atmospheres).  Before such identifications can be made, however,
the challenge is to find brown dwarfs---after their first few million
years they have cooled so much that they are both faint and emit the
bulk of their radiation in thermal infrared wavelengths, and hence are
difficult to detect.

Giant extrasolar planets also fall into the same class of objects as
brown dwarfs.  While their origins are likely quite different (they
are formed in the disk surrounding a young stellar object vs.\ a
separate gravitational condensation site), physically they are
identical objects (although environmental differences, e.g.\ proximity
to a star that provides an external heating source, may result in some
different observational characteristics).  The study of brown dwarfs
therefore overlaps with that of planetary science, and will provide
insight into the latter.

The study of brown dwarfs is thus a fundamental part of the study of
stars, since they can be formed by the same processes with the only
difference being the mass of the resulting object.  The future
evolution of brown dwarfs is, however, radically different to that of
stars, being that of a degenerate body without a central fusion
luminosity source.  It is less than a decade since the first clear
evidence of a brown dwarf was reported (Gliese 229B; Oppenheimer et
al.\ 1995) and there remains much to be learnt about their properties
and evolution.  While it is now clear that brown dwarfs do not provide
for the missing mass needed to account for galactic rotation curves,
it is also clear that their number density in the solar neighbourhood
is comparable to that of M dwarfs.  The initial mass function is still
rising as the mass falls below the fusion-edge defining the main
sequence.  Isolated brown dwarfs have also been found to be numerous
in several galactic clusters.  Of particular interest is the
occurrence of brown dwarfs in binary systems, for which only limited
information is available.  Binary frequency decreases with stellar
mass, from $\sim$60\% for solar-type stars to $\sim$35\% for M dwarfs.
Furthermore, there are few sub-stellar-mass companions found to
solar-type stars in radial velocity surveys targeting planets (the
so-called ``brown dwarf desert''; Marcy \& Butler 1998).
Understanding why this is the case is important for the theory of star
formation, and the cause of the initial mass function, but there
are limited statistics so far to provide a clear picture of the degree
of binarity in its low mass end.  Brown dwarfs also
display weather -- the atmospheric chemistry in their photospheres,
which is a strong function of temperature, and changes radically
during evolution as the brown dwarf cools.  Photometric and
spectroscopic variations can be expected due to presence of cloud
systems as the sources rotate.  Such changes will also be a function
of spectral type.

All these facets of brown dwarf behaviour are little understood, as
there are still relatively few sources known (a few hundred, dominated
by the hotter and younger sources).  PILOT, with its sensitivity in
the thermal-IR L and M bands, is well suited to extend the current
surveys, both to greater distances and to the detection of cooler (and
thus older, and presumably far more common) brown dwarfs.  In
particular, a survey at L band (3.8\micron) to a depth of 14.8 mags
(as could be reached in 1 minute with PILOT) would provide a similar
sensitivity to that reached by the 2MASS survey (Kirkpatrick et al.\
1999) in the K band, but be capable of detecting cooler and redder
sources.  Table~\ref{table:bds} (adapted from Burrows et al.\ 2001)
gives an indication of how far away a $\rm 15\, M_{Jupiter}$ brown
dwarf could be found, as a function of its age, in the K, L and M
bands.  While such an object can readily be detected when it is young,
by the time it has reached $10^9$\,years in age it can only be found
if it is relatively close to the Sun.  Note, however, that PILOT would
be as sensitive to detecting such objects in the M band as it is in
the K band, despite the greatly reduced sensitivity, because brown
dwarfs this old are so cool.  PILOT could therefore be used to conduct
the first extensive survey for old, cool brown dwarfs within a few
tens of parsecs of the Sun.

\begin{table}
\caption[]{Distance to which a 15\,M$\rm_{Jupiter}$ Brown Dwarf could be detected}
\label{table:bds}
\begin{center}
\begin{tabular}{ccccccccc}
\hline
Age & T$_{eff}$ & Spectral & K Flux & L Flux & M Flux & \multicolumn{3}{c}{Distance (pc)} \\
(yrs) & (K) & Type & (Jy) & (Jy) & (Jy) & K & L & M \\
\hline
10$^7$ & 2,225 & M & 3 (-1) & 3 (-2) & 3 (-2) & 6,000 & 300 & 200 \\
10$^8$ & 1,437 & L & 5 (-3) & 1 (-3) & 5 (-3) & 700 & 50 & 70 \\
10$^9$ &   593 & T & 1 (-5) & 1 (-5) & 5 (-4) & 30 & 5 & 20 \\
\hline
\end{tabular}
\end{center}
Fluxes in Jy of a 15 M$_{Jupiter}$ object at 10pc distance, as a
function of waveband (K, L, M $\equiv$ 2.2, 3.8 and 4.6\micron,
respectively) and age (adapted from Burrows et al.\ 2001).  The
effective surface temperature is also indicated, as is the spectral
class, which passes from M to L to T as the brown dwarf ages.  The
distance away that PILOT could detect such an object ($5 \sigma$, 1
hour) is also indicated.  The brown dwarfs themselves can be
identified through their colours, though imaging through additional
filters for on- and off- the methane bands, at 1.7 and 3.3\micron,
would be required for accurate spectral classification.
\end{table}

The primary advantage of Antarctica for this experiment is the high
sensitivity in the L and M bands compared to temperate sites.  The
project described here is a survey project aimed at detecting old
($\sim 10^9$\,years) brown dwarfs, though it is just one of many
possible investigations of brown dwarfs that could be conducted. To
detect all such brown dwarfs within a distance of $\sim 20$\,pc in a 1
square degree survey area, as described in Table~\ref{table:bds},
would take $\sim 1$ month per filter (assuming a 25\% observing
efficiency over that time). It is necessary to survey in at least 3
filters (K, L, M) and ideally also to obtain follow-up measurements in
methane on-band and off-band filters (the methane bands are at 1.7 and
3.3\micron) to cleanly spectrally type the brown dwarfs. This project
would therefore benefit greatly from the use of a dichroic
beamsplitter and multiple arrays.  Shallower surveys, covering greater
areas of sky, could be conducted of younger (and hotter) brown dwarfs;
for instance by spending 1 minute per position instead of 1 hour,
brown dwarfs of age $\sim 10^7$\,years could be detected to distances
of 2,000, 100 and 60\,pc, in the K, L and M band respectively, within
a 50 square degree survey area.

PILOT's good near-IR image quality over wide fields of view will
also allow it to make headway in the task of undertaking astrometry of
nearby objects.  Brown dwarfs are now regularly being discovered
through several survey programs. They represent a ``linking class'' of
objects, between the relatively well-understood low-mass stars, and
the (currently unobservable) extrasolar planets being inferred around
nearby stars. A key measurement required to quantify the physical
characteristics for these brown dwarfs is their distance, and the only
way to obtain it is via astrometric measurements of their parallax.

With its wide-field of view (providing many background reference stars
against which to measure positions) and image quality of a factor of
five better than that available on current leading astrometric
facilities (i.e.\ $0.2''$ {\it vs.}\ $1''$), PILOT would be able to
push the frontiers for high precision astrometry from the current
0.5~milli-arcsecond/epoch limits down to
0.1~milli-arcsecond/epoch. This is equivalent to measuring distances
five times further away, extending the reach from 50\,pc to 250\,pc.  This
is sufficient to measure a number of Southern star clusters,
and so make possible a census of their brown dwarf populations.

\subsubsection{Studies of pulsar wind nebulae}
A pulsar wind nebula (PWN) represents the interaction of the
electromagnetic wind from a young pulsar with the surrounding
environment. Some 99\% of the pulsar's spin-down energy is carried
away in this electromagnetic wind, which is only seen through the
shock features resulting from the interaction. One of the features
seen is unequal bipolar jets, extending along the inferred
pulsar spin axis. These jets often terminate in one or more bright
knots of emission. Other features seen include fibrous arcs, or wisps
as they are often called, which are cylindrically symmetric about the
spin axis.

Furthermore, these features are dynamic. The Crab Pulsar Nebula is the
prototype of an active PWN\@. The knots, wisps and jets have been
observed (Hester et al.\ 2002) to vary on timescales of a month at
both optical and X-ray energies, and a good correspondence has been
found between images in both energy ranges (with HST and
Chandra). Faster variability, on timescales of days, has also been
observed in the near-infrared by Melatos et al.\ (2004) using adaptive
optics on Gemini. The latter images have allowed a good determination
of the spectral slopes of the different features, and have provided
evidence for different emission processes in the wisps compared to the
knot at the base of the jet.

The most promising advantage offered by PILOT is the possibility of
high-angular resolution imaging at optical wavelengths, allowing one
to probe physical scales previously only achievable with HST\@. An
interesting possibility would be to examine variability of both the
non-thermal shock-related features and the emission-line features,
such as the H$\alpha$ filaments in the surrounding environment.

Diffraction-limited imaging in the near-IR over a full field, without
the distortions introduced by AO systems, would be desirable for this work. One
of the problems encountered by Melatos et al.\ was good
characterisation of the changing PSF over the AO-corrected field, a
problem what would be mitigated somewhat by the large isoplanatic
angle available at Dome C\@.

There are two approaches that could be taken for the observations. One
would be to take high-resolution images through a range of filters,
from the optical to near-infrared, to determine the spectral shapes,
and hence the emission processes. The second approach would be to look
for variability in the features over a period of months. This would
require imaging once per week (depending somewhat on the individual
PWN being studied), for several months, building up a lightcurve and
looking for movement of features. If done at near-IR wavelengths,
however, the frequency may need to be higher, due to the potentially
faster variability.

\subsubsection{Searching for obscured AGB stars in the Magellanic Clouds}
The chemical enrichment of the ISM is largely provided by mass loss
from evolved asymptotic giant branch (AGB; i.e.\ post-helium burning)
stars.  Though many such stars have been identified in optical and
near-infrared sky surveys, those with the highest mass loss rates
($\rm \ge 10^{-5} \, M_{\odot}$/yr) are often too obscured to be
seen optically.  Their spectra peak between 2 and 5\micron. In many
ways these are the most interesting sources, yet because they are only
bright in the infrared and in molecular lines, they remain a poorly
studied population.  This is especially the case in the Magellanic
Clouds (MCs), due to their greater distance and the limited
sensitivity and resolution of the IRAS satellite through which they
have been identified.  However, the well-constrained distances to the
MCs, and the relative lack of Galactic foreground absorption or
emission, provide a key advantage: stellar luminosities can be
determined more accurately.  This has been used, for example, to
demonstrate differences in the period-luminosity relation between
obscured and optically visible AGB stars (Wood 1998).  In addition,
the low metallicity of the MCs provides a unique environment in which
to study the process of `early' metal enrichment as would have
occurred in the young Universe (see e.g.\ van Loon 2000).

Our knowledge of the AGB population in the LMC, based for many years
principally on IRAS data, has improved significantly with the
completion of the 2MASS and MSX surveys (Egan, van Dyk \& Price 2001).
The MSX A band (8\micron) provides better source positions and
sensitivity than the IRAS 12\micron\ band.  Egan et al.\ (2001) showed
that inclusion of the MSX data can lift the severe colour
degeneracies, e.g.\ between planetary nebulae and HII regions, that
are present in near-IR data from 2MASS\@.  However, the sensitivity
of MSX was limited to magnitudes brighter than 7.5, whereas a survey
with PILOT could reach a limiting sensitivity of 14.8~mags at
3.8\micron, 13.4~mags at 4.8\micron\ and 7.6~mags at 11.5\micron,
over many square degrees (see Table~\ref{table:surveys}), and deeper
if just selected regions of the sky were chosen.  This would be
particularly valuable for identifying the dust-enshrouded post-AGB
stars that are the progenitors to planetary nebulae.  These are
expected to be quite rare due to the short duration ($\sim
10^3$~years) of this phase of stellar evolution, so well-calibrated
data with $\lambda > 3$\micron\ are essential to distinguish them from
the far more numerous red giant population.  Follow-up studies could
include photometric monitoring to study their variability,
spectroscopy to look for nebular emission lines, and infrared and
(sub-)millimetre spectroscopy to study their dusty, molecule-rich
envelopes.

Antarctica is the best place on earth to view the Magellanic Clouds.
Around latitude $-30^{\circ}$, where most southern telescopes are
located, the MCs reach only moderate elevations of 40--50$^{\circ}$,
and then only in the summer if one wishes to observe at night.  In
Antarctica the MCs can be observed continuously at high elevation.
There are other advantages that PILOT has for this work.  The wide
field helps to conduct blind surveys of the MCs for sources missed by
IRAS\@. The high angular resolution helps to resolve sources in star
clusters and crowded fields.  The wide spectral coverage, from the
visible into the thermal infrared, helps for detailed spectral typing.

\subsubsection{Stellar populations and near-field cosmology}
\label{sec:stellarpops}
Our understanding of galaxy formation and evolution has largely rested
to date on the measurement of global quantities like galaxy
luminosity, mass, colour and type. New measurements, however,
involving deep optical and near infrared imaging of stellar
populations in a few nearby galaxies, have begun to show that these
global estimates hide a richer past.

In the Local Group, there are two dominant galaxies (the Milky Way and
M31) and more than 40 (mostly dwarf) galaxies. The star formation
histories of all dwarfs reveals an old stellar population, followed by
a complex and chequered history. While there may be some dependence of
the populations on the radial distance of a dwarf to the nearest large
galaxy, existing data are not sufficient to provide strong support for
such assertions.

\begin{figure}
\plotone{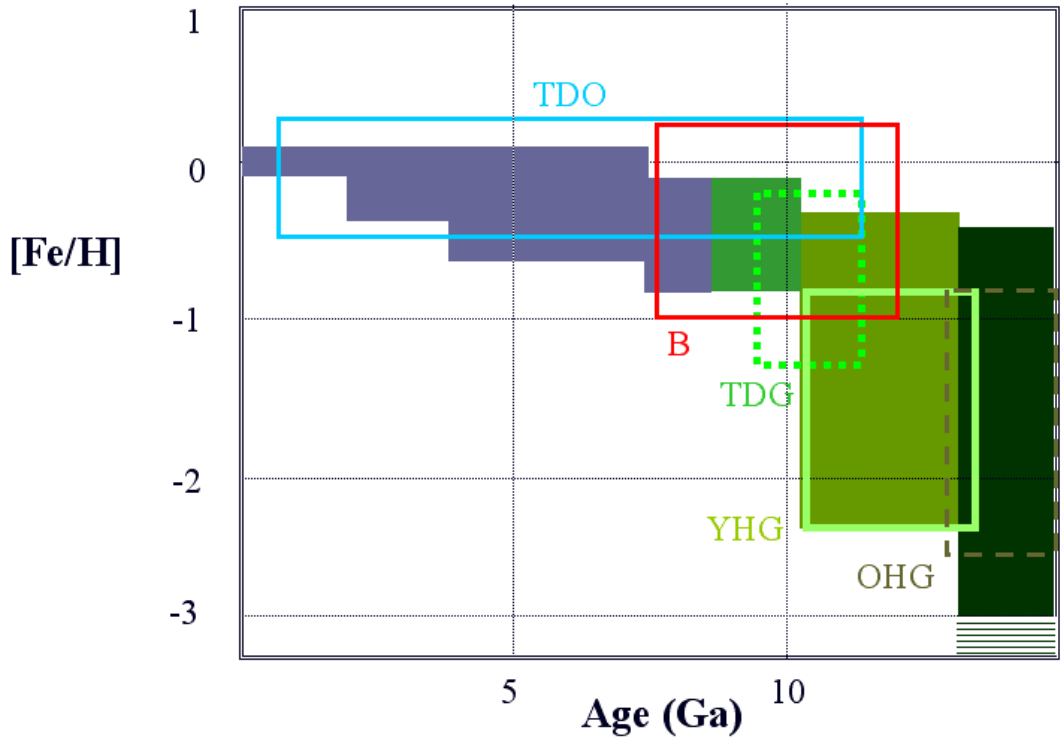}
\caption{The age-metallicity relation of the Galaxy for its different 
component: TDO = thin disk open clusters; TDG = thick disk globular
clusters; B = bulge; YHG = young halo globular clusters; OHG = old
halo globular clusters. The blue corresponds to thin disk field stars,
the green to thick disk field stars and the black shows the
distribution of halo field stars extending down to [Fe/H] = $-5$. From
Freeman \& Bland-Hawthorn, 2002.}
\label{fig:agemetal}
\end{figure}

For the Galaxy, we know a great deal. For instance,
Fig.~\ref{fig:agemetal} shows the age-metallicity distribution for all
components of the Galaxy. This complex plot, however, defies any
simple interpretation. The stellar bulge, halo and thick disk are all
dominated by old stellar components.

The deepest colour-magnitude diagram to date has come from the Hubble
Space Telescope (Brown et al.\ 2003). A small patch of the outer M31
halo was imaged in two bands for a total of 100 orbits. The
data reach down to V=31 (with 50\% completeness), well below the main
sequence turn off at the distance of M31\@.  These data clearly
demonstrate the important contribution of an intermediate age
(7--9\,Gyr) population in M31, a stellar population not seen in the
Galactic halo.  This emphasizes the different accretion histories that
are possible within two neighbouring galaxies.

Unfortunately, the HST is likely to be decommissioned before this
important work can be completed. There is an opportunity for PILOT to
fill this gap, through imaging galaxies in the local group in V and K
bands to resolve their stellar populations. It is necessary to study a
variety of galaxies in different Local Group density fields, in order
to examine how the accretion histories depend on both the halo or the
local density field.  Around 20 galaxies should be studied, covering
four different group density regimes.

Important inroads into this issue can be made out to 2\,Mpc with
PILOT\@. In V band, reaching to 29 magnitudes, it is possible to
extend below the main sequence turn off point for an old stellar
population. Equivalently, it is necessary to reach K=25 or J=25
magnitudes. While the best colour baseline will include V and K, the
project can be tackled, to a lesser extent, through I--K colours, and
even just J--K colours. The main source of confusion comes from signal
fluctuations due to unresolved stars. A $0.2''$ PSF will be
adequate for this project, fully-sampled over $5'$ fields of view. To
go this deep in V band, will require $\sim 100$\,hours of
integration to achieve a SNR=5. To go beyond the Local Group,
however, for instance to reach the Virgo cluster, will require the
successor telescope to PILOT\@.

\subsubsection{Stellar streams and dark matter halos}
A fundamental prediction of cold dark matter (CDM) cosmologies is that
the dark matter halos of massive galaxies, like our own Milky Way,
should be significantly flattened.  The shape of the dark matter halo
influences the motion of orbiting satellites, with asphericities
introducing strong torques which act to precess the orbit.  Until
recently testing this has proven to be problematic due to the lack of
suitable kinematic tracers beyond the Galactic disk.

The Sagittarius dwarf galaxy is being slowly dismembered by the tidal
forces of the Milky Way. Throughout its demise, stars have been torn
from the body of the dwarf and now litter the orbit, and this tidal
stream of stars now completely encircles the Galaxy.  With its
extensive range through the dark matter halo, it was realised that the
morphology and kinematics of this stream provided an ideal tracer of
the underlying mass distribution of the halo. Several analyses of the
stream, however, have concluded that the dark matter must be
essentially spherical, at odds with theoretical expectation.

While not completely ruled out by CDM, the apparent spherical form of
the halo of the Milky Way raises questions as to the nature of dark
matter in general. Clearly, it is important to determine the shapes of
the halos of other galaxies.  While tidal streams tend to be extremely
faint ($\sim$29--30 mags/arcsec$^2$), and hence difficult to detect
when unresolved, they stand out morphologically when considering the
distribution of individual stars (Ibata et al.\ 2001, McConnachie et
al.\ 2004).  As pointed out in \S\ref{sec:stellarpops}, PILOT can play
an important role in the detection of resolved stellar populations
within the Local Group.  With a wide-field camera, it will allow an
expansion of this program, allowing global stellar population
properties to be determined (Ferguson et al.\ 2002). Mapping the halos
of Local Group galaxies and nearby groups (such as the Sculptor Group)
will clearly reveal any tidal streams associated with disrupting
systems, calibrating the current rate of accretion. The morphology of
any tidal streams will provide a measure of the mass and shape of the
dark matter halo, allowing a catalogue of halo shapes to be
determined.

To fully characterise the shape of the dark matter halo, the detailed
orbital properties of a dwarf undergoing disruption need to be
determined.  To do so, the kinematic properties of the stellar streams
are required.  Current studies of the extensive stellar stream
discovered in M31 reveal that, with the 10\,m Keck telescope, stellar
kinematics can be determined to an accuracy of 10~km/s for a star with
m$_I \sim 21$~mags in a 1 hour integration (Ibata et al.\ 2004).
Since orbital velocities are of the order 250~km/s, this velocity
resolution is sufficient to accurately determine the kinematics of the
stream.  While PILOT may be too small to compete with Keck in this
measurement, a larger Antarctic telescope equipped with a spectrometer
would readily provide kinematic measurements of tidal stream stars in
various galaxies, and hence place significant constraints on the shape
of their dark matter halos.

This program, to image at high resolution and sensitivity in the V and
K bands the halos of galaxies, takes advantage of the high angular
resolution and low sky background afforded by Antarctica.  It
requires imaging with $0.1''$ pixel scale, and a wide field in order
to include the halos. It is focussed on Local Group galaxies; i.e.\
those within $\sim 3$\,Mpc of the Sun.

\subsection{Our Universe and its evolution}
\subsubsection{A survey for supernovae in starbursts}
Core-collapse supernovae (CCSNe) are responsible for generating, then
liberating, the bulk of the light metals in the Universe.  As such,
all models for the chemical evolution of galaxies have the CCSNe rate
as a fundamental input parameter.  Despite the number of amateur and
robotic supernova searches now underway, these can only deliver a
lower limit to the actual rate.

The CCSNe rate will of course be greatest in the regions where the
star formation rate is high, i.e.\ starbursts, Luminous (LIRG) and
Ultra-Luminous (ULIRG) Infrared Galaxies. Unfortunately, these regions
are also heavily obscured by dust (A$_v > 10$), and crowded with young
star clusters, making early detection of CCSNe in these regions
extremely difficult. Near-infrared, diffraction-limited observations
with PILOT would yield the spatial resolution required ($\rm 0.2''
\equiv 100$\,pc at 100\,Mpc) to reveal $\sim$5 new CCSNe per year
(based on current models; Mattila \& Meikle 2001) against a crowded
and dusty background in 15 LIRGs/ULIRGs, that would be missed by
existing searches.  The ability to re-visit each galaxy at any time
during the 4 month winter increases the odds of a discovery, as well
as enabling a light curve to be built up.  Adaptive optics systems
such as NAOS--Conica (NACO) on the ESO--VLT do not perform as well as
had been hoped when guiding on the bright, but partly-resolved ULIRG
nucleus, so truly diffraction-limited imaging with a 2\,m telescope in
Antarctica should be competitive with this facility.

Observations in the K band (2.2\micron) are best suited for
discovering, then confirming CCSNe, while J and H (1.25, 1.65\micron)
images assist in determining the line-of-sight extinction and in
confirming the type of supernova found.  Observations are needed of
15 LIRGs/ULIRGs three times per year for 20 minutes each. The
opportunity for discovery increases with the frequency of visits, as
does the quality of the light curves and hence the accuracy of the
sub-classification.  Each CCSNe discovered allows us to probe the
circumstellar medium, which contains a fossil record of the
progenitor's mass-loss history and/or a binary companion, as in the
case of SN~2001ig (Ryder et al.\ 2004).

\subsubsection{Time delays in gravitational lenses}
One of the more elegant ways to measure the Hubble Constant, H$_0$, is
to utilise strongly-lensed quasars. When a quasar is lensed by
intervening matter into two or more images, the distance that the
light travels to form each image is different. This path length
difference will translate into a time delay measured by the observer
between the images. If the quasar shows significant photometric
variations, the light curves for each image will be shifted relative
to each another. One can obtain a measurement of $H_0$ since the time
delay depends on the geometric distances to the lens and the source,
which in turn depends on the value of $H_0$\@. Such a measurement of
$H_0$ is independent of systematics found in determinations that use
the ``cosmic distance ladder''. Recent reviews of the topic can be
found in Kochanek \& Schechter (2004) and Kochanek (2002).

In practice, the measurements are more complicated than simply
measuring a time delay. It is also necessary to obtain good astrometry
of the lensed images, and good quality imaging of the entire
system. The astrometry is required to determine by how much the images
have been deflected. Good quality imaging is important for the
construction of the lens model, to constrain the nature of the lensing
object (i.e.\ is it a galaxy or a cluster?), and to determine the
location of the lensed images. Images with good spatial resolution are
thus crucial for accurate measurements. Such studies are currently
done with ground-based telescopes but often require some degree of
deconvolution, which carries with it potential systematic effects.

The evaluation of $H_0$ from gravitational lenses allows comparison of
measurements made in the local universe, such as the value of
$H_0=72\pm8$~km/s/Mpc from the HST Key Project on the extragalactic
distance scale (Freedman et al.\ 2001). Recent lens determinations
yield a similar value ($71\pm3$~km/s/Mpc, Kochanek \& Schechter 2004)
only under the assumption that the lensing galaxies have a constant
mass-to-light ratio. The more likely case (based on theory and other
observations), of an isothermal density profile in these galaxies,
results in a significantly lower value of $48\pm3$~km/s/Mpc, raising
the possibility that local determinations of $H_0$ are too high.
These results are based on just four systems, however, so more lensed
systems need to be investigated. A telescope such as PILOT provides an
opportunity to address this problem, and offers two important
advantages for observing programs over other telescopes:

\begin{enumerate}
\item  {\it High spatial resolution at optical and near-infrared
wavelengths.} Since PILOT will provide near-diffraction limited images
from the optical to the near-IR, the performance makes it comparable
to the HST\@.  Good angular resolution will enable accurate
modelling of the lens system, crucial for correct interpretation of
the time-delays and a measurement of $H_0$.

Diffraction-limited imaging in the near-infrared is a niche not being
exploited elsewhere. As well as high angular resolution, K band
(2.2\micron) imaging will be important for studies of dusty, reddened
lensed systems, where extinction in the lensing galaxy makes the
quasar images fainter, increasing the photometric errors for a given
exposure time. Additionally, near-infrared light is more
representative of the total stellar mass of a galaxy, so K band
imaging of the lens galaxy will allow better mass density
determination.

\item {\it Ability to monitor continuously for long periods.} 
For good determination of the time-delay, good temporal sampling is
necessary, particularly when the time-scale of photometric variations
is not known, a priori. Furthermore, the time-scale can range from
days to months, and so a long baseline may well be needed to determine
the delays.

\end{enumerate}

A typical observing scheme that can be envisaged would be a
``snapshot''-style program of relatively short exposures, repeated
regularly to build up a light curve. These exposures could thus fit
into other scheduled programs being executed on the telescope. An
example program would be an image every 12--24 hours, over the
winter. This would build up a dense lightcurve that would enable a
quite precise time-delay measurement.

Known quasar lenses are rarely more than a few arcseconds in size
(given by the separation of the images), but the lensing system will
often be larger. The field of view of the strawman AO-corrected
optical imager, or the high-resolution near-IR imager, will be
sufficient for these purposes.

\subsubsection{Galaxy evolution: a deep near-IR extra-galactic survey}

This program addresses three key areas studied extensively with recent
Hubble Space Telescope (HST) deep field observations, through a
wide-field near-IR survey:

\begin{enumerate}


\item {\it I band drop out galaxies at $z>6$}. With the release of the
HST ultra-deep field (UDF), attention has focused on the possibility
of $z'$--band drop outs and higher redshift objects.  However, such
work requires IR photometry in order for candidates to `drop out' of
the $z'$ band.  The scarcity of such objects requires wider field
observations than are practical with HST/NICMOS or 8\,m class
telescopes.  Using the PILOT near-IR camera it would be possible to
take this technique to the next level, with observations in the ZYJHK
bands replacing the classical Lyman break selection using UBVRI at
lower redshifts ($z<4$).  The wide field of PILOT makes this project
competitive in this field with an investment of some tens of hours of
telescope time.

\item {\it New classes of object.} Dickinson et al.\ (2000) discovered 
   a quite extraordinary objects (HDF--N J123656.3+621322) in the
   NICMOS observations of the HDF--N\@.  This object was detected only
   in the NICMOS H band image and then in follow-up ground-based K
   band imaging.  Recently, Yan \& Windhorst (2004) reported several
   similar extremely red objects in the NICMOS UDF image.  However,
   their nature and space density is currently unknown, with plausible
   explanations ranging form $z>10$ galaxies to extreme carbon stars.
   A wide field survey will make it possible to address problems
   associated with small number statistics and define a plausible
   spectral energy distribution for such types of object.

\end{enumerate}

A third program, studying galaxy evolution by examining the morphology
of galaxies at $z>1$ through their rest frame optical light
red-shifted into the near-IR, is described in more detail in
\S\ref{sec:galmass}.

A 2\,m telescope at Dome C offers a critical combination of
depth and area for these survey projects\@.  The deep optical HST
surveys (HDF--N/S, GOODS, UDF) have shown what can be achieved with
excellent quality data. The near-IR surveys undertaken with the
HST/NICMOS camera, however, are limited in their angular coverage.
PILOT, however, could deliver images to depths and resolutions on a
par with NICMOS, over fields of view that are far larger.

Unfortunately, the southerly location means that PILOT will struggle
to observe the southern HST GOODS field, centered on the Chandra Deep
Field South.  However, the HDF--S is well placed for study, providing
excellent optical data to use in tandem with the PILOT observations.

One could envisage a series of survey projects, with different depths
and area coverage, to address complementary science goals.  An
excellent model for such a survey would be the three planned UKIRT
extragalactic near-IR surveys, which provide a sequence of increasing
depth for smaller areal coverage (viz.\ the UKIDSS Large Area Survey
(LAS), Deep Extragalactic Survey (DXS) and Ultra Deep Survey
(UDS)\footnote{See www.ukidss.org.}).  The key opening for PILOT is
improved spatial resolution, which will uniquely allow morphological
studies, for a large sample of galaxies, in the rest frame optical
bands over the redshift range $z=$1--3.

To tackle such projects requires the use of large-format near-IR
cameras, with a $0.1''$ pixel scale to adequately sample the PSF at
the shortest wavelengths. The filter set should include ZYJH and
K\@. With a field of view of $5'$, this would be well matched to the
HST deep fields. With a 10 hour exposure a sensitivity of 22.3
magnitudes per square arcsecond would readily be obtained in K band,
on a par with those of the NICMOS HDF--S surveys.  The resolution and
area combination will allow ground breaking studies. The basic program
therefore requires 50 hours of observation (10 hours in each of 5
filters) on one field.

\subsubsection{Emission-line mapping of the high redshift universe}
While classical imaging and long-slit spectroscopy have provided a
wealth of astronomical data, our understanding of the detailed
properties of objects in the high redshift universe requires spatially
resolved spectroscopy.  The good seeing and low sky background
available in Antarctica immediately lends itself to integral field
spectroscopy.  Two potential astrophysical studies include:

\begin{enumerate}
\item {\it Gravitational lensing.}
Massive galaxies can split the light from distant quasars into a
number of images, separated by 0.5--1$''$ on the sky. As well as
producing multiple images, these gravitational lenses can also induce
significant magnification, producing gross distorted images of normally
unresolved features, such as the Broad Emission Line Region (BLR;
Mediavilla et al.\ 1998; Motta et al.\ 2004). With gravitational lens
inversion, the fine detail of such emission regions can be exquisitely
mapped (e.g.\ see Wayth et al.\ 2005).  Furthermore, additional
differential magnification effects are expected to occur, resulting in
differing variability for continuum and emission sources (Lewis et
al.\ 1998). For such a study, spectroscopic monitoring is required,
but it is also important for the emission from various lens images to
be clearly separated. Hence, an integral field unit taking advantage
of the good seeing afforded by an Antarctic telescope is ideal.

\item {\it Star-forming  galaxies.}
In recent years, the sub-mm / infrared view of the high redshift
universe has revealed that much of the star formation in young systems
is hidden, buried in dust cocoons that re-radiate the intense UV
produced by massive young stars at much longer wavelengths.  Sub-mm
imaging has uncovered a substantial population of these galaxies at at
redshifts of $z>1$, and integral field spectroscopy of the brightest
examples have revealed complex structure and dynamics, interpreted as
violent interactions and star burst induced superwinds (e.g.\ Swinbank
et al.\ 2005).  Furthermore, similar star forming populations in the
early universe have now been uncovered by the space infrared telescope
Spitzer.  Integral field spectroscopy of these galaxies will be vital
in uncovering the physical properties driving star formation in these
young systems.

\end{enumerate}

Of these two studies, gravitational lens systems provide a simpler
observational challenge, requiring integration times of order 1 hour
with current integral field spectrographs on 4\,m class telescopes in
order to make detailed maps. The high angular resolution, however, is
needed to ensure that the resultant continuum subtracted images
reflect real emission features, rather than poor PSF subtraction. The
study of star forming galaxies is more technically challenging, with
Ly$\alpha$ imaging of the brighter Spitzer sources requiring 8 hours
on current 8\,m class telescopes. Both Ly$\alpha$ and H$\beta$ imaging
would take advantage of the significantly lower sky background in
Antarctica, so bringing these projects into the grasp of PILOT\@.

To conduct this program requires using a lenslet array with $0.1''$
spatial resolution.  Widths of typical lines are from 500--5000~km/s,
targeting Ly$\alpha$ at $z>2.2$ ($\lambda > 3900$\AA) or H$\beta$ in
the near-IR (for $z>1$).

\subsubsection{The complete star formation history of the early universe}
\label{sec:sfhistory}
The rise of the Universal star-formation rate (SFR) from $z=0$ to
$z=1$ is well established from a variety of indicators (ultraviolet,
Balmer lines, far-IR emission). The history of the Universal SFR for
$z>1$ is poorly determined; it appears to peak and decline from $z=1$
to $z=6$ (the infamous `Madau-Lilly diagram'; Madau et al.\ 1996,
Lilly et al.\ 1996). However this result comes from the measured
ultraviolet light of high-redshift star-forming galaxies (`Lyman Break
Galaxies' or LBGs) and there is a big problem: dust (see e.g.\ Steidel
et al.\ 1999, Glazebrook et al.\ 1999).

Interstellar grains are thick in star-forming regions and are very
effective at absorbing ultraviolet (UV) light from young stars. For
example, typical visual extinctions {\em observed} in star-forming
galaxies locally predict large extinctions ($> 2$\, mags) in the UV\@.
In fact, there is now extensive evidence that this is also true at
high-redshift. The slopes of the UV spectra of LBGs are consistent
with around 2 mags of extinction.  There is also a population of
sub-mm sources (the `SCUBA galaxies') which can only be explained by
very dusty starbursts. Typically when plotting a SFR {\it vs.\ z}
diagram constant extinction is assumed for lack of anything better. The
majority, 90\%, of the UV light is absorbed.

Paradoxically, it turns out that the best place to measure the SFR of
a galaxy is in the deep red. This is because this is where the
H$\alpha$ emission line comes out. The H$\alpha$ line strength is a
direct measure of the number of ionizing photons in a galaxy and hence
the number of young stars. Being emitted at 6563\AA\ it is relatively
little affected by dust, compared to Ly~$\alpha$, with a mean A$_v
\sim 1$\,mag.\ through the galaxies. 
At $z > 1$ H$\alpha$ is observed in the near-IR, and some limited
spectroscopy has been obtained from $z=1$ galaxies (Glazebrook et al.\
1999, Doherty et al.\ 2004) and $z=3$ LBGs (Erb et al.\ 2003). These
typically show the SFRs to be several times larger than was calculated
from the UV light, consistent with most of it being absorbed by
dust. However, only a handful of galaxies have been studied due to the
extreme difficulty of obtaining near-IR spectroscopy, even on 8\,m
class telescopes.

To truly measure the evolution of the cosmological SFR we need deep,
H$\alpha$-selected samples to allow the construction of the H$\alpha$
luminosity function from $z=1$ to $z=4$. The only existing dataset 
which comes close to doing this
is from HST, using the NICMOS camera in a slitless grism mode in J \&
H. Because of the low space background, slitless images can be used to
directly search for lines down to a limiting H$\alpha$ flux. The
resulting H$\alpha$ luminosity function has been measured at $z=1.5$
(Yan et al.\ 1999, Hopkins et al.\ 2000) for only a few dozens of
objects over just tens of square arcminutes.

The `cosmologically dark' K band window from PILOT would allow us to
extend this work to high-redshift and backwards in time -- in particular to
$z=3$ and the era of Lyman break galaxies. Deep imaging in the K band
at a series of redshifts from $z=2.0$ to $z=3.0$ with narrow band
filters is required. In conjunction with deep broad--K images,
narrow-line excess objects can be identified using standard
techniques.

Is star-formation really declining from $z=1$ to $z=3$? Such a long
baseline (2--6\,Gyr after the Big Bang) will reveal the epoch of peak
galaxy formation. The measured SFR-distribution function will probe
how it depends on galaxy size, and when combined with the deep
broad--K observations discussed in \S\ref{sec:galmass}, on galaxy
mass. In fact, the combination of direct SFR measurements and stellar
mass measures of the same galaxies has proved powerful at $1 < z <2$
in the Gemini Deep Deep Survey -- direct evidence for `downsizing'
(the motion of peak SFR to smaller mass galaxies with time) is
seen (Juneau et al.\ 2005). Both these measurements at $z > 2$ are
ideally suited for deep K band observations.

This program requires narrow-band observations which are most
sensitive in the darkest windows; ie the K--dark window is ideal and
confers a considerable advantage for this type of work.  Diffraction limited
imaging (0.3") over a wide field, only possible from Antarctica, makes
it possible to search for extremely compact dwarf star-forming
galaxies at these redshifts. The number of galaxies found
is the determining factor for luminosity function
calculations, and so the uniquely wide field is an advantage.

Narrow band filters, of 1\% bandwidth, will be required from 2 to
2.6\micron. A 10 hour exposure through a 1\% filter at
2.2\micron\ will reach a line flux of $10^{-17}$\,ergs cm$^{-2}$
s$^{-1}$ for PILOT parameters. At $z=2.35$ this is L(H$\alpha$) = $4
\times 10^{41}$\,ergs/s, or a SFR of $2.8 M_{\odot}$ per year, an exceptionally
low level to be sensitive to.

The line limit corresponds to an equivalent width of 20\AA\ in the
{\it rest frame} if we match the galaxies detected through their
emission lines in the narrow band filter to those that would be seen
in a deep (i.e.\ 100 hour, K$<23.5$ mags) broad band image (as discussed in
\S\ref{sec:galmass}). This would result in complete emission line 
identification of the K-selected sample (the limit would detect 80\%
of star-forming galaxies locally and even more when extrapolated to
high redshift) with corresponding unambiguous photometric redshift
identification from the line detections $+$ broad-band colour
constraints. Comparing the measurements of stellar mass and SFR would
allow us to study the `downsizing' of galaxy assembly at $z>2$ when
the most massive ones form.

The number of objects that would be found can be estimated using the
luminosity function of Hopkins et al.\ and transposing it to $z=2.35$
(i.e.\ with no evolution). For $S > 10^{-17}$\,ergs cm$^{-2}$ s$^{-1}$
we calculate a H$\alpha$ source density of 300 per $20'$ diameter
PILOT field, an ample number for eliminating cosmic variance which
plagues smaller surveys. [At z=2.35 the survey size is $\sim 40 \times
35 \times 35$ comoving Mpc, which is many clustering scales.]

One might also want to cover a larger area ($3 \times 3$ pointing)
grid with shallower surveys as in the `galaxy stellar mass' survey
discussed in \S\ref{sec:galmass}. This two tier approach would better sample
rare, brighter objects as well as the more common, fainter ones.

With such deep narrow-band observations there might be many
unidentified lines, some with objects too faint for reliable
photometric redshifts and some which only appear in the line (i.e.\ no
continuum). These would be excellent candidates for follow-up
multi-object spectroscopy -- the flux limit of
$10^{-17}$\,ergs\,cm$^{-2}$\,s$^{-1}$ is easily in reach of telescopes
such as Gemini with forthcoming new instruments
(a typical calculation shows S/N=5 at the limit in 10
hrs). An Antarctic ELT would do considerably better. Spectroscopy at
moderate resolution (R=3000) could distinguish between various
possibilities: one would observe H$\alpha$ + [NII] or H$\beta$ +
[OIII] or the [OII] doublet or the asymmetric Ly$\alpha$ line (at
z=17). Spectroscopy with wide wavelength coverage could directly
measure the dust extinction (from H$\alpha/$H$\beta$) and metallicity
(via R23 or O3N2 indices) of the $z>2$ star-forming gas.

Finally we can speculate on the serendipitous discovery of $z=17$
Ly$\alpha$ emission line galaxies. Is the
possibility even reasonable? The serendipitous potential
of really high-redshift searches ($z > 10$) from {\it any} deep K band
observations is high. It is unexplored parameter space.
The Ly$\alpha$ luminosity at $z=17$ would be $>
4 \times 10^{43}$\,ergs/s, which is a few times that of the $z=6.5$
object that Rhoads et al.\ (2004) found in a narrow band optical
search in a survey of comparable sky area. Of course the SFR at $z >
10$ is unknown, so its measurement is a powerful test of the
hypothetical Population III and re-ionization. One point to bear in
mind is that low-metallicity Population III galaxies would have much
higher Ly$\alpha$ equivalent widths than their low-z counterparts. One
might want to pursue this with a deeper narrow-band survey of $>
100$\,hours exposure.

\subsubsection{The evolution of galaxy mass and morphology: high resolution 
imaging beyond the Hubble limit -- a PILOT ultra deep field}
\label{sec:galmass}
The Hubble Space Telescope has revolutionized our view of the
high-redshift Universe with its deep, high-resolution
images. Galaxies paint the sky in numbers up to $10^7$ per square
degree and star-formation rates are much higher than they are
today. The picture of the $1 < z < 4$ Universe is one of great
disturbance: many anomalous looking galaxies abound with little sign
of the regular Hubble sequence.

However HST Deep Fields (HDFs) are inherently biased: they are
predominantly taken in the optical and have very tiny fields of view
of only a few arcmin. At $z > 1$ the optical samples the rest-frame
ultraviolet---in such pictures only the scattered star-forming regions
containing young UV luminous OB stars can be seen. Young `Lyman-break
galaxies' (LBGs) with prodigious star-formations rates dominate the
picture. Older, redder more regular stellar components cannot be
seen. HST does have J and H band imaging with the NICMOS camera, but
here the field is even smaller and so no large surveys are possible.

To truly characterise the high-redshift Universe ultra-deep surveys
are needed in the near-infrared, ideally in the reddest possible band
to pick up more normal galaxies at high-redshift. It is also necessary
to cover much larger areas---the typical cosmic variance on the scale
of the Hubble Deep Fields is 100\%, leading to large uncertainties in
measurements.

Tantalizing results from deep ground-based near-IR surveys have revealed
populations of massive galaxies beyond $z > 1$ with much more regular
morphologies. The Gemini Deep Deep Survey has shown spectroscopically
that massive, old galaxies exist to $z=2$ at K$\sim 20$ (GDDS;
Glazebrook et al.\ 2004), and images from the ACS instrument on the
HST reveal regular elliptical and spiral galaxies. Franx et al.\
(2003) have found a substantial population of `Distant Red Galaxies'
(DRGs) with J--K $> 2.3$, $z > 2$ and K$< 22.5$; these appear to be
massive and much redder/less UV luminous than LBGs. One has an HST
NICMOS image showing a classical bulge plus disk.

Why is the near-IR so important? This is where the bulk of the light
from old stars come out, and so to weigh a galaxy by stellar mass it
is necessary to measure the light at wavelengths greater than that of
the 4000\AA\ break. For $z > 2$ this means that the K band is
absolutely essential to measure stellar mass at high-redshift, rather
than the transient UV bright episodes of star-formation. Massive
galaxies tend to be highly clustered as well; there are zero at $z >
2$ in the HDF--N and 3 in the HDF--S (as determined from deep VLT K
band imaging) for example.

Ultra-deep, wide-area K band imaging will obtain resolved images over
substantial areas. The goal is to image large numbers of $z > 2$ red
galaxies and measure their morphologies. Are they ancestors to spiral
galaxies? Do they have disks? How far back in time can we see disks or
genuine elliptical galaxies?  Such galaxies are the most interesting;
galaxies which are already old and at $z=3-4$ constrain the epoch of
first star-formation in the Universe, pushing it back to $z >> 6$.

Quantitative morphological measures (bulge/disk decomposition,
concentration, asymmetry, `Gini' coefficients) will be
applied. Comparison with deep optical images will establish
photometric redshifts. Stellar mass functions will be established as a
function of redshift and morphology, which will measure the nature of
the growth of galaxies (`hierarchical assembly' or `down-sizing'?) and
hence test galaxy formation models. By covering large areas we will be
able to measure clustering and hence obtain constraints on the dark
mass of their halos.

This project takes advantage of the `cosmological window' at
2.4\micron, where the sky is very dark, to enable ultra deep fields to
be obtained.  It also takes advantage of the wide diffraction-limited
field offered by Dome C.  The southern latitude means
most of the southern sky is circumpolar and visible all winter, there
is no daylight to interrupt observing. This is ideal for the
accumulation of long exposures on deep fields. PILOT is the only
facility conceived which can go to the required depth, over a large
enough area.

Wide-field, K band imaging, with a resolution of $0.3"$ or better, is
required. J and H images are also desirable and perhaps V and I images (though these
images could be obtained elsewhere, albeit not in as good seeing).

To study rare bright objects as well as more common fainter objects
a two-tier approach is needed:

\begin{enumerate}

\item A 100 hour exposure will reach K = 23.5 mags for extended sources, thus
K$<22.5$ mags galaxies will have S/N$ > 25$, suitable for determining the
morphology of resolved objects with $0.3"$ resolution. A $20'$
diameter FOV will contain 1,000 galaxies of the type studied by GDDS
and Franx et al.\ and an unknown number of fainter objects yet to be
studied. The field would represent a 100 fold improvement in
cosmological volume on any comparable large telescope imaging (e.g.\
the FIRES VLT field).

\item A 10 hour exposure reaches K = 22.3; a $3 \times 3$ pattern of 
such fields will cover $1^{\circ} \times 1^{\circ}$ on the sky. A
$1^{\circ}$ scale corresponds to a transverse size of 110 comoving Mpc
at $z=3$, this field will be ample for studying the clustering and
large-scale structure of $1 < z < $5 red galaxies (with correlation
lengths $\sim 10$\,Mpc).

\end{enumerate}

The brightest galaxies (K$<21$ mags) found at high-redshift would be followed
up spectroscopically with Gemini (FLAMINGOS-2, GNIRS) to study stellar
populations and velocity dispersions. Fainter galaxies could one day 
be followed up by JWST.

Such a uniquely wide-area deep K band survey has the potential to
detect new populations of objects at very high redshift ($z>13$) which
are invisible at shorter wavelengths. There is quite possibly a second
epoch of star-formation in Population III objects at these redshifts
(see \S\ref{sec:GRB}).  If this is the case then the number of
galaxies that would be bright enough to be seen is unknown. The K =
23.5 mags limit corresponds to a rest frame unobscured UV luminosity from a
SFR of 50~M$_{\odot}$ per year at $z=15$ and would be a factor 2--3
times more luminous than the brightest LBGs at $z=3$. Of course, the
abundance of $z=15$ objects is a subject of educated speculation; for
example these UV luminosities assume a Salpeter IMF, and a Population
III IMF would result in even more UV output. These deep fields would
lay the ground work for the cosmological science case for any future
Antarctic ELT or future space telescopes such as JWST.

\subsubsection{Probing the First Light in the universe with gamma ray bursts}
\label{sec:GRB}
Gamma Ray Bursts (GRBs) are the most powerful, energetic explosions in
the Universe. For a period of a few days they are 100--1000 times more
luminous than Quasars. Current satellite missions are capable of
detecting the gamma ray flux of GRBs to $z=20$ and the
SWIFT mission (launched in 2004) will be able to reach $z=70$ (Lamb
\& Reichart 2001). Note that $z=20$ is 180~Myr after the Big Bang
(just over 1\% of the current age of the Universe), $z=70$ is 28~Myr
(0.2\% of now).

We now know that GRBs are associated with star-formation in galaxies.
They occur in off-nuclear star-forming disks.  Spectra have been
obtained to redshifts as high as $z \sim 4$. The best theoretical
model is that GRBs represent a `hypernova' associated with the core-collapse of
a super-massive star directly in to a black hole.

However 20--40\% of GRBs are `dark bursts'; i.e.\ they have no optical
counterpart. Given that GRBs are detectable to very high-redshifts the
natural conclusion is these dark bursts have $z > 7$. At these
redshifts all optical light ($\lambda <$~Ly~$\alpha$ in the rest
frame) is removed by neutral hydrogen absorption in the IGM.

This would imply that a considerable amount of star-formation occurred in
the Universe at $z > 7$, an epoch not yet probed by any observations,
whether from ground or space. This would address a fundamental problem
in cosmology: the re-ionization of the early Universe.

After its early fireball phase, the Universe consisted of neutral
hydrogen until the first stars were formed at some, currently unknown,
time. This is often referred to as the time of `First Light'. These
early stars would have produced ultraviolet radiation which would then ionize the
Universe, some time between $z=20$ and $z=7$. Observations of galaxies
and quasars at $z=6$ show that the amount of ultraviolet produced at this
`late' time is insufficient to ionize the Universe and further, that
the Universe is already almost completely ionized so that this event must
have happened earlier. Observations of the Cosmic Microwave Background
are consistent with a range of $z = $7--20.

The early universe contained no heavy elements -- these are made in
stars. Calculations of the likely modes of star-formation in pristine
material shows that the very first generation of stars was likely to
be have been very different from current day star formation, which
occurs in a `dirty' interstellar medium. The first stars are expected
to be much more massive, on average, than stars born today. They are
known as `Population III' stars. Being massive, they would produce
considerable ultraviolet radiation and hence be capable of ionizing
the Universe. Many models predict an early peak of star-formation due
to this Population III at $z>10$ (e.g.\ Cen 2003).

Because GRBs are ultra-luminous and trace massive star-formation, they
can be used to trace Population III stars and re-ionization. The
opportunity is timely with the imminent launch of SWIFT, which will be
rapidly followed by further, more powerful, gamma-ray satellites.

PILOT can provide rapid JHKLM imaging of GRB locations to search for
afterglows; the position of the break will be revealed by colours and
hence the redshift can be determined. Redshifts at $z > 20$ can be
probed. The M band can reach $z=35$ if star-formation ever occurred
back then -- only 80\,Myr after the Big Bang.  Studying the redshift
distribution of GRBs will reveal any Population III stars and the
epoch of `First Light'.

The L and M bands can be superbly probed at cosmological distances from
Antarctica, where the sky is uniquely dark. The K band is also enormously more sensitive in the Antarctic than from
temperate sites, providing a window on $10 < z <15$ GRBs.  In other
words, the regime where the Antarctic has its greatest advantage is
exactly the redshift range that needs to be probed.
GRBs are also point sources, which means the good seeing is a major
advantage as it provides much better sensitivity.

The observations need rapid response, which rules out space telescopes
(such as Spitzer and JWST) and even large conventional ground-based
telescopes. The latter are not well suited to following up large
numbers of events. In contrast PILOT could follow-up hundreds. The
long polar night means any rapid response is not going to be
interrupted by any untimely daylight--GRB light-curves could be
followed continuously.  Thus, PILOT is a uniquely capable probe of the
$z > 20$ GRB regime due to its singular combination of sensitivity
and availability for frequent rapid response.

This project needs a $1'$ field of view and K, L and M band imaging
capability; J and H band imaging, in addition, would be useful (SWIFT
will produce $10''$ accuracy in positions, at the rate of about
100--200 per year, so that 10--80 would have $z > 7$). If the FOV was
as large as $10'$, it would be possible to work with the $\pm 4'$
SWIFT $\gamma$-ray burst alert positions --- it would be possible to
follow-up GRBs with weak or absent X-ray afterglows which has never
been done before.

Gamma ray bursts remain bright to very high-redshifts. Lamb \& Reichart
(2001) calculate L=M=$15\,\mu$Jy for a $z>20$ GRB\@. One beautiful reason for
this is that time dilation makes the fading slower at high-redshift, so
if one observes at a fixed time after the event, the time dilation
effect approximately cancels the cosmological dimming.

The Lamb \& Reichart number above refers to 24 hours post-burst. A
typical afterglow fades at $t^{-4/3}$, so one hour after the burst
these would be 70 times brighter (i.e.\ $\sim 1000\,\mu$Jy).

These fluxes are easily detectable with PILOT.  A 5 minute exposure
reaches a noise limit of $300\,\mu$Jy in L; if it is not detected a 1
hour M band exposure would reach $200\,\mu$Jy. This would correspond to
a z=36 object.  For the `easy' $z=10$ case, we estimate that the GRB
would be $700\,\mu$Jy in K band, whereas a 1 minute exposure in this
band reaches to $15\,\mu$Jy.  A filter sequence, JHKLM, would provide the desired
combination of observations for find the GRBs.

We note there is considerable variation in GRB properties -- many would be a
lot brighter but some would be fainter. Part of the science objective
is to pin down in detail their elusive properties.

An obvious upgrade would be a spectrograph working in the K and L
bands. This will allow one to probe the gas in the Universe and the
host galaxies at $z > 10$. It would be possible to measure the
elemental abundances, the large-scale structure of the early Universe
from metal forest absorption lines, and the amount of re-ionization
from the shape of the Ly$\alpha$ edge.

Spectroscopy at R=100--300 would be possible even on a 2\,m telescope.
Such a telescope is also an obvious pathfinder for an Antarctic ELT, which would be
capable of making very high spectral and spatial resolution
observations of Galaxies at $z > 10$.

%
%
%
%
%

\subsubsection{Cosmic shear}
Light rays follow geodesics, which bend in the presence of matter.  It
follows that a coherent shape distortion is imprinted in the
distribution of distant background galaxies by mass fluctuations in
the intervening cosmic web.  This pattern of cosmic shear is a
powerful cosmological probe.  It is directly sensitive to the dark
matter distribution predicted by theory, and does not depend on the
details of how galaxy light traces mass.

The most serious limitation for ground-based optical cosmic shear
experiments is the systematic variation in the point-spread function
(PSF) which arises from inevitable changes in atmospheric seeing and telescope
properties with position and time (e.g.\ Kaiser, Squires \& Broadhurst
1995).  These PSF distortions are typically an order of magnitude
larger than the cosmological shear that we wish to measure.  The
outstanding natural seeing performance (i.e.\ stable atmosphere)
obtainable at Dome C is therefore advantageous for
controlling systematic PSF variability and, equally importantly,
resolving high-redshift galaxy shapes.

In addition to excellent image quality, cosmic shear surveys demand
high galaxy surface densities (10--100 per square arcmin) to reduce
statistical noise.  This requirement is much more important than
measuring accurate shapes for individual galaxies, because there is an
irreducible experimental scatter in the shape information owing to the
unknown galaxy ellipticity before shear.  Galaxy shapes must be
resolved, but are not required to be measured at high signal-to-noise.

Cosmic shear experiments may be broadly divided into two categories
(for useful reviews see Hoekstra, Yee \& Gladders 2002; Refregier
2003):

\begin{enumerate}

\item {\it Experiments mapping `blank fields' to detect the shear pattern
resulting from gravitational lensing by the cosmic web of large-scale
structure.}  Current state-of-the-art surveys cover a few square
degrees and have measured the amplitude of mass fluctuations (denoted
$\sigma_8$, where the subscript refers to a scale of $8$ Mpc) to 10\%
(e.g.\ Bacon et al.\ 2003).  In 10--15 years, cosmic shear surveys
will cover the whole sky and will have the power to characterize the
properties of the dark energy and matter as a function of redshift.

\item {\it Experiments targeting the most massive clusters where the shear
distortions are the largest.}  The shear pattern can be used to
reconstruct the cluster gravitational potential (Kaiser \& Squires
1993), that can be compared with CDM (cold dark matter) theory.
Currently this has been performed for tens of clusters and also from
space using the Hubble Space Telescope.

\end{enumerate}

The required magnitude limits in optical or near-infrared wavebands to
deliver the minimum source density of 10 arcmin$^{-2}$ (of galaxies
that are usable for lensing studies) are R$ \sim 24$ or K$ \sim 21$
mags.  To date, most cosmic shear experiments have been performed in
the optical -- with existing facilities, the attainable surface density
of background galaxies (in a fixed integration time) is roughly an
order of magnitude higher with optical observations than with near-IR
imaging.  Moreover, advances in infrared detector technology have
lagged significantly in terms of the available field-of-view.

However, given the greatly reduced near-IR background available
in Antarctica, it is timely to reassess this situation.  Mapping
cosmic shear in infrared wavebands offers a number of advantages: (i)
galaxy shapes are smoother, tracing older stellar populations rather
than knots of star formation, thus ellipticities are easier to
measure; (ii) background galaxies at high redshift ($z > 1$) are more
readily detected; (iii) the availability of infrared colours greatly
enhances the efficacy of photometric redshifts, allowing a more
accurate determination of the redshift distribution of the background
galaxies; (iv) for observations targeting cosmic shear behind galaxy
clusters, the availability of K band imaging enables the
`contamination' from foreground objects and cluster members to be
greatly reduced (e.g.\ King et al.\ 2002); and (v) combining the K
band luminosity of such clusters with the weak lensing mass estimate
yields a mass-to-light ratio.

Let us assess the competitiveness of some specific cosmic shear
observational programmes possible with PILOT:

\begin{itemize}

\item {\it `Blank-field' cosmic shear survey over $\sim$ 100 square
degrees.}  This would be competitive if completed within $\sim
5$\,years (the CFHT Legacy `wide' survey is intending to map $\approx
170$ sq. deg.).  PILOT can reach the required detection sensitivity of
K$\approx 21$ in a $\sim 1$ hour pointing; the $20'$ field-of-view of
the near-IR wide-field imager implies that $100$ deg$^2$ can be mapped
in $\sim 900$ hours of integration time.  The strawman optical imager
has a much smaller field-of-view and so would not be competitive for
this type of survey.

\item {\it Targeted observations of the outskirts of galaxy clusters.}
PILOT could map the outer regions of clusters (or bright galaxies)
where cosmic shear observations can discriminate between different
dark matter halo profiles (see Hoekstra et al.\ 1998; Hoekstra, Yee \&
Gladders 2004).  The viability of the cluster shear observational
technique in infrared wavebands has recently been demonstrated by King
et al.\ (2002).

\item {\it Targeted observations of superclusters.}  By mapping known
superclusters, a cosmic shear survey can quantify the degree of
filamentary structure present in the web of dark matter (see Gray et
al.\ 2002), a critical observable discriminating between theories of
structure formation.

\end{itemize}

PILOT could also make a valuable contribution in the regime of {\it
strong} gravitational lensing, via observations of lensed arcs in the
vicinity of high-redshift clusters.  It has been shown that the
probability of giant arc formation due to galaxy clusters is a
sensitive measure of global cosmic parameters, particularly for the
dark energy model (Bartelmann et al.\ 1998). Intriguingly, numerical
simulations of clusters in the currently-favoured cosmological
constant model fall short by an order of magnitude in reproducing the
observed abundance of large gravitational arcs (Bartelmann et al.\
2003).  However, another study suggests that simulations using
realistic redshift distributions for background galaxies resolves this
discrepancy in the observed arc statistics (Wambsganss et al.\ 2004).
Also, possible modifications to our current understanding of cluster
sub-structure and evolution may be necessary to help resolve this
difference in theory and observation.

A recent project addressing these issues is the `Red-Sequence Cluster
Survey', which mapped 90~sq.~deg.\ in R and $z$-bands, discovering 8
strong gravitational arcs (Gladders et al.\ 2003).  Near-infrared
imaging is a pre-requisite for pursuing these studies to higher
redshift.  Hence we suggest another program addressing strong
gravitational lensing:

\begin{itemize}

\item {\it Survey for strongly lensed gravitational arcs in galaxy clusters.}
A 100~sq.~deg.\ survey to K$\sim 21$ would yield 100--500 clusters
with velocity dispersions in excess of 700~km~s$^{-1}$ in the interval
$1.0 < z < 1.5$. Doing this type of wide-field, near-IR `arc survey'
of galaxy clusters would address the above questions and obtain an
accurate, independent measurement of cosmological parameters.

\end{itemize}

\section{An observing program for PILOT}
This paper has outlined an extensive range of challenging science
projects that a telescope like PILOT could be used to tackle, from
imaging of our planetary neighbours to searching for signatures from
the first stars to form in the Universe.  The projects also require a
diverse range of instrumentation, from high resolution optical imagers
using AO, through to sub-mm spectrometers.  Some projects involve
monitoring, making periodic measurements of a few hours
duration every few weeks, while others are dedicated surveys that will
require several months to accomplish. 
We anticipate that PILOT could be fully operational within three years of funding approval. 
While it is indeed possible that
these projects could all be tackled with the same telescope, it is
also unlikely that this would happen in practice.  Indeed, given the
low cost of an Antarctic 2\,m telescope compared to substantially
larger telescopes at temperate sites, it is quite possible that once
PILOT is operating, with just one or two of the strawman instruments,
the demand will be such that another similar-sized telescope would be
built equipped with a different instrument complement, perhaps
dedicated to a single project.

\begin{table}
\caption[]{Strawman Observing Program for PILOT}
\label{table:observe}
\small
\begin{center}
\begin{tabular}{lclll}
\hline
Program	& Cat.	& Instrument	& Type of Observation	& Time Needed \\
\hline
Orbital Debris & P & VRI WF & Monitor & 50 hours, twilight \\
&& JHK WF && \\
Planetary Imaging & P &	VRI HR & Monitor &Several days continuously  \\
&& JHK HR && (Opp. Venus, Conj. Mars) \\		
Exo-planet Transits & P & VRI HR & Follow-up detections	& 10--20 hours per candidate,  \\
&& JHK HR && spread over $\sim 1$ week \\		
Disks & P & KLM & Selected sources  & 100 hours \\
&&& -- mini surveys \\
Planetary Microlensing & P & VRI HR & Follow-up detections & 1--2 days continuous,  \\
&&&-- overrides & $\sim 6$ times per year \\
Asteroseismology & P & VRI WF & Monitor & 5--10 days continuous, \\
&&& selected sources & once per year \\
Massive Protostars & G & NQ & Selected sources & 50 hours \\
&& Sub-mm \\
Early SF Census & G & Sub-mm & Selected sources & 50 hours \\
Galactic Ecology & G & KLM & Selected sources & 30 hours \\
Molecular Core Spect. & G & Sub-mm & Selected sources & 50 hours \\
Brown Dwarfs & P+G & KLM & Survey & 3 months \\
Pulsar Wind Nebulae & G & VRI HR & Monitoring &	Few minutes, weekly \\
AGB Stars in LMC & G & KLM & Survey + Monitor & Few hours, weekly \\
Stellar Populations & G & VRI WF & Survey & 1 month per field  \\
&& JHK HR && selected (up to 10 fields) \\	
Stellar Streams	& G & VRI WF & Survey & 1 month per field selected \\
&& JHK HR \\		
SNe in Starbursts & C & JHK HR & Monitor & 8 hours, 3 times per year \\
Time Delays in Lenses & C & VRI WF & Monitor & Daily images \\
&& JHK HR \\		
SF History Early Univ. & C & JHK WF & Survey & 50 hours \\
Ultra Deep Field & C & JHK WF &	Survey & 100 hours \\
First Light via GRBs & C & JHK HR & Follow-up detections &2 hours per override \\
&& KLM  && \\ 		
Cosmic Shear & C & JHK WF & Survey +  & 900 hours \\
&&& Selected sources \\
Strong Lensing & C & JHK WF & Survey & 150 hours \\
\hline
\end{tabular}
\end{center}
\normalsize
Potential science programs for PILOT, as described in this paper (see
\S\ref{sec:scienceprograms}), their instrument requirements and the
approximate observing time they need.  The programs are divided into
the three major science categories (P = planetary, G = galaxy
environment, C = cosmology) used here. The instruments are as
described in Table~\ref{table:strawman}, with HR = High Resolution and
WF = Wide Field. The type of observation that each program requires is
listed in the fourth column (e.g.\ survey, monitoring, follow-up from
other observations, selected individual sources). An approximate time
requirement to accomplish the program is given in the last column.
\end{table}

Nevertheless, it is still a useful exercise to examine the overall
time requirements for the projects discussed in this document.  These
projects are summarised in Table~\ref{table:observe}.  Taken together,
about 3--4 years of telescope time is needed to undertake them all. In
practice, of course, as with all observing programs, this requires a
detailed assessment of their individual needs, based on the final
performance specifications achieved by the facility, and modified by
the experience gained as the program is undertaken.  Several programs,
however, would take the lions share of the observing time. Surveys for
old brown dwarfs, the determination of stellar populations in Local
Group galaxies and in tidal streams, the ultra deep field in the
near-IR searching for high redshift galaxies, and mapping of the
cosmic web of dark matter through gravitational lensing, each require
several months to be devoted to them.  A few programs need continuous
monitoring, but only over short periods of time (e.g.\ orbital debris,
planetary imaging, asteroseismology).  Most of the monitoring programs
could be simultaneously executed with the surveys by setting aside
2--3 hours each `day' for their conduct (e.g.\ exo-planet transits,
pulsar wind nebulae, SNe in starbursts, time delays in gravitational
lenses).  A few programs will need to make use of program overrides to
be undertaken, most notably the follow up of GRB detections to search
in the IR for signatures of the First Light at extremely high
redshift, and to monitor planetary microlensing candidates. Programs
which require mid-IR and sub-mm instrumentation, which are generally
the most complex to operate in view of their cryogenic requirements,
can however be conducted during daylight and so could be undertaken
during the summer period.  This particularly applies to studies of the
early stages of star formation and the chemical environment of
molecular cores.  In addition, monitoring of global atmospheric
changes in Mars and Venus also can be conducted during daylight hours,
particularly in twilight.  Characterizing orbital debris is also best
undertaken in the twilight period.  Finally, there are a number of
smaller projects, generally studying selected sources at thermal-IR
wavelengths, that would require smaller time allocations (e.g.\
searches for proto-planetary disks, studying the Galactic ecology,
signatures of the early stages of massive star formation).


\section{Beyond PILOT: ELTs and interferometers}
PILOT would be capable of a wide range of science but, as its
name suggests, the telescope is primarily envisaged as a pathfinder
for more powerful facilities to follow, able to fully-exploit the most
advantageous conditions available for ground-based astronomy on earth.
PILOT may lead to larger telescopes, from the size of the current 8\,m
class telescopes to the era of ELTs (Extremely Large Telescopes, with
apertures of 20\,m or more).  As well, PILOT may provide a path to
networks of telescopes, as envisaged with the proposed KEOPS (Vakili et al.\
2004) and Antarctic
Planet Interferometer (API; Swain et al.\ 2003), several 2--4\,m class
telescopes placed over baselines of up to one kilometre at Dome C and
operating in the thermal-IR, and the API's own pathfinder facility,
the API Science Demonstrator (API--SD; Coud\'{e} du Foresto et al.\
2003).  We briefly consider here what such facilities might be able to
accomplish.

As indicated by the performance calculations in
\S\ref{sec:sensitivity}, an 8\,m telescope at Dome C would be
more sensitive than any temperate-latitude 8\,m
telescope, by between 1 to 3 magnitudes, depending on the part of the
optical and infrared bands being compared.  It would also provide
superior spatial resolution on account of the better seeing, and a
wider wavelength coverage through opening windows in the mid-IR\@.
The performance of an Antarctic 8\,m can also be compared to that of a
temperate latitude ELT, in much the same way that we compared the
performance of the 2\,m PILOT to a temperate 8\,m telescope in
\S\ref{sec:sensitivity}.  For instance, the sensitivities of the
Antarctic 8\,m and a temperate 30\,m ELT would typically
differ by less than a magnitude (c.f.\ Tables~\ref{table:point} and
\ref{table:extended}), and the same advantages of wavelength coverage
and good seeing would apply to the Antarctic telescope as before.
Adaptive optics is critical to the future success of ELTs, for their
science drivers require not just extraordinary sensitivity, but
superlative spatial resolution as well.  As discussed in
\S\ref{sec:resolution}, the good seeing, wider isoplanatic angle and
longer coherence times at Dome C all serve to make AO operation there
significantly easier than at temperate locations, for the same level
of performance.

The science cases for these grand design facilities are predicated on
two major objectives.  The first is to conduct ultra deep field
surveys in the infrared, able to probe back through the entire history
of the Universe to the epoch of CMBR formation, in particular to the
time when the first stars formed.  The second is the detection of
exo-earths, planets like our own orbiting other stars.

The sensitivity of a telescope operating in background limited
conditions at the diffraction limit, $S_{\lambda}$, is proportional to
$D^2 \sqrt(\frac{B}{t})$ where $D$ is the telescope diameter, $B$ the
background flux and $t$ the integration time. A figure of merit of its
power to survey a field of solid angle $\Omega$ (larger than the field
of view of the telescope) in total time $T$ is thus inversely
proportional to $S_{\lambda}$ and given by $D^2 \sqrt(\frac{T
\Omega}{B})$ (see also Angel 2004).  Two performance comparisons of
relevance to the discussion may be made here.  

The first is when comparing an Antarctic 8\,m to a temperate ELT\@.
Whereas the background reduction is equivalent to needing a telescope
of half the diameter to complete the survey in the same time, it is
also notable that it is significantly easier to provide a wide field
of view on the smaller facility, and this will be reflected through
the increased cost of instrumentation for the larger facility.  

The second comparison is with future planned space facilities, such as
the 6\,m JWST (see Angel, Lawrence \& Storey 2004).  At shorter
infrared wavelengths, where the further background reduction from
Antarctica to a space-based location is less than two orders of
magnitude, a 20\,m at Dome C will be competitive to the JWST for
spectroscopic applications for $\lambda < 4$\micron.  It will also
provide far easier access for the necessarily sophisticated
instrumentation requirements for spectroscopy.  Such follow-up of
ultra deep field sources is an essential aspect of interpreting the
data that surveys from the JWST will produce.

The task of detecting exo-earths is even more formidable.  For a
sun-like star 10\,pc from the Earth, for which there are about a
dozen, an exo-earth would be separated from its parent star by
$0.1''$.  It would be $\sim 10^{-10}$ times as bright in the visible,
rising to a maximum contrast of $\sim 10^{-6}$ times around 20\micron.
It would also induce a wobble in the star's position of about 1
micro-arcsecond with a period of 1 year.  Detecting the signature of
the planet flux from within the point spread function of the star is
the challenge for ELTs. A crucial aspect of being able to do so is
extremely good correction of the atmospheric turbulence in order to
minimise the size of the point spread function (including any
scattered light) of the AO system.  The technology to accomplish such
a feat has yet to be developed, but it is possible to calculate the
performance of prospective systems.  By virtue of the superior
conditions for AO correction at Dome C, a telescope capable of
detecting exo-earths can have smaller size if built there than at a
temperate site.  Indicative calculations by Lardi\`{e}re et al.\
(2004), for example, suggest that a 15\,m telescope at Dome C might
suffice for such a detection in the visible and near-IR, whereas a
30\,m telescope would be needed from Mauna Kea.

While ELTs will provide superb sensitivity, many of the most topical
astrophysical research frontiers entail observation of matter in the
close environment of stars or in highly luminous cores -- i.e\ they
require exquisite spatial resolution to conduct. Studies of
exo-planets, star formation, stellar winds and active galactic nuclei
are all limited by the extreme dynamic ranges, high angular
resolutions and high measurement precision needed to discriminate the
faint signals against the glare of the central star.  These
considerations drive optical designs towards an interferometer,
however at mid-latitude sites there is a heavy penalty caused by the
seeing in the turbulent atmosphere.  Quantifying just how strong this
penalty will be depends in detail on the experiment, however
interferometer performance will in general be highly sensitive to all
three fundamental seeing parameters: $r_0$ (spatial), $t_0$ (temporal)
and $\theta_0$ (angular) coherence properties of the incoming
wavefront. From the Antarctic plateau, all three of these atmospheric
properties attain their most favorable values on the surface of the
earth, as described in \S\ref{sec:conditions}.

For techniques such as nulling interferometry and astrometric
interferometry, the improved atmospheric conditions should result in
orders of magnitude increase in sensitivity on top of gains already
available from the extreme low temperature and water vapour.  For
instance, as discussed in \S\ref{sec:conditions}, the mean square
astrometric error of a dual-beam interferometer depends on the
strength of the turbulence, weighted by the square of the height at
which that turbulence occurs. At temperate sites turbulence sets a
limits of around 100 micro-arcseconds in the accuracy to which
positions can be measured (Shao \& Colavita 1992).  While this may be
sufficient to detect Jupiter-type systems, it is two orders of
magnitude away from the precision needed to detect exo-earths.  The
necessary precision is within the range of an instrument operating at
Dome C (Lloyd et al.\ 2002).

Imaging nulling interferometers may not only be able to detect
exo-earths, but be able to analyse their atmospheres. This would
permit, for example, the signature of non-equilibrium chemistry, that
is readily apparent in the earth's atmosphere and induced by the
presence of biota in the environment (as indicated by the simultaneous
presence of CO$_2$, H$_2$O and O$_3$ in the mid-IR spectral window;
see Angel \& Woolf, 1997), to be detected.

The potential for Antarctic interferometry is fundamentally different
from that at a conventional site.  The proposed Antarctic Plateau
Interferometer could be operating before the immensely more expensive
space missions, and so begin the task of characterising exo-planets
and searching for objects down to earth-masses (see also Storey et
al.\ 2002, Storey 2004).  Furthermore, a wealth of other stellar
astrophysics could be addressed using such a device. Observations of
disks around young stellar objects should reveal substructure such as
spiral density waves and gap clearing due to planet growth.  Mass loss
phenomena in evolved stars, and substructure such as disks and jets
within dusty compact micro-quasars, will be within reach of imaging
observations. Studies of fundamental stellar properties -- sizes,
effective temperatures, distances and (using binary stars) masses --
will mean that almost every branch of stellar astronomy will benefit.

The roadmap towards the construction of a fully-capable Antarctic
interferometer is being articulated through the proposal to build the
API--SD\@. The first phase aims to ensure that the required
infrastructure for an interferometer is constructed at Dome C\@.  In
the second phase, the modest initial 2-element 40\,cm optics of the
Science Demonstrator would be upgraded to a number (up to six)
2--4\,m-sized telescopes.  PILOT would have direct and immediate
utility to the API project.  It could be used as one array element on a
part-time basis for dedicated projects, or used to perform initial
sky-testing and sensitivity measurements before the upgrade path to
larger apertures is finalized. Furthermore, PILOT and API have many
common requirements apart from being winterized 2\,m
telescopes. Nulling interferometry, for example, will require delivery
of high and stable Strehl ratios from the adaptive optics system, a
feature which can be verified with PILOT\@.

\section{Conclusions}
Modest-sized telescopes, built for optical and infrared imaging and
placed at Dome C, would be able to perform a wide range of
competitive science for a fraction of the cost incurred by larger
facilities that are built at temperate locations.  Wide-field infrared
imaging surveys, particularly in the K, L and M bands (from 2.3 to
5\micron), conducted with near-diffraction limited resolution, provide
the greatest performance gains for Antarctic telescopes.
Near-diffraction limited imaging could also be obtained at optical
wavelengths with a 2\,m-sized telescope, a regime not attempted from
temperate sites because of their adverse site seeing characteristics.  An
Antarctic telescope can also explore the cosmos through new
ground-based windows, from 17 to 40\micron\ and at 200\micron.

The performance characteristics we describe for such an Antarctic
telescope assist science projects aimed at studying planetary
systems, star formation within the Milky Way, the formation and
evolution of the galaxies in our Local Group, the star formation
history of the Universe from the time of formation of the first
massive stars, and the structure of the cosmic web of dark matter.

PILOT, the Pathfinder for an International Large Optical Telescope,
the 2\,m-sized telescope described in this paper as a facility able to
tackle this diverse range of science, is, however, only designed as a
pathfinder.  Antarctica offers an opportunity to pursue projects that
would require either much larger facilities if they were sited
elsewhere, or need considerably more expensive space facilities.
Before such projects can be contemplated, it has to first be
demonstrated that it is indeed possible to operate the complex
instrumentation needed for the science from Antarctica. PILOT's
principal aim is to do this, by showing that the special conditions of
the Antarctic plateau can indeed be utilised for the conduct of
sophisticated experiments.  PILOT may indeed serve as a prototype
towards the construction of large optical and infrared telescopes in
Antarctica, even the so-called ELTs.  However PILOT will equally well
serve as a technology demonstrator towards the construction of
infrared interferometers, operating with kilometre-sized baselines.
Both these kinds of facilities would be able to further two of the
grand design astrophysical projects being pursued today: the imaging
of ultra deep fields to uncover the complete star formation history of
the Universe from the time of the First Light, and the search for, and
subsequent characterization of, other earth-like planets in the
Galaxy.

\section*{Acknowledgements}
Many of our colleagues have contributed to this paper through vigorous
scientific discussions, and we wish to acknowledge them for the ideas
this has help generated.  Our thanks go to David Bennett, Paolo
Calisse, Jessie Christiansen, Jon Everett, Suzanne Kenyon, Will
Saunders and Peter Wood.  We also thank the Australian Research
Council for funding support through the Discovery Grant and Australian
Postdoctoral Fellowship programs, the Australian Antarctic Division
for a postgraduate scholarship and the University of New South Wales
for its continued support of the Antarctic astronomy program.  We also
thank the referee, Gerry Gilmore, for his helpful comments on the
paper.

\section*{References}

Abe, F. et al.\ (35 authors) 2004, Science, 305, 1264

\reference Andr\'e, P., Motte, F. \& Bacmann, A. 1999, ApJL, 513, L57

\reference Angel, J.R.P. \& Woolf, N.J. 1997, ApJ, 475, 373

\reference Angel, R., 2004, Proc.\ SPIE, Astronomical Telescopes and 
Instrumentation Symposium, Glasgow, Scotland, June 2004 , in press

\reference Angel, R., Lawrence, J. \& Storey, J.,  2004, Proc.\ SPIE,
5382, 76

\reference Aristidi, E., Agabi, K., Azouit, M., Fossat, E., Vernin, J.,
Travouillon, T., Lawrence, J.S., Meyer, C., Storey, J.W.V., Halter,
B., Roth, W.L. \& Walden, V. 2005, A\&A, 430, 739

\reference Aristidi, E., Agabi, K., Vernin, J., Azouit, M., Martin, F., 
Ziad, A. \& Fossat, E. 2003, A\&A, 406, L19

\reference Ashley, M.C.B., Burton, M.G., Calisse, P.G., Phillips, A. \&
Storey, J.W.V. 2004a, in IAU Pub. Highlights in Astronomy, 13, 
eds. O. Engvold \& M. G. Burton (Ast.\ Soc.\ Pac.), in press

\reference  Ashley, M.C.B., Burton, M.G., Lawrence, J.S. \& Storey, J.W.V.
2004b. Astron. Nachr., 325, 619

\reference Ashley, M.C.B., Burton, M.G., Storey, J.W.V., Lloyd, J.P., 
Bally, J., Briggs, J.W. \& Harper, D.A. 1996, PASP, 108, 721

\reference Bacon, D.J., Massey, R.J., Refregier, A.R. \& Ellis,
R.S. 2003, MNRAS, 344, 673

\reference Bailey, J., Chamberlain, S., Walter, M. \& Crisp, D. 2004, 
Proc.\ 3rd European Workshop in Exo-Astrobiology, ESA-SP 545,
7. eds. A. Harris \& L. Ouwehand

\reference Baldwin, J.E., Tubbs, R.N., Cox, G.C., Mackay, C.D., Wilson, R.W. \&
Andersen, M.I. 2001, A\&A, 368, L1

\reference Bartelmann, M., Huss, A., Colberg, J.M., Jenkins, A. \&
Pearce, F.R. 1998, A\&A, 330, 1

\reference Bartelmann, M., Meneghetti, M., Perrotta, F., Baccigalupi,
C. \& Moscardini, L. 2003, A\&A, 409, 449

\reference Bedding, T.R. \& Kjeldsen, H. 2003, PASA 20, 203

\reference Benjamin, R.A. et al.\ (20 authors) 2003, PASP, 115, 953


\reference Bennett, D.P. et al.\ (24 authors) 2004, Proc.\ SPIE, 
Astronomical Telescopes and Instrumentation Symposium, Glasgow,
Scotland, June 2004, in press

\reference Bergin, E.A., Alves, J., Huard, T. \& Lada, C.J.  2002, ApJ, 570, L101

\reference Bond, I.A. et al.\ (30 authors) 2002, MNRAS, 333, 71

\reference Bond, I.A. et al.\ (32 authors) 2004, ApJ, 606, L155

\reference Bonnell, I.A. \& Bate, M.R. 2002, MNRAS, 336, 659

\reference Brown T.M. 2003, ApJ, 593, L125

\reference Brown, T.M., Ferguson, H.C., Smith, E., Kimble, R.A., Sweigart, A.V.,
Renzini, A., Rich, R. \& VandenBerg, D.A. 2003, ApJ, 592, L17

\reference Burrows, A., Hubbard, W.B., Lunine, J.I. \& Liebert, J. 2001,
Rev.\ Mod.\ Phys., 73, 719

\reference Burton, M.G., Storey, J.W.V. \& Ashley, M.C.B. 2001, PASA, 18, 158

\reference Burton, M.G. et al.\ (20 authors) 1994, PASA, 11, 127

\reference Burton, M.G., et al.\ (14 authors) 2000, ApJ, 542, 359

\reference Caldwell, D.A., Borucki, W.J., Showen, R.L., Jenkins, J.M., Doyle, L.,
Ninkov, Z. \& Ashley, M. 2004, Bioastronomy 2002: Life Among the
Stars, Proc. IAU Symp.\ 213, 93, eds. R. Norris \& F. Stootman

\reference Calisse, P.G., Ashley, M.C.B., Burton, M.G., Phillips, M.A, 
Storey, J.W.V., Radford, S.J.E. \& Peterson, J.B. 2004, PASA, 21, 256

\reference Candidi, M. \& Lori, A. 2003, Mem.\ Soc.\ Ast.\ It., 74, 29

\reference Cen, R. 2003, ApJ, 591, 12

\reference Chamberlain, M.A., Ashley, M.C.B., Burton, M.G., Phillips, M.A.,
\& Storey, J.W.V. 2000, ApJ, 535, 501

\reference Chamberlain, S., Bailey, J. \& Crisp, D. 2004, PASA, submitted

\reference Chamberlin, R.A., Lane, A.P. \& Stark, A.A. 1997, ApJ, 476, 428

\reference Chamberlin, R.A., Martin, R.N., Martin, C.L. \& Stark, A.A. 2003,
Proc.\ SPIE, 4855, 9

\reference Chanover, N.J., Glenar, D.A. \& Hillman, J.J. 1998, JGR, 103, 31355

\reference Christensen-Dalsgaard, J. 2002, Rev.\ Modern Phys.\ 74, 1073

\reference Clough, S.A. \& Iacono, M.J. 1995, JGR, 100, 16519

\reference Coud\'{e} du Foresto, V., Swain, M., Schneider, J. \& Allard, F. 2003,
Mem.\ Soc.\ Ast.\ It.\ Supp., 2, 212

\reference Dantowitz, R.F., Teare, S.W. \& Kozubal, M.J. 2000, AJ, 119, 2455

\reference Dempsey, J.T., Storey, J.W.V. \& Phillips, M.A. 2005, PASA, 22, 91

\reference Dickinson, M. et al.\ (14 authors) 2000, ApJ, 531, 624

\reference Doherty, M., Bunker, A., Sharp, R., Dalton, G., Parry, I.,
Lewis, I., MacDonald, E., Wolf, C. \& Hippelein, H. 2004, MNRAS, 354,
7

\reference Drake, A.J. \& Cook, K.H. 2004, ApJ, 604, 379

\reference Egan, M.P., van Dyk, S.D. \& Price, S.D. 2001, AJ, 122, 1844

\reference Erb, D.P., Shapley, A.E., Steidel, C.C., Pettini, M., Adelberger,
K.L., Hunt, M.P., Moorwood, A.F.M. \& Cuby, J. 2003, ApJ, 591, 101

\reference Ferguson, A.M.N., Irwin, M.J., Ibata, R.A., Lewis, G.F. \& Tanvir, N.R.
2002, AJ, 124, 1452

\reference Forget, F., Hourdin, F., Fournier, R., Hourdin, C., Talagrand, O.,
Collins, M., Lewis, S.R., Read, P.L. \& Huot, J.-P. 1999, JGR-Planet,
104, 24155

\reference Fossat, E. 2003, Mem.\ Soc.\ Ast.\ It.\ Sup., 2, 139

\reference Fowler, A. M., Sharp, N., Ball, W., Schinckel, A. E. T., Ashley, 
M. C. B., Boccas, M. Storey, J. W. V., Depoy, D., Martini, P., Harper
D. A. \& Marks, R. D., 1998, Proc.\ SPIE, 3354, 1170
        
\reference Franx, M., Labb, I., Rudnick, G., van Dokkum, P.G., Daddi, E.,
Schreiber, N.M., Moorwood, A., Rix, H., Rottgering, H., van de Wel,
A., van der Werf P. \& van Starkenburg, L. 2003, ApJ, 587, L79

\reference Freedman, W.L. et al.\ (15 authors) 2001, ApJ 533, 47

\reference Freeman, K. \& Bland-Hawthorn, J. 2002, ARAA, 40, 487

\reference Gibb, E.L. et al.\ (11 authors) 2000, ApJ, 536, 347

\reference Gladders, M.D., Hoekstra, H., Yee, H.K.C., Hall, P.B. \&
Barrientos, L.F. 2003, ApJ, 593, 48

\reference Glazebrook K., Abraham R.G., McCarthy P.J., Savaglio S., Chen H.,
Crampton D., Murowinski R., Jorgensen I., Roth K., Hook I., Marzke
R.O. \& Carlberg R.G. 2004, Nature, 430, 181

\reference Glazebrook K., Blake C., Economou F., Lilly S. \& Colless M. 
1999, MNRAS, 306, 843

\reference Gould, A. \& Loeb, A. 1992, ApJ, 396, 104

\reference Gray, M.E., Taylor, A.N., Meisenheimer, K., Dye, S., Wolf,
C. \& Thommes, E. 2002, ApJ, 568, 141

\reference Griest, K. \& Safizadeh, N. 1998, ApJ, 500, 37

\reference Hereld, M. 1994, in Astrophys. \& Sp.\ Sci.\ Lib., 190,
`Infrared Astronomy with Arrays, the Next Generation', ed. 
I. McLean (Kluwer Acad.\ Pub.), 248

\reference Hester, J.J. et al.\ (9 authors) 2002, ApJ, 577, L49

\reference Hidas, M.G., Burton, M.G., Chamberlain. M.A. \& Storey, J.W.V., 
2000, PASA, 17, 260

\reference Hidas, M.G., Irwin, M., Ashley, M.C.B., Webb, J.K., Phillips, A., 
Toyozumi, H., Derekas, A., Christiansen, J., Crothers, S. \& Nutto, C.
2004, in preparation

\reference Hoekstra, H., Franx, M., Kuijken, K. \& Squires, G. 1998,
ApJ, 504, 636

\reference Hoekstra, H., Yee, H.K.C. \& Gladders, M.D. 2002, NewAR, 46,
767

\reference Hoekstra, H., Yee, H.K.C. \& Gladders, M.D. 2004, ApJ, 606,
67

\reference Hopkins A.M., Connolly A.J. \& Szalay A.S. 2000, AJ, 120, 2843

\reference Ibata, R., Irwin, M., Lewis, G., Ferguson, A.M.N. \& Tanvir, N. 2001,
Nature, 412, 49
	    
\reference Ibata, R., Chapman, S., Ferguson, A.M.N., Irwin, M., Lewis, G. \&
McConnachie, A.W. 2004, MNRAS, 351, 117

\reference Indermuehle, B.T., Burton, M.G. \& Maddison, S.T. 2005, PASA, 22, 73

\reference Ivezi\'{c}, Z. \& Elitzur, M. 1997, MNRAS, 287, 799

\reference Juneau S., Glazebrook K., Crampton D., Abraham R.G., McCarthy P.J.,
Savaglio S., Chen H., Murowinski R., Jorgensen I., Roth K., Hook I.,
Marzke R.O. \& Carlberg R.G. 2005, ApJ, 619, L135

\reference Kaiser N. \& Squires G., 1993, ApJ, 404, 441

\reference Kaiser, N., Squires, G. \& Broadhurst, T. 1995, ApJ, 449, 460

\reference King, L.J., Clowe, D.I., Lidman, C., Schneider, P., Erben,
T., Kneib, J.-P. \& Meylan, G. 2002, A\&A, 385, 5

\reference Kirkpatrick, J.D., Allard, F., Bida, T., Zuckerman, B., Becklin, E.E., 
Chabrier, G. \& Baraffe, I. 1999, ApJ, 519, 802

\reference Kjeldsen H. \& Frandsen S. 1992, PASP, 104, 413

\reference Kochanek, C. 2002, ApJ 578, 25

\reference Kochanek, C. \& Schechter, P. 2004, in `Measuring and Modeling the
Universe', Carnegie Ob.\ Ast. Ser., p117, C.U.P., ed. W.L. Freedman

\reference Lada, C. 1999a, in NATO ASIC Proc.\ 540, The Origin of Stars and Planetary
Systems, p143

\reference Lada, E. 1999b, in NATO ASIC Proc.\ 540, The Origin of Stars and Planetary
Systems, p441

\reference Lamb D.Q. \& Reichart D.E. 2001, Proc.\ 20th Texas Symp.\
on Relativistic Astrophysics, AIPC, 586, 605

\reference Lane, A.P. 1998, ASP Conf.\ Proc.\ 141, 289. eds. G. Novak \&
R.H. Landsberg

\reference Lane, A.P. \& Stark, A.A. 1997, Ant.\ J.\ US, 30, 377

\reference Lardi\`{e}re, O., Salinari, P., Jolissaint, L., Carbillet, M.,
Riccardi, A. \& Esposito, S. 2004, Proc.\ SPIE, 5382, 550

\reference Lawrence, J. 2004a, PASP, 116, 482

\reference Lawrence, J. 2004b, Applied Optics, 43, 1435

\reference Lawrence, J.S., Ashley, M.C.B., Tokovinin, A. \& Travouillon, T.,
2004, Nature, 431, 278

\reference Lewis, G.F. \& Belle, K.E. 1998, MNRAS, 297, 69

\reference Lewis, G.F., Irwin, M.J., Hewett, P.C. \& Foltz, C.B. 1998, MNRAS, 295, 573

\reference Lilly S.J., Le Fevre O., Hammer F. \& Crampton D. 1996, ApJ, 460,  L1

\reference Lloyd, J.P., Oppenheimer, B.R. \& Graham, J.R. 2002, 
PASA, 19, 318

\reference Longmore, S.N., Burton, M.G., Minier, V. \& Walsh, A.J. 2005, 
MNRAS, submitted

\reference Madau P., Ferguson H.C., Dickinson M.E., Giavalisco M., Steidel
C.C. \& Fruchter A. 1996, MNRAS, 283, 1388

\reference Maercker, M., Burton, M.G. \& Wright, C. 2005, A\&A, submitted

\reference Mao, S. \& Paczynski, B. 1991, ApJ, 374, L37

\reference Marcy, G.W. \& Butler, R.P. 1998, ARAA, 36, 57

\reference Marks, R.D. 2002, A\&A, 385, 328

\reference Marks, R.D., Vernin, J., Azouit, M., Briggs, J.W., Burton, M.G., 
Ashley, M.C.B. \& Manigault, J.F. 1996, A\&AS, 118, 385

\reference Marks, R.D., Vernin, J., Azouit, M., Manigault, J.F. \& 
Clevelin, C. 1999, A\&AS, 134, 161

\reference Martin, C.L., Walsh, W.M., Xiao, K., Lane, A.P., Walker, C.K., 
\& Stark, A.A. 2004, ApJS, 150, 239

\reference Mattila, S. \& Meikle, W.P.S. 2001, MNRAS, 324, 325

\reference McConnachie, A.W., Irwin, M.J., Lewis, G.F., Ibata, R.A., Chapman,
S.C., Ferguson, A.M.N. \& Tanvir, N.R. 2004, MNRAS, 351, 94

\reference McKee, C.F. \& Tan, J.C. 2002, Nature, 416, 59

\reference Mediavilla, E. et al.\ (11 authors) 1998, ApJ, 503, 27

\reference Melatos, A., Scheltus, D., Whiting, M., Eikenberry, S., Romani,
R., Rigaut, F., Spitovsky, A., Arons, J. \& Payne, D. 2004, ApJ, submitted

\reference Minier, V., Burton, M.G., Hill, T., Pestalozzi, M.R., Purcell, C., 
Longmore, S., Garay, G. \& Walsh, A. 2005, A\&A, 429, 945

\reference Motta, V., Mediavilla, E., Munoz, J.A. \& Falco, E. 2004, ApJ, 613, 86

\reference Nguyen, H.T., Rauscher, B.J., Severson, S.A., Hereld, M., Harper,
D.A., Lowenstein, R.F., Morozek, F. \& Pernic, R.J. 1996, PASP, 108,
718

\reference Oppenheimer, B.R., Kulkarni, S.R., Matthews, K. \& van Kerkwijk, M.H.
1995, Science, 270, 1478

\reference Paczynski, B. 1986. ApJ, 304, 1

\reference Payne, J.M. 2002, ASP Conf. Ser., 278. Single-Dish Radio
Astronomy: Techniques and Applications, 453


\reference Phillips, M.A., Burton, M.G., Ashley, M.C.B., Storey, J.W.V., 
Lloyd, J.P., Harper, D.A. \& Bally, J. 1999, ApJ 527, 1009

\reference Pojmanski, G. 2002, Acta Astronomica, 52, 397

\reference Ragazzoni, R. et al. (14 authors) 2004, Proc.\ SPIE, 5489, 481

\reference Refregier, A.R. 2003, ARAA, 41, 645

\reference Rhie, S.H. et al.\ (41 authors) 2000, ApJ, 533, 378

\reference Rhoads, J.E. et al.\ (9 authors) 2004, ApJ, 611, 59

\reference Ruhl, J. et al.\ (27 authors), 2004. Proc.\ SPIE, 5498, 11

\reference Ryder, S.D., Sadler, E.M., Subrahmanyan, R., Weiler,
K.W., Panagia, N. \& Stockdale, C. 2004, MNRAS, 349, 1093

\reference Shao, M. \& Colavita, M.M. 1992, A\&A, 262, 353

\reference Staguhn, J.G. et al.\ (14 authors) 2003, Proc.\ SPIE, 4855, 100

\reference Stark, A.A. 2002, AIP Conf.\ Proc., 616, Experimental Cosmology 
at Millimetre Wavelengths, 83

\reference Stark, A.A. 2003, The Future of Small Telescopes In The New
Millennium, Vol.\ II: The Telescopes We Use, 269

\reference Stark, A.A., Bolatto, A.D., Chamberlin, R.A., Lane, A.P., 
Bania, T.M.,
Jackson, J.M. \& Lo, K.--Y. 1997, ApJ, 480, L59

\reference Steidel C.C., Adelberger K.L., Giavalisco M.,
Dickinson M. \& Pettini M. 1999, ApJ, 519,  1

\reference Storey, J.W.V. 2004, Proc.\ SPIE, 5491, 169

\reference Storey, J.W.V., Ashley, M.C.B., Lawrence, J.S. \& Burton,
M.G. 2003, Mem.\ Soc.\ Ast.\ It., 2, 13

\reference Storey, J.W.V., Burton, M.G. \& Ashley, M.C.B. 2002, Proc.\ SPIE, 4835, 110

\reference Swain, M.R., Coud\'{e} du Foresto, V., Fossat, E. \& Vakili, F. 2003,
Mem.\ Soc.\  Ast.\ It.\ Supp, 2, 207

\reference Sudbury Neutrino Observatory Collaboration (179 authors) 2002,
Phys.\ Rev.\ Lett., 89, 011301

\reference Swinbank, A.M., Smail, I., Bower, R.G., Borys, C., Chapman,
S.C., Blain, A.W., Ivison, R.J., Howat, S.R., Keel, W.C. \& Bunker,
A.J. 2005, MNRAS, in press

\reference Tingley, B. 2004, A\&A, 425, 1125

\reference Travouillon, T., Ashley, M.C.B., Burton, M.G., Storey, 
J.W.V. \& Loewenstein, R.F. 2003a, A\&A, 400, 1163

\reference Travouillon, T., Ashley, M.C.B., Burton, M.G., Storey, 
J.W.V., Conroy, P., Hovey, G., Jarnyk, M., Sutherland, R. \&
Loewenstein, R.F. 2003b, A\&A, 409, 1169

\reference Vakili, F. et al.\ 2004, Proc.\ SPIE, 5491, 1580

\reference van Dishoeck, E.F. 2004, ARAA, 42, 119

\reference van Loon, J. Th. 2000, A\&A, 354, 125

\reference Walden, V.P., Town, M.S., Halter, B. \& Storey, J.W.V. 2005, PASP,
117, 300

\reference Wambsganss, J., Bode, P. \& Ostriker, J.P. 2004, ApJ, 606, 93

\reference Ward-Thompson, D., Andr\'e, P. \& Kirk, J.M. 2002, MNRAS, 329, 257

\reference Wayth, R.B., Warren, S.J., Lewis, G.F. \& Hewett, P.C. 2005, 
MNRAS, submitted

\reference Wood, P.R. 1998, A\&A, 338, 592

\reference Yan, L., McCarthy, P.J., Freudling, W., Teplitz, H.I., 
Malumuth, E.M., Weymann, R.J. \& Malkan, M.A. 1999, ApJ, 519, L47

\reference Yan, H. \& Windhorst, R.A. 2004, ApJ, 612, 93

\vfill\eject

\end{document}